\pdfoutput=1
 
\newcommand*{\ATLASLATEXPATH}{}
\documentclass[UKenglish, cernpreprint, texlive=2016]{\ATLASLATEXPATH atlasdoc}
 
\usepackage[backend=biber]{\ATLASLATEXPATH atlaspackage}
\usepackage{\ATLASLATEXPATH atlasphysics}

\graphicspath{{logos/}{figures/EBplots_eps/}{figures/}}

\usepackage{ANA-STDM-2017-34-PAPER-defs}
\usepackage{caption}
\usepackage{graphicx}
\usepackage{verbatimbox}
\usepackage{microtype}
 
\usepackage{multirow}
 
\usepackage{epstopdf}

\AtlasTitle{
Measurement of jet-substructure observables in top quark, $W$ boson and light jet production in proton-proton collisions at $\sqrt{s}=13$ TeV with the ATLAS detector
}

\author{The ATLAS Collaboration}
 
\usepackage{authblk}
 
\makeatletter
\renewcommand\AB@affilsepx{, \protect\Affilfont}
\makeatother

\renewcommand\Affilfont{\itshape\small}

\PreprintIdNumber{CERN-EP-2019-011}
\AtlasRefCode{STDM-2017-34}

\AtlasJournalRef{JHEP 08 (2019) 033}
\AtlasDOI{10.1007/JHEP08(2019)033}
 
\AtlasAbstract{
A measurement of jet substructure observables is presented
using data collected in 2016  by the ATLAS experiment at the LHC with proton-proton collisions at $\sqrt{s}=13$~TeV.
Large-radius jets groomed with the trimming and soft-drop algorithms are studied.
Dedicated event selections are used to study jets produced by light quarks or gluons,
and hadronically decaying top quarks and $W$ bosons.
The observables measured are sensitive to substructure,
and therefore are typically used for tagging large-radius jets from boosted massive particles.
These include the energy correlation functions
and the $N$-subjettiness variables.
The number of subjets and the Les Houches angularity are also considered.
The distributions of the substructure variables, corrected
for detector effects, are compared to the
predictions of various Monte Carlo event generators.
They are also compared
between the large-radius jets originating from light quarks or gluons,
and hadronically decaying top quarks and $W$ bosons.
}

\hypersetup{pdftitle={Measurement of jet-substructure observables with the ATLAS detector at $\sqrt{s} = 13$ TeV},pdfauthor={}}

\begin{document}
 
\maketitle

\section{Introduction}
\label{sec:intro}

Increasing the centre-of-mass energy of proton--proton ($pp$) collisions from 7 and 8~\TeV\ in Run 1 to 13~\TeV\. in
Run 2 of the Large Hadron Collider (LHC) leads to a larger fraction of heavy particles such as top quarks,
vector bosons and Higgs bosons being produced with large transverse momenta.
This large transverse momentum leads to collimated decay products.
They are usually reconstructed in a large-radius jet, whose internal (sub)structure shows interesting features that can be
used to identify the particle that initiated the jet formation~\cite{Abdesselam:2010pt,Aaboud:2018psm}.
 
This is relevant for a host of measurements and searches,
which involve identifying the large-radius jets coming from top
quarks~\cite{Aaboud:2017fha, Aaboud:2018eqg, Aaboud:2017mae, Aaboud:2019roo, Aaboud:2018juj}.
or Higgs bosons~\cite{Aaboud:2017rss, Aaboud:2018zhk, Aaboud:2018knk, Aaboud:2018zhh}, for example in Run 2 in ATLAS.
Usually a two step procedure is employed.
In the first step, termed grooming, the effect of soft, uncorrelated radiation contained in the large-radius jet in reduced.
Then jet substructure observables, which describe the spatial energy distribution inside the jets, are used to
classify the jets originating from different particles. This process is called jet tagging and the algorithms are referred to as taggers.
 
Most of the grooming algorithms and jet substructure observables were
developed on the basis of theoretical calculations or Monte Carlo (MC) simulation programs and then t
hey are applied to data. Given that often large differences have been seen between predictions from MC and data,
large correction factors need to be applied to simulation results.
Additionally, taggers suffer from large systematic uncertainties as the modelling of the
substructure observables is not well constrained~\cite{Gras:2017jty,Aaboud:2018psm}.
Most of these variables have never been measured in data, and
performing a proper unfolded measurement is a common request from the theory community.
Measuring these observables will help in optimising and developing current and future substructure taggers,
as well as tuning hadronization models in the important but still relatively unexplored regime of jet substructure.
The choice of variables measured in this paper prioritized jet shapes commonly used in jet tagging, as well as those most useful for model tuning.

The ATLAS Collaboration has performed measurements of jet mass and substructure variables at the $pp$
centre-of-mass energies of $\sqrt{s}=7$, $8$
and $13$~\TeV~\cite{ATLAS:2012am, Aad:2012meb, TOPQ-2014-09, STDM-2014-17, Aad:2016oit, Aaboud:2017qwh, Aaboud:2018ibj}
in inclusive jet events, and
the CMS Collaboration has performed measurements of jet mass and substructure in dijet, $W$/$Z$ boson,
and \ttbar events~\cite{Chatrchyan:2012mec, Chatrchyan:2013vbb, Sirunyan:2017tyr, Sirunyan:2018xdh, Sirunyan:2018asm} at
$\sqrt{s}=7$, $8$ and $13$~\TeV.
This paper presents measurements of substructure variables in large-radius jets produced in
inclusive multijet events and in \ttbar events at $\sqrt{s}=13$~\TeV\ using
$33$~fb$^{-1}$ of data collected in 2016 by the ATLAS experiment.
In this analysis, the lepton+jets decay mode of \ttbar events is selected, where
one $W$ boson decays into a muon and a neutrino, and the other $W$
boson decays into a pair of quarks. Then the large-radius jets are
separated into those that contain all the decay products of a hadronically top quark
and those containing only hadronic $W$ boson decay products.

The contents of this paper are organised as follows. First, a description of the ATLAS detector
is presented in \SecRef{sec:atlasdet} and then the MC samples used in the analysis are discussed in \SecRef{sec:mc}.
In \SecRef{sec:sel},
event and object selections are summarised. The measured jet substructure observables are
defined in \SecRef{sec:obs}. The background estimation is described in \SecRef{sec:bg} and
the systematic uncertainties are assessed in \SecRef{sec:syst}.
In \SecRef{sec:det-res}, detector-level mass and \pT distributions corresponding to selected
large-radii jets are shown, and the unfolding is described in \SecRef{sec:unf}.
Finally, the unfolded results are presented in \SecRef{sec:res}, and the conclusions in \SecRef{sec:conc}.

\section{ATLAS detector}
\label{sec:atlasdet}

The ATLAS experiment uses
a multipurpose particle detector~\cite{:2008zzm, Capeans:1291633} with a forward--backward symmetric
cylindrical geometry and a near $4\pi$ coverage in solid angle.\footnote{
ATLAS uses a right-handed coordinate system with its origin at the nominal interaction point (IP)
in the centre of the detector and the $z$-axis along the beam pipe.
The $x$-axis points from the IP to the centre of the LHC ring, and the $y$-axis points upwards.
Cylindrical coordinates $(r,\phi)$ are used in the transverse plane, $\phi$ being the azimuthal angle around the $z$-axis.
The pseudorapidity is defined in terms of the polar angle $\theta$ as $\eta = -\ln \tan(\theta/2)$.
An angular separation between two objects is defined as $\Delta R \equiv \sqrt{(\Delta\eta)^{2} + (\Delta\phi)^{2}}$,
where $\Delta\eta$ and $\Delta\phi$ are the separations in $\eta$ and $\phi$.
Momentum in the transverse plane is denoted by \pT.}
It consists of an inner tracking detector (ID) surrounded by a thin superconducting solenoid providing a $2$~T axial
magnetic field, electromagnetic (EM) and hadron calorimeters, and a muon spectrometer.
The ID consists of silicon pixel, silicon microstrip, and straw-tube transition-radiation tracking detectors, covering
the pseudorapidity range $|\eta| < 2.5$.
The calorimeter system covers the pseudorapidity range $|\eta| < 4.9$.
Electromagnetic calorimetry is performed with barrel and endcap
high-granularity lead/liquid-argon (LAr) sampling calorimeters, within the region $|\eta| < 3.2$.
There is an additional thin LAr presampler covering $|\eta| < 1.8$, to correct for
energy loss in material upstream of the calorimeters.
For $|\eta| < 2.5$, the LAr calorimeters are divided into three layers in depth.
Hadronic calorimetry is performed with a steel/scintillator-tile calorimeter,
segmented into three barrel structures within $|\eta| < 1.7$, and two copper/LAr hadronic endcap calorimeters,
which cover the region $1.5 < |\eta| < 3.2$.
The forward solid angle up to $|\eta| = 4.9$ is covered by copper/LAr and tungsten/LAr calorimeter
modules, which are optimised for energy measurements of electrons/photons and hadrons, respectively.
The muon spectrometer consists of separate trigger and high-precision tracking chambers
that measure the deflection of muons in a magnetic field generated by superconducting air-core toroids.

The ATLAS detector selects events using a tiered trigger system~\cite{Aaboud:2016leb}.
The first level is implemented in custom electronics.
The second level is implemented in software running on a general-purpose processor
farm which processes the events and reduces the rate of recorded events to 1~kHz.

\section{Monte Carlo samples}
\label{sec:mc}

Simulated events are used to optimise the event selection, correct the data for detector effects and
estimate systematic uncertainties.
The predictions of different phenomenological
models implemented in the Monte Carlo (MC) generators are compared with the data corrected
to the particle level (i.e.\ observables constructed from final-state particles within the detector acceptance).
 
The generators used to produce the samples are listed in \TabRef{tab:mc}.
The dijet (to obtain multijet events), \ttbar and single-top-quark samples are considered to be signal processes in this analysis,
corresponding to the dedicated selections.
The background is estimated using $Z/W$+jets and diboson samples.
The $t\overline{t}$ samples are scaled to next-to-next-to-leading order (NNLO)
in perturbative QCD, including soft-gluon resummation to next-to-next-to-leading-log order
(NNLL)~\cite{Czakon:2011xx} in cross-section,
assuming a top quark mass $m_{t} =172.5$~\GeV.
The \powheg model~\cite{Frixione:2007nw} resummation damping parameter, $h_{\textrm{damp}}$,
which controls the matching of matrix elements to parton showers and regulates
the high-\pT radiation, was set to $1.5m_{t}$~\cite{ATL-PHYS-PUB-2016-020}.
The single-top-quark~\cite{Re:2010bp,Alioli:2009je,Frederix:2012dh, Kidonakis:2011wy,Kidonakis:2010ux,Kidonakis:2010tc}
and $W/Z$ samples~\cite{Anastasiou:2003ds} are scaled to the NNLO
theoretical cross-sections.

The predicted shape of jet substructure distributions depends on the
modelling of final-state radiation (FSR), and fragmentation and hadronisation, as well
as on the merging/matching between matrix element (ME) and parton shower (PS)
generators. The \pythiaeight and the \sherpa generators use a dipole shower ordered in transverse momentum,
with the Lund string~\cite{Andersson:1983ia} and cluster hadronisation model~\cite{Marchesini:1991ch}
respectively.
The \herwigseven generator uses an angle-ordered shower, with the cluster hadronisation model.
For comparison purposes in dijet events, a sample was generated with \sherpa using
the string hadronisation model.

\begin{table*}[ht]
\begin{center}
\scriptsize
\begin{tabular}{llllll}
\toprule
Process & Generator &  Version &  PDF & Tune  & Use \\
\midrule
Dijet & \pythiaeight~\cite{Sjostrand:2007gs,Sjostrand:2014zea} &  8.186 &  NNPDF23LO~\cite{Ball:2012cx} & A14~\cite{ATL-PHYS-PUB-2014-021} & Nominal for unfolding   \\
& \sherpa~\cite{Gleisberg:2008ta} &  2.2.1 & CT10~\cite{Pumplin:2002vw}  & Default & Validation of unfolding\\
&                                                     &           &                                                 &              & (with two different hadronisation models) \\
& \herwigseven~\cite{Bellm:2015jjp}  & 7.0.4  & MMHT2014  & H7UE~\cite{Bellm:2015jjp} & Comparison \\
\midrule
$t\overline{t}$  & \powheg~\cite{Alioli:2010xd} & v2 & NNPDF30NLO &  & Nominal for unfolding\\
& ~ + \pythiaeight  & 8.186  & NNPDF23LO & A14 & \\
&   \powheg & v2 & CT10 &  & Validation of unfolding \\
& ~ +\herwigpp~\cite{Bahr:2008pv} & 2.7 & CTEQ6L1 & UE-EE-5 tune~\cite{Seymour:2013qka} \\
&   \powheg & v2 & CT10 &  & Comparison \\
& ~ +\herwigseven & 7.0.4  & MMHT2014  & H7UE &\\
& \amcnlo~\cite{Alwall:2014hca} & 2.6.0 & NNPDF30NLO &  & Comparison\\
& ~ + \pythiaeight  & 8.186  & NNPDF23LO & A14 & \\
&   \sherpa  &  2.2.1 & CT10   & Default & Comparison \\
Single top & \powheg & v1 &  CT10 &  & Nominal for unfolding \\
& ~ + \pythiasix~\cite{Sjostrand:2000wi,Sjostrand:2006za}  & 6.428  &  CTEQ6L1~\cite{Pumplin:2002vw} &   Perugia2012~\cite{Skands:2010ak} \\
\midrule
$Z$+jets  & \sherpa & 2.2.1& CT10 & Default  & Background estimation \\
$W$+jets & \sherpa & 2.2.1 & CT10 & Default  & Background estimation (nominal)\\
$W$+jets &  \amcnlo & 2.2.5 & CT10 &   & Background estimation (cross-check) \\
& ~ + \pythiaeight  & 8.186  & NNPDF23LO & A14 & \\
Diboson  & \sherpa &  2.2.1 & CT10 & Default  & Background estimation \\
\bottomrule
\end{tabular}
\caption{Main features of the Monte Carlo models used to simulate signal and background samples,
and to produce predictions to be compared with data.
The nominal samples listed are used for comparisons with corrected data at particle level as well.
For convenience, \amcnlo is referred to as MG5\_aMC in Figures~\ref{fig:res-nsub}--\ref{fig:res-tau}.}
\label{tab:mc}
\end{center}
\end{table*}

The MC samples were processed through the full ATLAS detector simulation~\cite{:2010wqa}
based on Geant4~\cite{Agostinelli:2002hh},
and then reconstructed and analysed using the same procedure and software that are used for the data.
Additional $pp$ collisions generated by \pythiaeight, with parameter values set to the A2 tune~\cite{ATL-PHYS-PUB-2011-014}
and using the MSTW2008~\cite{Martin:2009iq} PDF set, were overlaid to simulate the effects of additional collisions from
the same and nearby bunch crossings (pile-up), with a distribution of the number of extra collisions
matching that of data.

\section{Object and event selection}
\label{sec:sel}

This analysis uses $pp$ collision data at $\sqrt{s} =13$~\TeV\ collected
by the ATLAS detector in 2016,
that satisfy a number of criteria to ensure that the ATLAS detector was in good operating condition.
All selected events must
have at least one vertex with at least two associated tracks with \pT $> 400$~\MeV.
The vertex with the highest $\sum p_{\textrm{T,track}}^{2}$,
where $p_{\textrm{T,track}}$
is the transverse momentum of a track associated with the vertex, is chosen as the primary vertex.
 
Jets are reconstructed from the EM-scale or locally-calibrated topological energy
clusters~\cite{ATLAS-CONF-2015-037} in both the EM and
hadronic calorimeters using the \akt algorithm~\cite{Cacciari:2008gp} with a
radius parameter of $R = 0.4$ or $R=1.0$,
referred to as small-radius and large-radius jets respectively.
These clusters are assumed to be massless when computing the
jet four-vectors and substructure variables.
A trimming algorithm~\cite{Krohn:2009th} is employed for the large-radius jets
to mitigate the impact of initial-state radiation, underlying-event activity, and pile-up.
Trimming removes subjets of radius $R_\mathrm{sub}  = 0.2$ with $\pT^{i}/\pT^{\textrm{jet}} < f_\mathrm{cut}$,
where $\pT^{i}$ is the transverse momentum of the $i^{\textrm{th}}$ subjet, $\pT^{\textrm{jet}}$ is the transverse momentum
of the jet under consideration,
and $f_\mathrm{cut} = 0.05$.
All large-radius jets used in this paper are trimmed
before applying the selection criteria.
The energies of jets are calibrated by applying \pT- and rapidity-dependent corrections derived from Monte
Carlo simulation with additional correction factors for residual non-closure in data
determined from data~\cite{ATLAS-CONF-2015-037, Aaboud:2017jcu}.

In order to reduce the contamination by small-radius jets originating from pile-up,
a requirement is imposed on the output of the Jet Vertex Tagger (JVT)~\cite{Aad:2015ina}.
The JVT algorithm is a multivariate algorithm that uses tracking
information to reject jets which do not originate from the primary vertex, and
is applied to jets with $\pT < 60$~\GeV\ and $|\eta| < 2.4$.
Small-radius jets containing $b$-hadrons are tagged using a
neural-network-based algorithm~\cite{Aad:2015ydr, ATL-PHYS-PUB-2016-012, Aaboud:2018xwy}
that combines information from the track impact parameters, secondary vertex location, and
decay topology inside the jets.
The operating point corresponds to an overall $70\%$ \btag efficiency in simulated \ttbar events, and to a
probability of mis-tagging light-flavour jets of approximately $1\%$.
 
Muons are reconstructed from high-quality muon spectrometer track segments matched to ID tracks.
Muons with a transverse momentum greater than 30~\GeV\ and within $|\eta| < 2.5$
are selected if
the associated track has a longitudinal impact parameter
$|\zzsth| < 0.5$~mm and a transverse
impact parameter significance $|\dzero|/\sigma(\dzero)| < 3$.
The impact parameter $d_0$ is measured relative to the beam line.
The muon candidates are also required to be isolated from nearby hadronic activity~\cite{Aad:2016jkr}.
The muon isolation criteria remove muons that lie a distance
$\Delta R(\mu,\mathrm{jet}) < 0.04 +10\mathrm{~\GeV}/p_{\mathrm{T},\mu}$
from a small-radius jet axis, where \ptmu is the \pT of the muon.
Since muons deposit energy in the calorimeters, an overlap removal procedure is applied
in order to avoid double counting of leptons and small-radius jets.

Electrons are reconstructed from energy deposits measured in the EM calorimeter which are
matched to ID tracks. They are required to be isolated from nearby hadronic activity
by using a set of \pT- and $\eta$-dependent criteria based on calorimeter and track
information as described in Ref.~\cite{Aaboud:2019ynx}.
Their selection also requires $\pT > 30$~\GeV\ and $|\eta| < 2.5$,
excluding the region $1.37 < |\eta| < 1.52$ which
corresponds to the transition region between the barrel and end-cap calorimeters.
Photon candidates are reconstructed from clusters of energy deposited in the EM calorimeter, and must have
$\pT > 30$~\GeV\ and $|\eta| < 2.5$.
Photon identification is based primarily on shower shapes in the calorimeter~\cite{Aaboud:2018yqu}.

The missing transverse momentum, with magnitude
\met, is calculated as the negative vectorial sum of the transverse momenta
of calibrated photons, electrons, muons and jets associated with the primary vertex~\cite{Aaboud:2018tkc}.
The transverse mass of the leptonically decaying $W$ boson, \mtw, is defined using the absolute value of \met as
$\mtw=\sqrt{2\ptmu\met \big(1-\cos\Delta\phi(\mu, \met)\big)}$.
 
In order to examine large-radius jets originating from light quarks and gluons,
from top quarks and from $W$ bosons, three event selections are defined.
These are referred to as \textit{dijet}, \textit{top} and \textit{$W$ selections},
and are indicative of the origin of the large-radius jet.

In the dijet selection, the events are accepted by a single-large-radius-jet trigger
that becomes fully efficient for jets with $\pT > 400$~\GeV.
The offline dijet selection requires a leading trimmed large-radius jet with
$\pT > 450$~\GeV\ and $|\eta| < 1.5$, and at least one other trimmed large-radius jet
with \pt $>200$~\GeV\ and $|\eta| < 2.5$,
and rejects the event if an electron or muon is present.

For both the top and $W$ selections,
events are collected with a set of single-muon triggers that become fully efficient
for muon $\pT > 28$~\GeV.
The top quarks and the $W$ bosons are identified from their decay products.
A geometrical separation between the decay products of the two top quark candidates is required.
Additional requirements are applied to separate large-radius jets containing all decay products of the
top quark from those where the large-radius jet only contains the hadronic $W$ boson decays, with the
\btagged small-radius jet reconstructed independently.
These form the top selection and the $W$ selection respectively. The selections are described
in \TabRef{tab:sel}.
After these requirements the data sample contains about $3.2\times10^7$ events in the dijet selection, and
roughly 6800 and 4500 events in the top and $W$ selection respectively.

Particle-level observables in Monte Carlo simulation are constructed from stable particles, defined as
those with proper lifetimes $c\tau \gtrsim 10$~mm. Muons at particle level are dressed by
including contributions from photons with an angular distance $\Delta R <0.1$ from the muon.
Particle-level jets do not include muons or neutrinos.
Particle-level \btagging is performed by requiring a prompt
$b$-hadron to be ghost-associated~\cite{Cacciari:2008gn} with the jet.

\begin{table*}[ht]
\begin{center}
\begin{tabular}{l|l|l}
\toprule
& Detector level & Particle level \\
\midrule
\multicolumn{3}{l}{Dijet selection:}       \\
\midrule
\multirow{2}{*}{Two trimmed anti-\kt $R = 1.0$ jets} & $ \pT >  200 $ \GeV  & $ \pT >  200$ \GeV  \\
& $|\eta| <  2.5 $   & $|\eta| <  2.5$    \\
\midrule
Leading-\pT trimmed anti-\kt $R = 1.0$ jet  & \multicolumn{2}{c}{$\pT> 450$ \GeV} \\
\midrule
\multicolumn{3}{l}{Top and $W$ selections:} \\
\midrule
\multirow{3}{*}{Exactly one muon} & $ \pT >  30$ \GeV  & $ \pT >  30$ \GeV  \\
& $|\eta| <  2.5$   & $|\eta| <  2.5$   \\
&   $|\zzsth| < 0.5$ mm and $|\dzero/\sigma(\dzero)| < 3$ & \\
\midrule
\multirow{3}{*}{Anti-\kt $R=0.4$ jets} & $\pT >  25$  \GeV & $\pT >  25$ \GeV \\
& $|\eta|<4.4$  & $|\eta|<4.4$  \\
&   $\textrm{JVT~output} > 0.5$ (if $\pT< 60$ \GeV) & \\
\midrule
\multirow{2}{*}{Muon isolation criteria} & If $\Delta  R(\mu,\mathrm{jet})<0.04 + 10\mathrm{~\GeV}/\ptmu$:   &   \multirow{2}{*}{None} \\
& muon is removed, so the event is discarded & \\
\midrule
\met, \mtw & \multicolumn{2}{c}{ \met $>$ 20 \GeV, \met + \mtw $>$ 60 \GeV}   \\
\midrule
Leptonic top & \multicolumn{2}{c}{At least one small-radius jet with $0.4 < \Delta R(\mu, \mathrm{jet}) < 1.5$ }\\
\midrule
\multicolumn{3}{l}{Top selection:}       \\
\midrule
\multirow{3}{*}{Leading-\pT trimmed anti-\kt $R = 1.0$ jet} &  \multicolumn{2}{c}{$| \eta | < 1.5$, $\pT> 350$ \GeV,  $\textrm{mass} > 140$ \GeV} \\
&  \multicolumn{2}{c}{$\Delta R(\mathrm{large \text - radius~jet},b\textrm{-tagged jet}) < 1$} \\
& \multicolumn{2}{c}{$\Delta \phi(\mu,\mathrm{large \text - radius~jet}) > 2.3$} \\
\midrule
\multicolumn{3}{l}{$W$ selection:}              \\
\midrule
\multirow{3}{*}{Leading-\pT trimmed anti-\kt $R = 1.0$ jet}  & \multicolumn{2}{c}{$| \eta | < 1.5$, $\pT> 200$ \GeV,  $\textrm{mass}  > 60$~\GeV\ and $\textrm{mass} < 100$~\GeV} \\
& \multicolumn{2}{c}{$1 < \Delta R(\mathrm{large \text - radius~jet},b\textrm{-tagged jet}) < 1.8$} \\
& \multicolumn{2}{c}{$\Delta \phi(\mu,\mathrm{large \text - radius~jet}) > 2.3$} \\
\bottomrule
\end{tabular}
\caption{
Summary of object event selections for detector-level and particle-level dijet and \ttbar events.
``Leptonic top'' refers to the top quark that decays into a leptonically decaying $W$ boson, while
``\btagged jet'' refers to small-radius jets that pass a \btagging requirement.
The top and $W$ selections are common up to the requirement on the leptonic top,
then they differ on the requirements on the leading-\pT trimmed large-radius jet.
All selections are inclusive, unless otherwise mentioned.
}
\label{tab:sel}
\end{center}
\end{table*}

\section{Definition of the jet observables}
\label{sec:obs}

All large-radius jets are trimmed before being used in the selections,
and subsequently only the leading trimmed large-radius is considered
in the analysis.
Then the large-radius jet constructed from the
original constituents of the selected jet before the trimming step
is groomed using the soft-drop algorithm, and
the jet substructure observables studies are constructed
from that soft-dropped large-radius jet.

Soft-drop~\cite{Larkoski:2014wba, Marzani:2017kqd}
is an extension of the original split-filtering technique~\cite{Butterworth:2008iy}
and relies on reclustering the jet constituents using the angle-ordered Cambridge--Aachen
jet algorithm and then sequentially considering each splitting in order to remove
soft and wide-angle radiation.
At each step the jet is split into two proto-jets.
The removal of proto-jets in a splitting is controlled by two parameters: a measure of the
energy balance of the pair, $z_{\textrm{cut}}$, and the significance of the angular
separation of the proto-jets, $\beta^{\textrm{SD}}$. These are used to define the soft-drop condition:
\begin{equation*}
\frac{ \mathrm{min}(p_{\mathrm{T}1}, p_{\mathrm{T}2}) }{  p_{\mathrm{T}1} + p_{\mathrm{T}2} } > z_{\textrm{cut}} \left( \frac{\Delta R_{12}}{ R } \right)^{\beta^{\textrm{SD}}}
\end{equation*}
where $R_{12}$ is the angular distance between the two proto-jets and $R$ is the radius of the large jet.
In this analysis, values of
$z_{\textrm{cut}}=0.1$ and $\beta^{\textrm{SD}}=0.0$ are used, based on previous ATLAS studies~\cite{Aaboud:2017qwh},
which is equivalent to modified mass drop tagger~\cite{Dasgupta:2013ihk}.
An important feature of soft-drop is that groomed observables are analytically calculable
to high-order resummation
accuracy~\cite{Frye:2016aiz, Marzani:2017mva, Kang:2018vgn}.

The following substructure variables are measured in this analysis:
 
\begin{itemize}
 
\item Number of subjets with \pT$>10$~\GeV, reconstructed from the selected large-radius jet constituents
using the \mbox{\kt algorithm~\cite{Cacciari:2005hq}} with $R=0.2$.
 
\item Generalised angularities defined as:
\begin{align*}
\lambda_{\beta^{\textrm{LHA}}}^{\kappa} &= \sum_{i \in J} z_{i}^{\kappa} \theta_{i}^{\beta^{\textrm{LHA}}},
\end{align*}
where $z_i$ is the transverse momentum 
of jet constituent $i$ as a fraction of the scalar sum of the $p_{\textrm{T}}$ of all constituents
and $\theta_{i}$ is the
angle of the $i^{\textrm{th}}$ constituent relative to the jet axis, normalised by the jet radius.
The exponents $\kappa$ and $\beta^{\textrm{LHA}}$ probe
different aspects of the jet fragmentation. The $(\kappa =1, \beta^{\textrm{LHA}} = 0.5)$ variant is termed the
Les Houches angularity (LHA)~\cite{Larkoski:2014pca} and used in this analysis.
It is an infrared-safe version of the jet-shape angularity, and provides a measure of the broadness of a jet.

\item
Energy correlation functions ECF2 and ECF3~\cite{Larkoski:2013eya}, and related ratios $C_\textrm{2}$,  $D_\textrm{2}$~\cite{Larkoski:2015kga}.
The 1-point, 2-point and 3-point energy correlation functions for a jet $J$ are given by:
 
\begin{align*}
{\mathrm{ECF1}}   = & \sum_{i \in J} p_{\mathrm{T}_{i}},  \\
{\mathrm{ECF2}}  (\beta^{\textrm{ECF}}) = & \sum_{i < j \in J} p_{\mathrm{T}_{i}} p_{\mathrm{T}_{j}} \left(\Delta R_{ij} \right)^{\beta^{\textrm{ECF}}},  \\
{\mathrm{ECF3}}  (\beta^{\textrm{ECF}}) = & \sum_{i < j < k \in J} p_{\mathrm{T}_{i}} p_{\mathrm{T}_{j}} p_{\mathrm{T}_{k}} \left(\Delta R_{ij} \Delta R_{ik} \Delta R_{jk}\right)^{\beta^{\textrm{ECF}}},
\end{align*}

where the parameter $\beta^{\textrm{ECF}}$ weights the angular separation of the jet constituents.
In the above functions,
the sum is over the $i$ constituents in the jet $J$, such that the 1-point correlation function
ECF1 is approximately the jet \pT.
Likewise, if one takes $\beta^{\textrm{ECF}} = 2$, the 2-point correlation functions scale as the mass of a
particle undergoing a two-body decay in collider coordinates.
In this analysis, $\beta^{\textrm{ECF}}=1$ is used, and for brevity, $\beta^{\textrm{ECF}}$ is not explicitly mentioned hereafter.
 
The ratios of some of these quantities (written in an abbreviated form) are defined as :
 
\begin{align*}
e_{2} = & \frac{\mathrm {ECF2} }{({\mathrm{ECF1}})^2},  \\[2ex]
e_{3} = & \frac{\mathrm{ECF3}  }{({\mathrm{ECF1}})^3}.
\end{align*}

The observables $e_{2}$ and $e_{3}$ are measured, and are later referred to as
$ECF2^{\textrm{norm}}$ and $ECF3^{\textrm{norm}}$.
These ratios are then used to generate the variable $C_{2}$~\cite{Larkoski:2013eya}, and its
modified version $D_{2}$~\cite{Larkoski:2014pca, Larkoski:2015kga},
which have been shown to be particularly useful in identifying two-body structures within jets~\cite{Larkoski:2014gra}.
The $C_2$ and $D_2$ variables as defined below are measured in this analysis:

\begin{align*}
C_{2} = & \frac{e_{3}}{(e_{2})^2},  \\[2ex]
D_{2} = & \frac{e_{3}}{(e_{2})^3}.
\end{align*}

\item Ratios of $N$-subjettiness~\cite{Thaler:2010tr}, $\tau_{21}$ and $\tau_{32}$.
The $N$-subjettiness describes to what degree the substructure of a given jet is compatible with being composed of $N$ or fewer subjets.
 
In order to calculate $\tau_{\mathrm{N}}$, first $N$ subjet axes are defined within
the jet by using the exclusive \kt algorithm, where the jet reconstruction continues until a desired number of jets are found.
The 0-, 1-, 2-,and 3-subjettiness are defined as:
 
\begin{subequations}
\begin{align}
\tau_0 (\beta^{\textrm{NS}}) = & \sum_{i \in J} p_{\mathrm{T}_{i}}  R^{\beta^{\textrm{NS}}}, \label{eq:tau0} \\[1ex]
\tau_1 (\beta^{\textrm{NS}}) = & \frac{1}{\tau_{0} (\beta^{\textrm{NS}})} \sum_{i \in J} p_{\mathrm{T}_{i}} \Delta R_{a_{1},i}^{\beta^{\textrm{NS}}}, \label{eq:tau1} \\[1ex]
\tau_2 (\beta^{\textrm{NS}}) = & \frac{1}{\tau_{0} (\beta^{\textrm{NS}})} \sum_{i \in J} p_{\mathrm{T}_{i}} \min(\Delta R_{a_{1},i}^{\beta^{\textrm{NS}}}, \Delta R_{a_{2},i}^{\beta^{\textrm{NS}}}),\label{eq:tau2} \\[1ex]
\tau_3 (\beta^{\textrm{NS}}) = & \frac{1}{\tau_{0} (\beta^{\textrm{NS}})} \sum_{i \in J} p_{\mathrm{T}_{i}} \min(\Delta R_{a_{1},i}^{\beta^{\textrm{NS}}}, \Delta R_{a_{2},i}^{\beta^{\textrm{NS}}}  \Delta R_{a_{3},i}^{\beta^{\textrm{NS}}} ),\label{eq:tau3}
\end{align}
\end{subequations}
 
where $\Delta R$ is the angular distance between constituent $i$ and the jet axis, $a_i$,
and $\Delta R_{a,n}$ is the angular distance between constituent $i$ and the axis
of the $n^{\mathrm{th}}$ subjet. The term $R$ in equation~\ref{eq:tau0} is the radius parameter of the jet.
The parameter $\beta^{\textrm{NS}}$ gives a weight to the angular separation of the jet constituents.
In the studies presented here, the value of $\beta^{\textrm{NS}} = 1$ is used.
In the above functions, the sum is performed over the constituents $i$ in the jet $J$,
and a  normalisation factor $\tau_{0}$ (Eq.~(\ref{eq:tau0}))
is used.
The ratios of the $N$-subjettiness functions, $\tau_{21} = \tau_{2}/\tau_{1}$ and
$\tau_{32} = \tau_{3}/\tau_{2}$  have been shown to be particularly useful in identifying two-body
and three-body structures within jets.
 
Studies presented in Ref.~\cite{Larkoski:2014uqa} have shown that an
alternative axis definition can increase the discrimination power of these variables.
The winner-takes-all (WTA) axis uses the direction of the hardest constituent in the subjet
obtained from the exclusive \kt algorithm
instead of the subjet axis, such that the distance measure $\Delta R_{a_{1},i}$ changes in the calculation.
In this analysis,
the same observables calculated
with the WTA axis definition, \TauTwoOneWTA\ and \TauThreeTwoWTA, are used.

\end{itemize}

\section{Data-driven background estimation}
\label{sec:bg}

The largest non-$t\bar{t}$ contributions to the $W$ and top selections
come from the $W$+jets and
single-top processes.
Additionally non-prompt and mis-reconstructed muons
are a separate source of background for the top and $W$ selections.
Contributions from other processes were considered
and found to be negligible.
A data-driven method, following Ref.~\cite{Aad:2015fna}, is used to estimate
the contribution from the $W$+jets process while the
single-top process is considered part of the signal.
 
At the LHC the production rate of $W^+$+jets events is larger than
that of $W^-$+jets due to the higher density of $u$-quarks than $d$-quarks in the proton.
This results in more events with positively charged leptons.
Other processes do not contribute significantly to this charge asymmetry.
The data are used to derive scale factors that correct the normalisation and flavour fraction
given by the MC simulation~\cite{Aad:2014zka}.

Normalisation scale factors are determined by comparing the
charge asymmetry in data with the asymmetry estimated by simulation.
Contributions to the asymmetry from other processes are
estimated by simulation and subtracted. A selection that contains the full top and
$W$ selection criteria without any \btagging requirements is initially used.
The total number of $W$+jets events in data, $N_{W^+} + N_{W^-} $, is given by
\begin{equation*}
N_{W^+} + N_{W^-}  = \left( \frac{ r_{\mathrm{MC}}+ 1 }{ r_{\mathrm{MC}} - 1} \right) ( D_+ - D_- )
\end{equation*}
where $r_{\mathrm{MC}}$ is the ratio of the number of events with positive muons
to the number of events with negative muons obtained from the MC simulation while
$D_+$ and $D_-$ are the number of events with positive and negative muons in data,
respectively, after using simulation to subtract the estimated background contribution of all processes other than $W$+jets.
From the above equation the scale factor $C_\mathrm{A}$ is extracted which is defined as the ratio of $W$+jets
events evaluated from data to the number predicted by the simulation
\begin{equation*}
C_{\text{A}} =  \left( \frac{r_{MC} +1}{r_{MC} -1} \right) (D_{+} - D_{-} ) \cdot \frac{1}{N_{W}^{\text{MC}}}
\end{equation*}
where $N_{W}^{\text{MC}}$ is the predicted number of $W$+jets events.

Scale factors correcting the relative fractions of $W$ bosons produced
in association with jets of different flavour are also estimated using data.
The fractions of $W+b\bar{b}, W+c\bar{c}, W+c$ and $W$+light-quark events
are initially estimated from simulation in a selection
without the \btagging requirements, which corresponds to the selection
mentioned in \TabRef{tab:sel} without the $\Delta R$ requirement imposed during the top and $W$ selections.
A system of three equations is used to fit the fractions estimated from simulation
to the selection with full \btagging requirements:
 
\begin{equation}  \label{eqn:wjets}
\left( \begin{matrix}
C_{\mathrm{A}}(N_{bb}^{-} + N_{cc}^{-})  &   C_{\mathrm{A}} N_{c}^{-}   &  C_{\mathrm{A}} N_{\mathrm{light}}^{-} \\
f_{bb} + f_{cc}  & f_{c} & f_{\mathrm{light}} \\
C_{\mathrm{A}}(N_{bb}^{+} + N_{cc}^{+})  &   C_{\mathrm{A}} N_{c}^{+}   &  C_{\mathrm{A}} N_{\mathrm{light}}^{+} \\
\end{matrix} \right)
\cdot
\left( \begin{matrix}
K_{bb,cc} \\
K_{c} \\
K_{\mathrm{light}}\\
\end{matrix} \right)
=
\left( \begin{matrix}
D_{W^-} \\
1\\
D_{W^+} \\
\end{matrix} \right),
\end{equation}
where $f_{bb}, f_{cc}, f_{c}$ and $f_{\text{light}}$ are flavour factors estimated from
simulation while $K_{bb}, K_{cc}, K_{c}$ and $K_{\text{light}}$  are the respective correction factors.
The corresponding number of events estimated by simulation with positive (negative) leptons
are given by $N^{+ (-)}_{bb},N^{+ (-)}_{cc}, N^{+ (-)}_{c}$ and $N^{+ (-)}_{\text{light}}$.
The terms $D_{W^{\pm}}$ are the expected numbers of  $W$+jets events with positively or negatively charged leptons in the data.
An iterative process is used to find the $K_{\mathrm{flavour}}$ correction factors which are used to correct the associated
$f_{\mathrm{flavour}}$ fractions used in the calculation of $C_\mathrm{A}$. The correction factors are determined
by inverting Eq.~(\ref{eqn:wjets}) and then the process is repeated with a new $C_\mathrm{A}$ calculated using
the corrected flavour fractions. This process is repeated 10 times and further iterations produce negligible
changes in $C_\mathrm{A}$.

This process is repeated individually for all variables in the top and $W$ selections since, depending on the substructure of the
selected large-radius jet, events can fall out of the acceptance for a subset of the variables.
The final calculated scale factors are, however, consistent across both selections and all variables.
These scale factors are $0.84 \pm 0.02$, where the uncertainty is statistical,  and the overall contribution to the final selections
is shown in Table~\ref{tab:bkg}.
In order to determine the uncertainty in the shape of the subtracted
$W$+jets distribution, the contribution from an alternative MC generator ({\amcnlo}+\pythiaeight as opposed to default \sherpa) was used.
Both MC samples were scaled to the estimated number of events
and the envelope of the shape difference was taken as an uncertainty.

There is also a contribution from events where a jet is misreconstructed as a muon
or when a non-prompt muon is misidentified as a prompt muon which satisfies
the selection criteria.
This contribution is estimated using the matrix method, comparing the yields of muons and non-prompt muons
that pass a loose selection with the yields of those that pass a tight selection.
The efficiency for real muon selection ($\varepsilon_{\mathrm{real}}$) is measured
using a tag-and-probe method
with muons from $Z \rightarrow \mu \mu$ events.
The efficiency for misreconstructed muon selection ($\varepsilon_{\mathrm{fake}}$)
is measured in control regions dominated by background from multijet processes,
after using simulation to subtract the contribution of other processes.
Event weights are computed using the above efficiencies, which are parameterised
in the kinematics of the event. The weight for event $i$, where the muons satisfy the loose criteria, is given by
\begin{equation*}
w_i = \frac{ \varepsilon_{\mathrm{fake}} } { \varepsilon_{\mathrm{real}} - \varepsilon_{\mathrm{fake}}} (\varepsilon_{\mathrm{real}}  - \delta_i )
\end{equation*}
where $\delta_i$ equals unity if the muon in event $i$ satisfies the tight criteria and zero otherwise.
The background estimate in a given bin is therefore the total sum of weights in that bin.
The estimated contributions to the yield from misreconstructed or non-prompt muons for the top and $W$ selections
are shown in Table~\ref{tab:bkg}.
These corrections have very little effect on the shape of
the distributions considered.
 
\begin{table*}[ht]
\begin{center}
\begin{tabular}{ccc}
\hline
Background & Top selection & $W$ selection\\
\hline
~ & \multicolumn{2}{c}{(Percent contributions)} \\
\hline
$W$+jets & 4.0 $\pm$ 0.1 & 2.6 $\pm$ 0.1 \\
\hline
Misreconstructed and non-prompt muons & 6.6 $\pm$ 0.1 &  5.5 $\pm$ 0.1 \\
\hline
\end{tabular}
\caption{Contributions from background processes which are subtracted in the top and $W$ selections.
The uncertainties are statistical only.}
\label{tab:bkg}
\end{center}
\end{table*}

\FloatBarrier

\section{Systematic uncertainties}
\label{sec:syst}
 
\subsection{Large-radius jet uncertainties}
 
As jets are built from topological clusters reconstructed in the calorimeter,
systematic uncertainties in the jet substructure observables are calculated using
a bottom-up approach applied to the clusters forming each jet~\cite{Aaboud:2017qwh}.
The following components of the uncertainty are considered:
 
\begin{itemize}
 
\item Cluster reconstruction efficiency (CE):
Accounts for low energy particles that fail to seed a cluster
based on the fraction of inner-detector tracks matched to no clusters in low $\mu$ data.
The uncertainty is the observed difference between simulation and data.
Since the efficiency reaches $100\%$ for cluster energy above $2.5$~\GeV,
no uncertainty is assumed above this value.
 
\item Cluster energy scale variation (CESu/CESd):
The cluster energy scale is determined by studying clusters matched to isolated tracks in data events with low pile-up.
A fit of the $E/p$ distribution is used to extract an overall energy scale.
The uncertainty in the scale is given
by taking the difference of the ratio of the scales calculated in data and simulation from unity. Clusters are
independently scaled up and down and the resulting variations in observables are added in quadrature.
 
\item Cluster energy smearing (CES):
The difference in quadrature of the width of the $E/p$ distribution measured in data and given by simulation is
defined as the uncertainty in the energy resolution. The cluster energies are smeared by this value
and the effect on the observables is taken as an uncertainty.
 
\item Cluster angular resolution (CAR):
The radial distance between clusters and their matched tracks (extrapolated to the corresponding calorimeter layer)
is measured in bins of $\eta$ and as a function of $E$, to account for the resolution in
various regions of the calorimeter.
A conservative uncertainty of $5$~mrad is used to smear cluster positions.
 
\end{itemize}
 
\begin{figure}[h]
\centering
\includegraphics[width=0.4\textwidth]{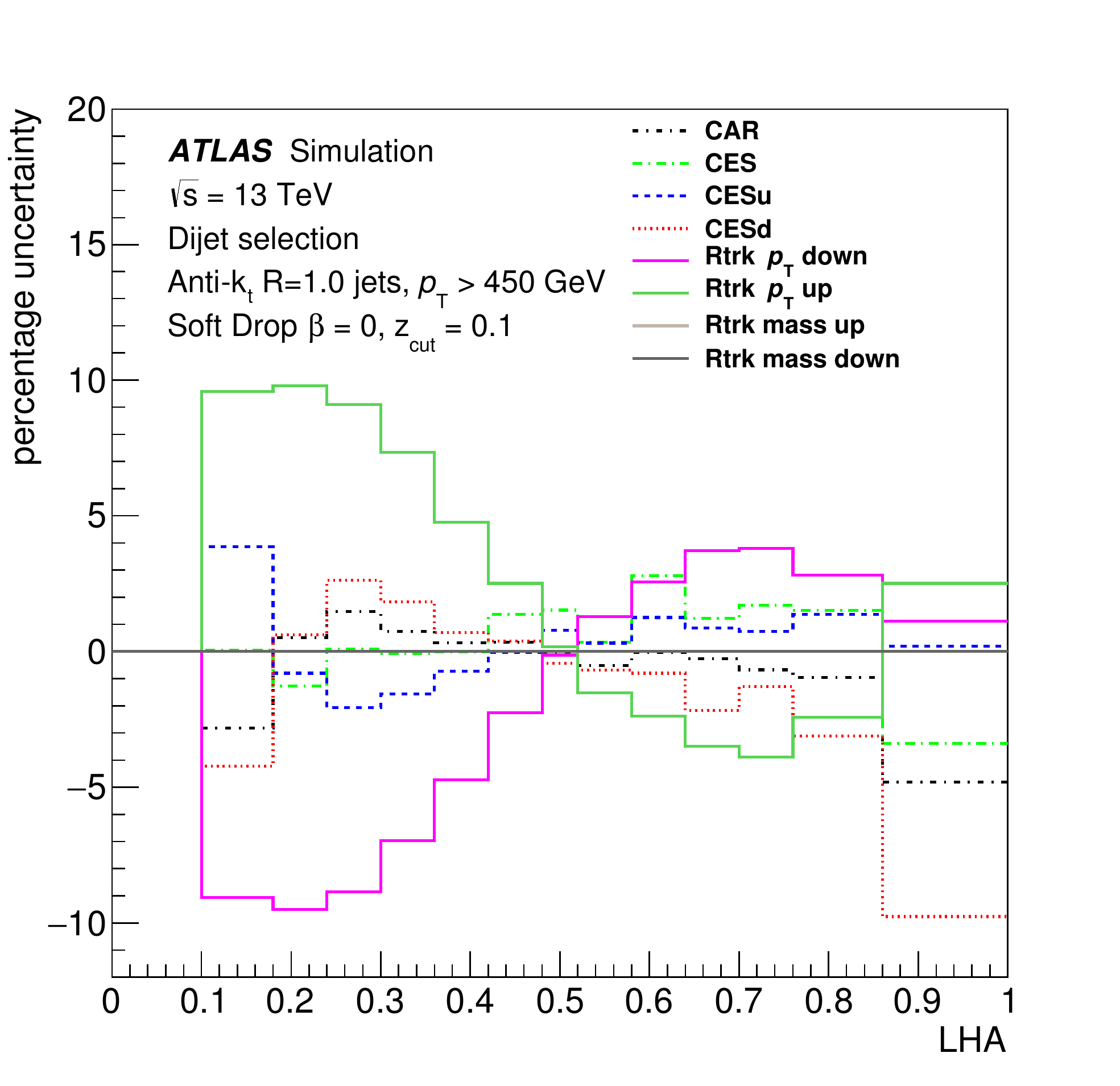}\qquad
\includegraphics[width=0.4\textwidth]{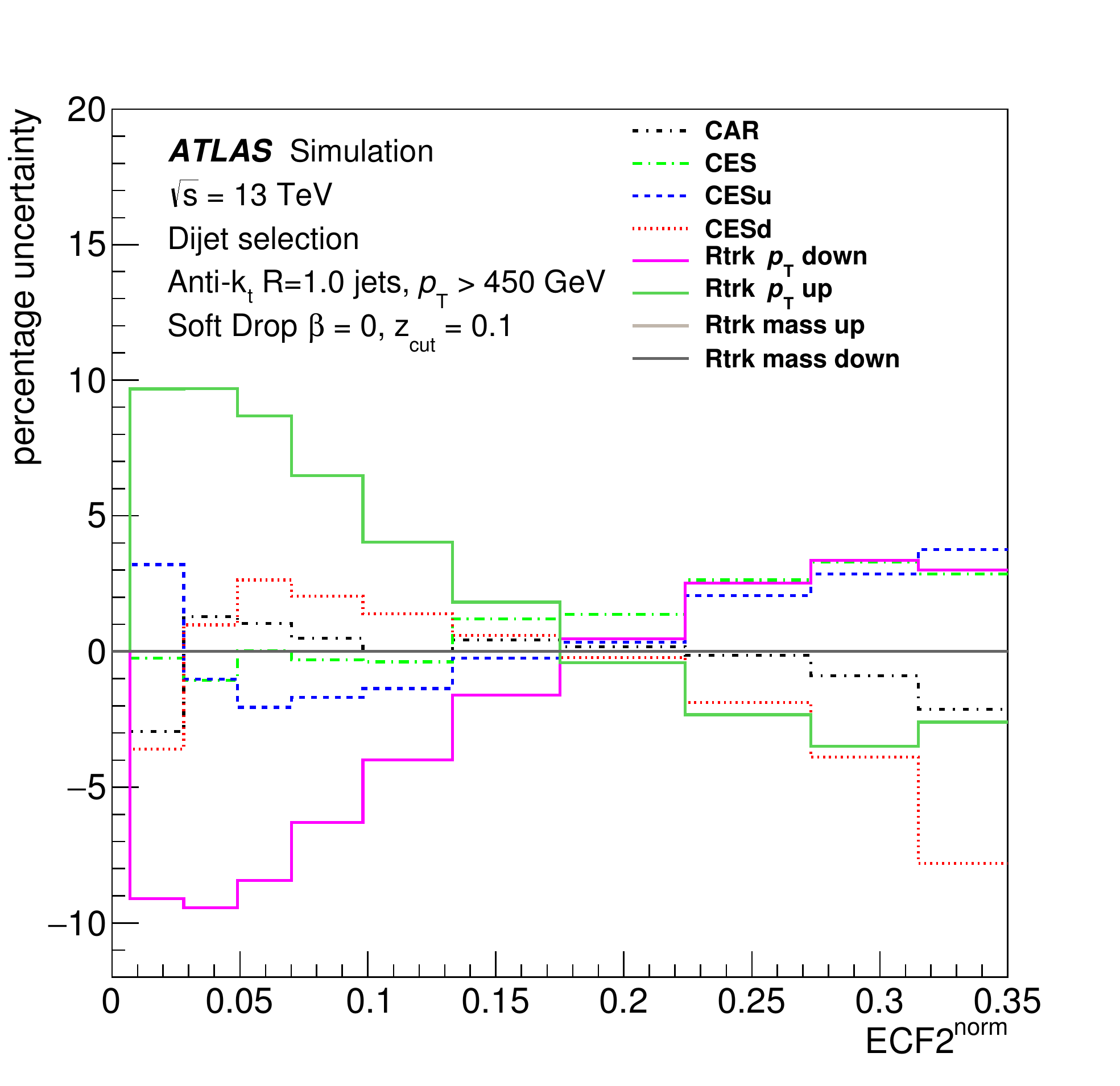}\\
\includegraphics[width=0.4\textwidth]{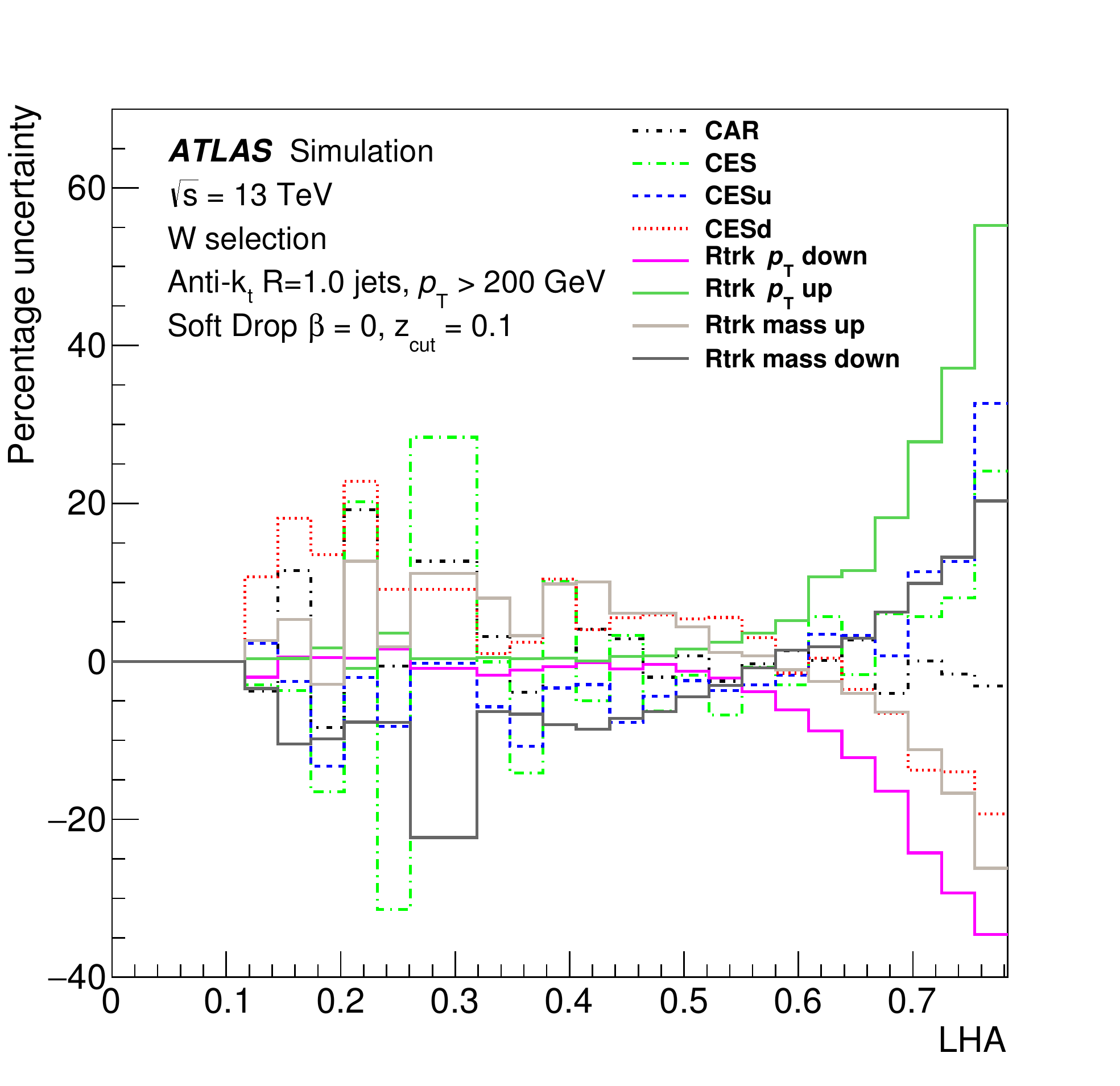}\qquad
\includegraphics[width=0.4\textwidth]{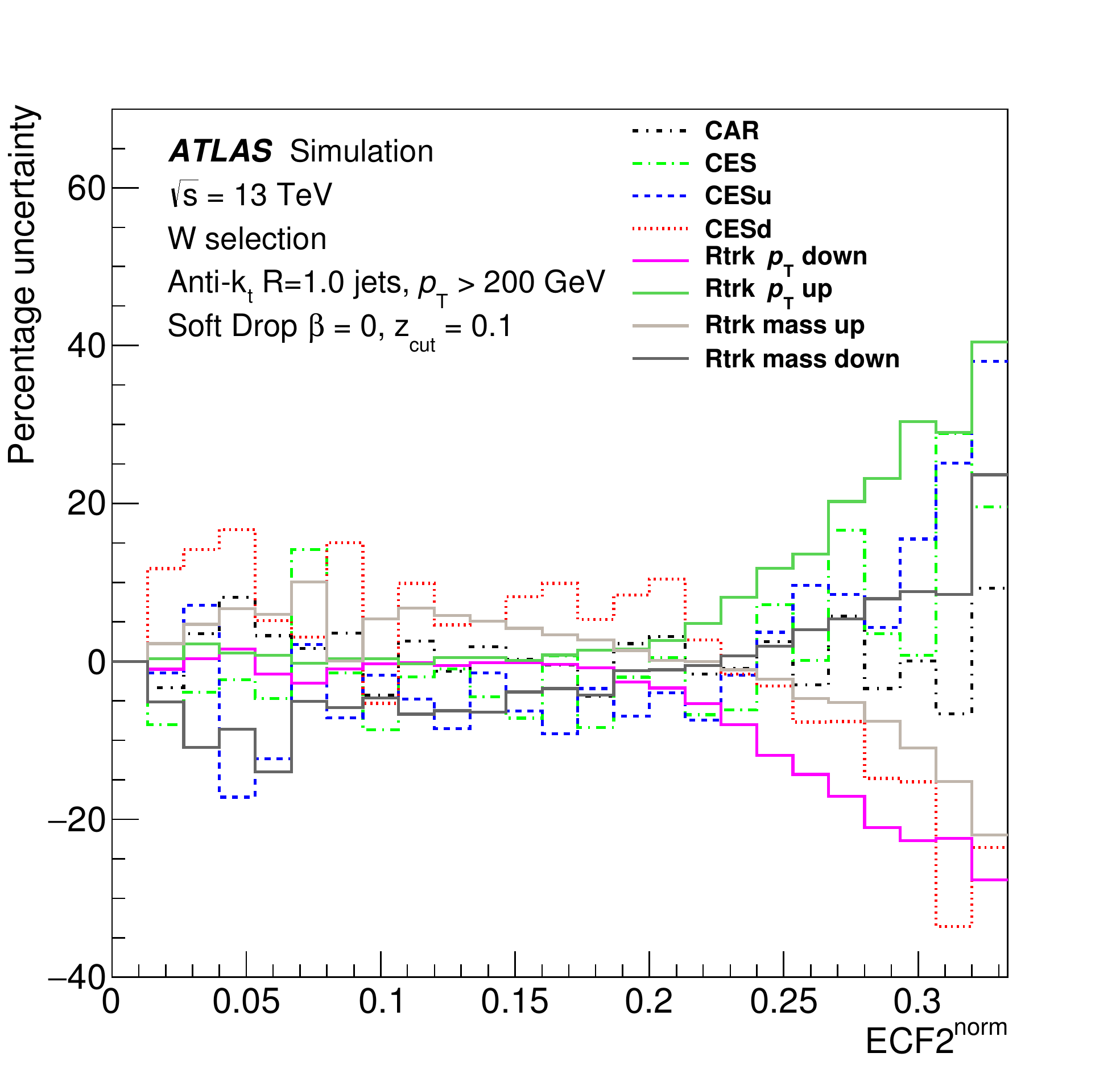}\\
\caption{Bin-by-bin systematic uncertainties
due to large-radius jet reconstruction uncertainties associated with cluster, Rtrk and jet mass calibrations
in the dijet (top) and $W$ (bottom) selections for the soft-drop groomed Les Houches angularity variable (left) and the normalised ECF2 variable (right).}
\label{fig:systPlots}
\end{figure}

Uncertainties in the jet \pt and mass are derived
by the
R$_{\text{trk}}$ method~\cite{Aaboud:2018kfi}, comparing the variables calculated using the energy deposited in the
calorimeter with those using the momenta of charged-particle tracks.
The largest effect on the majority of measured distributions comes from cluster energy smearing
for the top and $W$ selections, typically around $8\%$ but can be as high as 16\% in some regions.
The other cluster uncertainty components contribute between $1\%$ and $6\%$ in
the statistically significant part of the distributions for the top and $W$ selections.
For the dijet selection, the typical values are
between $2\%$ and $4\%$ for all observables, but reach 10\% in some bins.
The dominant large-radius jet uncertainties for a subset of variables are shown in Figure~\ref{fig:systPlots}.

In addition to the above uncertainties the sensitivity of the measured distributions to other detector effects was considered.
This are summarised as follows:
 
\begin{itemize}
 
\item Energy scaling correlation scheme: applying  the variations to clusters with
different kinematics and with different properties, assuming them to be uncorrelated.
 
\item Since the cluster energy calibration is based on pion energy deposition,
additional tests are carried out to account for the different energy deposited by non-pion hadrons,
such as $K_\textrm{L}$, and the impact on the distributions under study.
 
\item Cluster merging and splitting: topo-clusters can be split or merged during
the clustering procedure and this process can be sensitive to noise fluctuations.
 
\end{itemize}
 
In all cases, very conservative variations were applied in order to ensure that the distributions considered
were not sensitive to the above effects. For the majority of the distributions the observed
variations due to
other detector effects
were smaller than the cluster uncertainties.
However, it was found that $N$-subjettiness variables in the dijet selection had shifts of about 50\%
when some of the cluster merging and splitting variations were applied.
Using a different axis definition, rather than the WTA variant, did not sufficiently reduce the sensitivity of the variables to this effect.
While these variations were conservative, in order to ensure that no systematic
uncertainties are being underestimated the $N$-subjettiness variables and their ratios were not used in the dijet selection.

\subsection{Other sources of uncertainties}
 
Systematic uncertainties are also derived for other reconstructed objects which are considered in
the top and the $W$ selections~\cite{Aaboud:2018mjh}.
Uncertainties associated with small-radius jets, \btagged jets,
reconstructed muons and $E_{\mathrm{T}}^{\mathrm{miss}}$ are all considered and are found to be subdominant.
The theory normalisation uncertainties are also found to be negligible.
 
Finally, uncertainties in the shape of the subtracted $W$+jets component
are derived by comparing, for each variable, the shapes obtained using the nominal
MC sample and an alternative sample, as listed in \TabRef{tab:mc}.
The envelope is taken as an uncertainty in the subtracted shape, and results in uncertainties which are smaller than $1\%$.
The uncertainties due to signal modelling in MC generators are accounted
for in unfolding, as described in Section.~\ref{sec:unf}.

\FloatBarrier

\section{Detector-level results}
\label{sec:det-res}

\begin{figure}[h]
\centering
\includegraphics[width=0.4\textwidth]{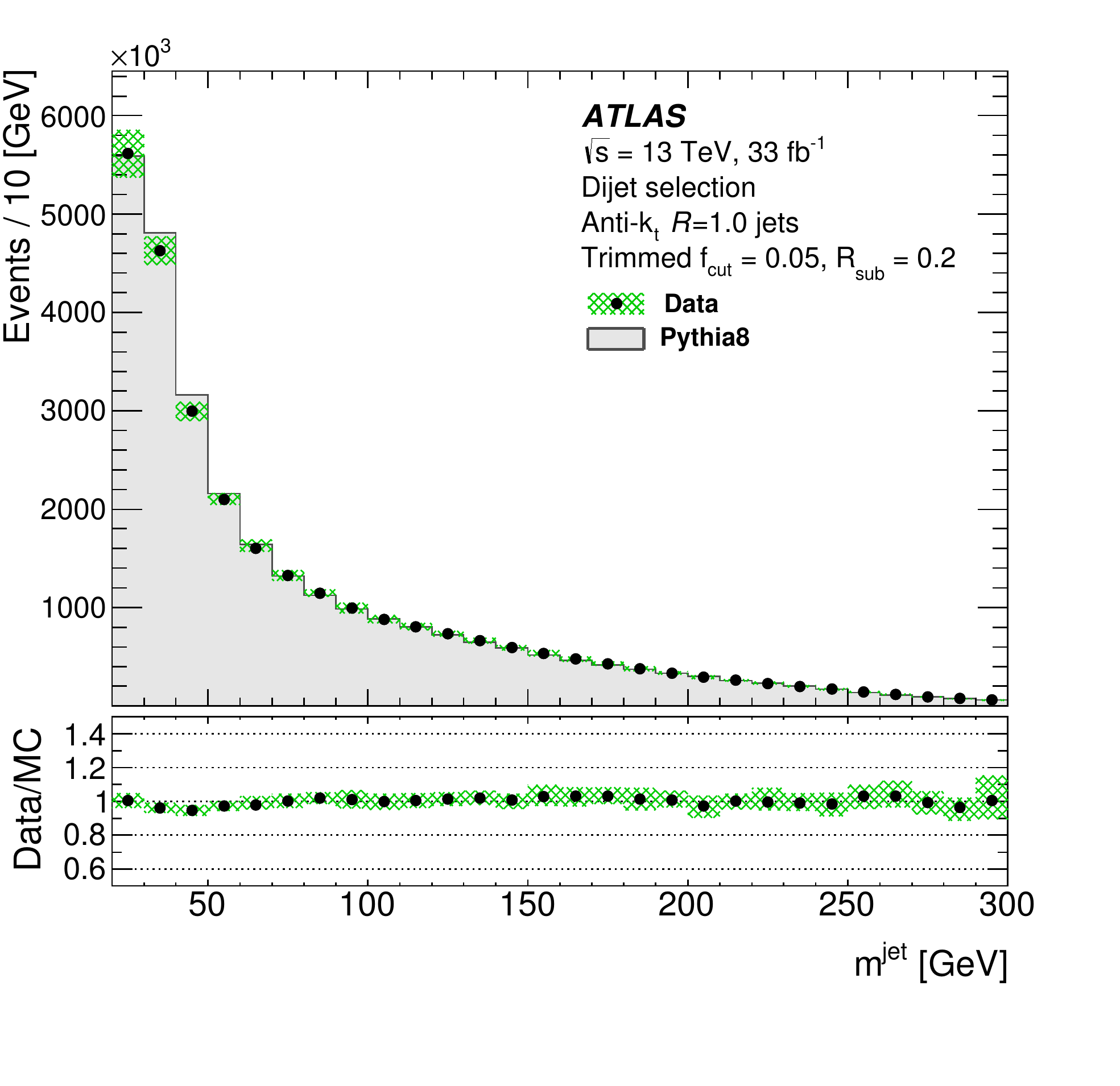}\qquad
\includegraphics[width=0.4\textwidth]{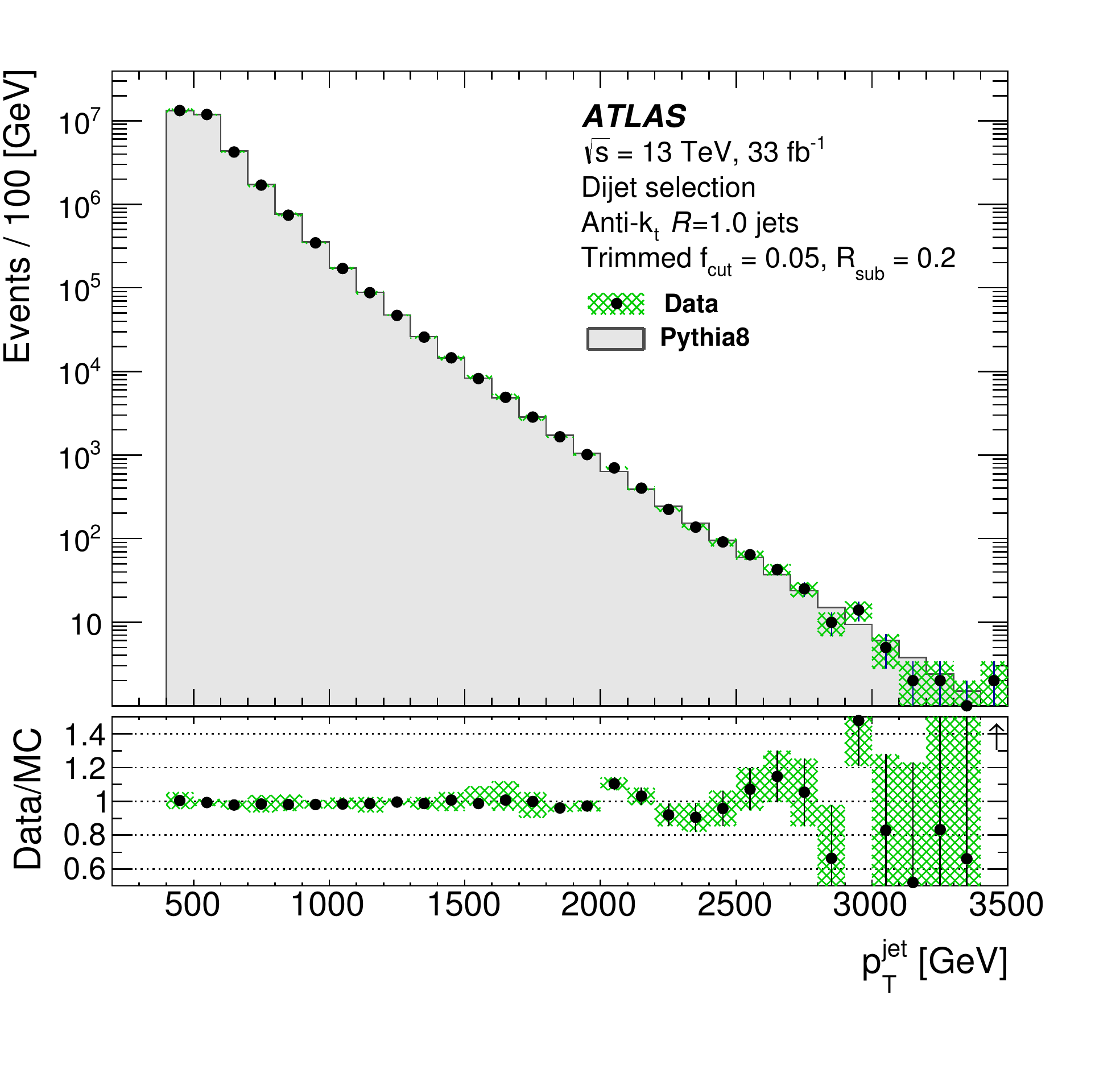}\\
\includegraphics[width=0.4\textwidth]{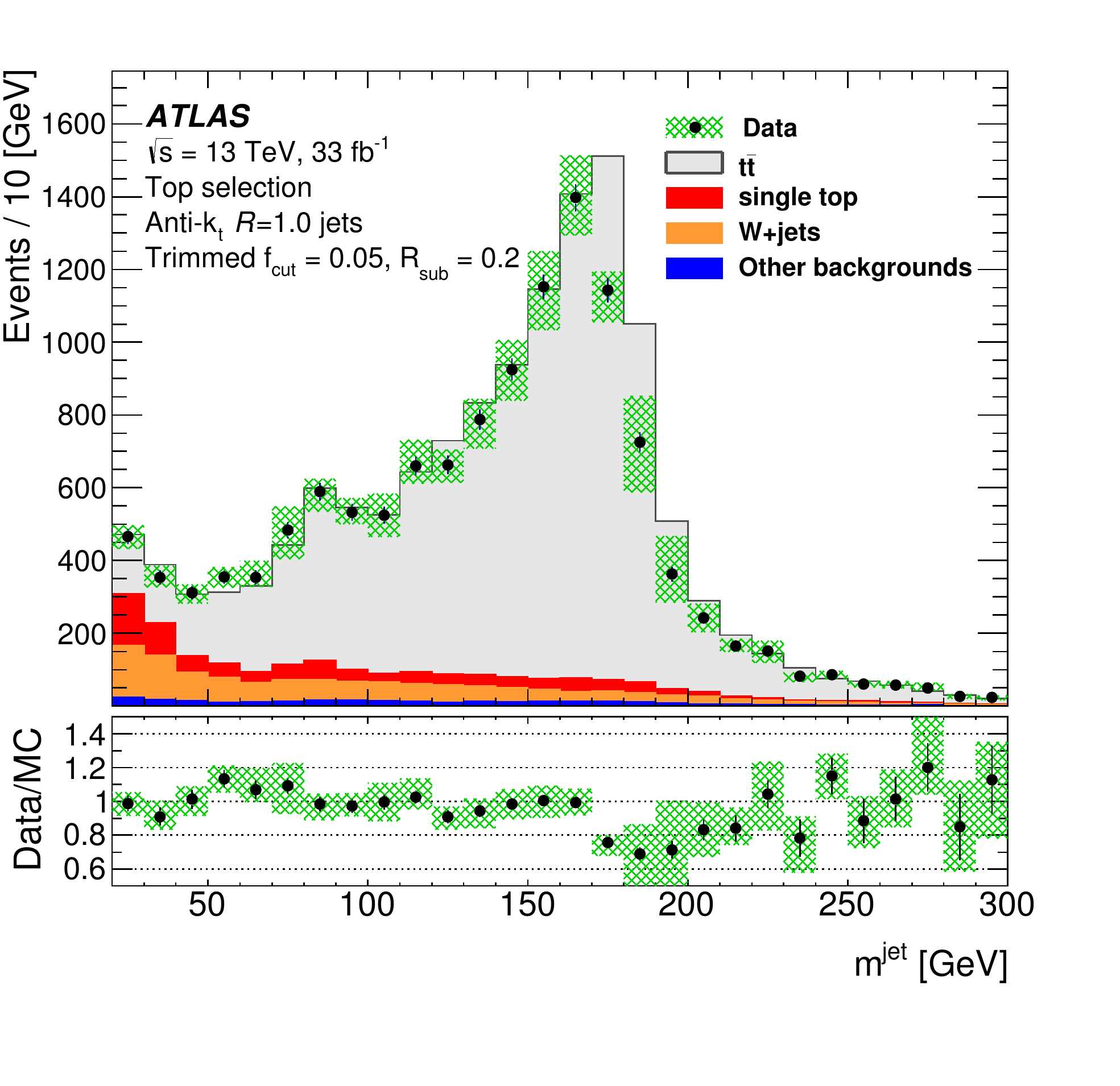}\qquad
\includegraphics[width=0.4\textwidth]{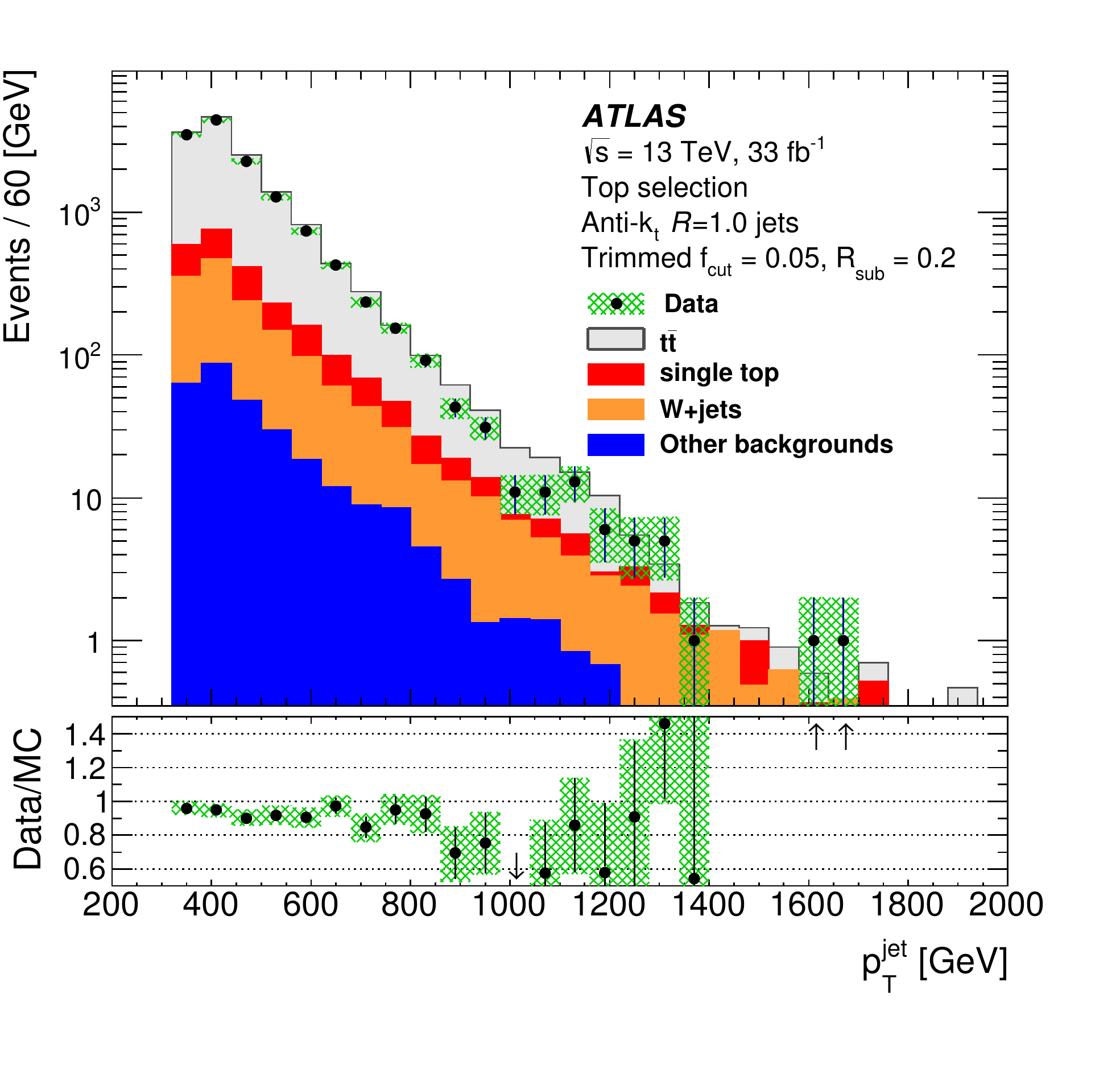}\\
\includegraphics[width=0.4\textwidth]{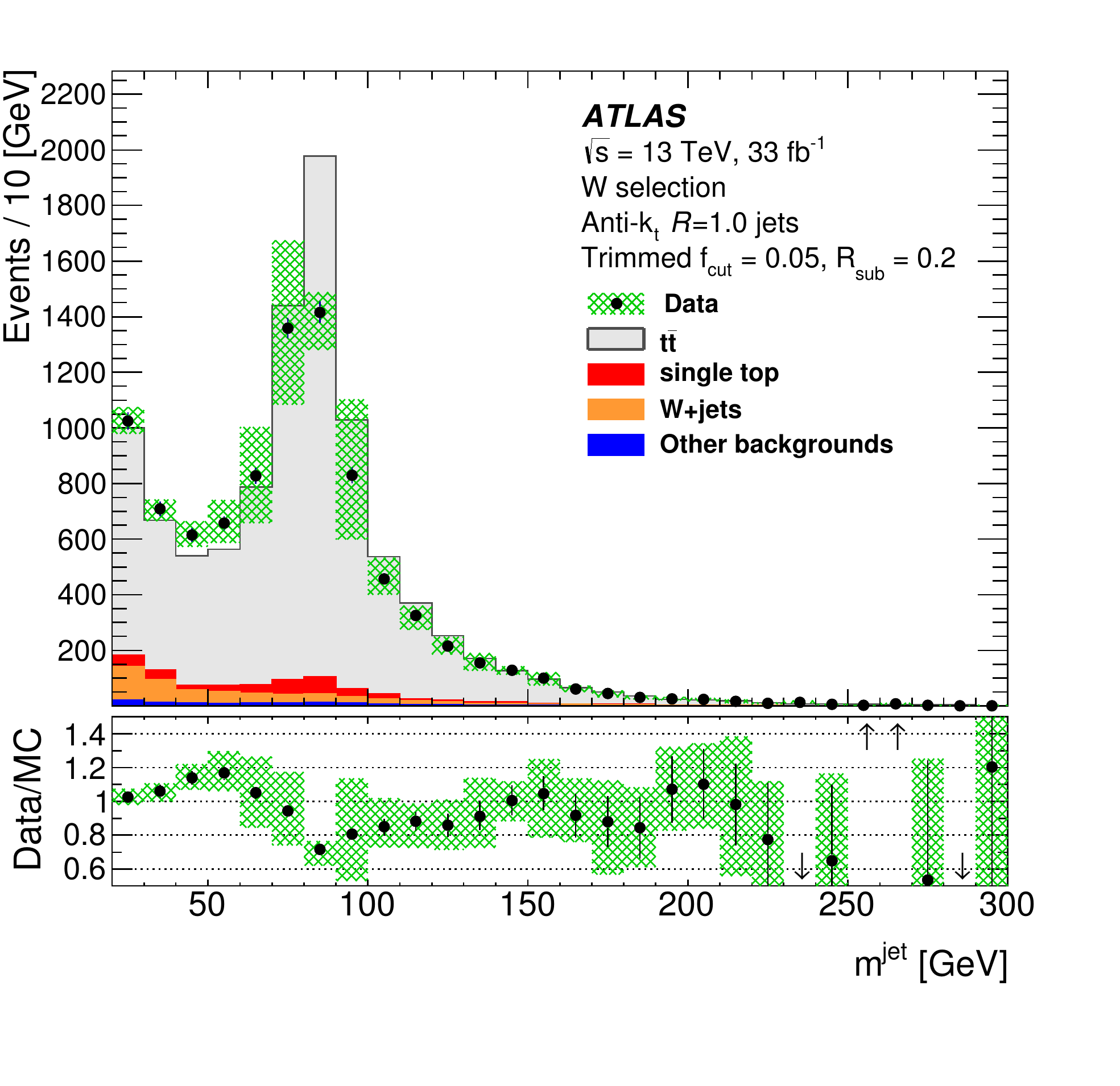}\qquad
\includegraphics[width=0.4\textwidth]{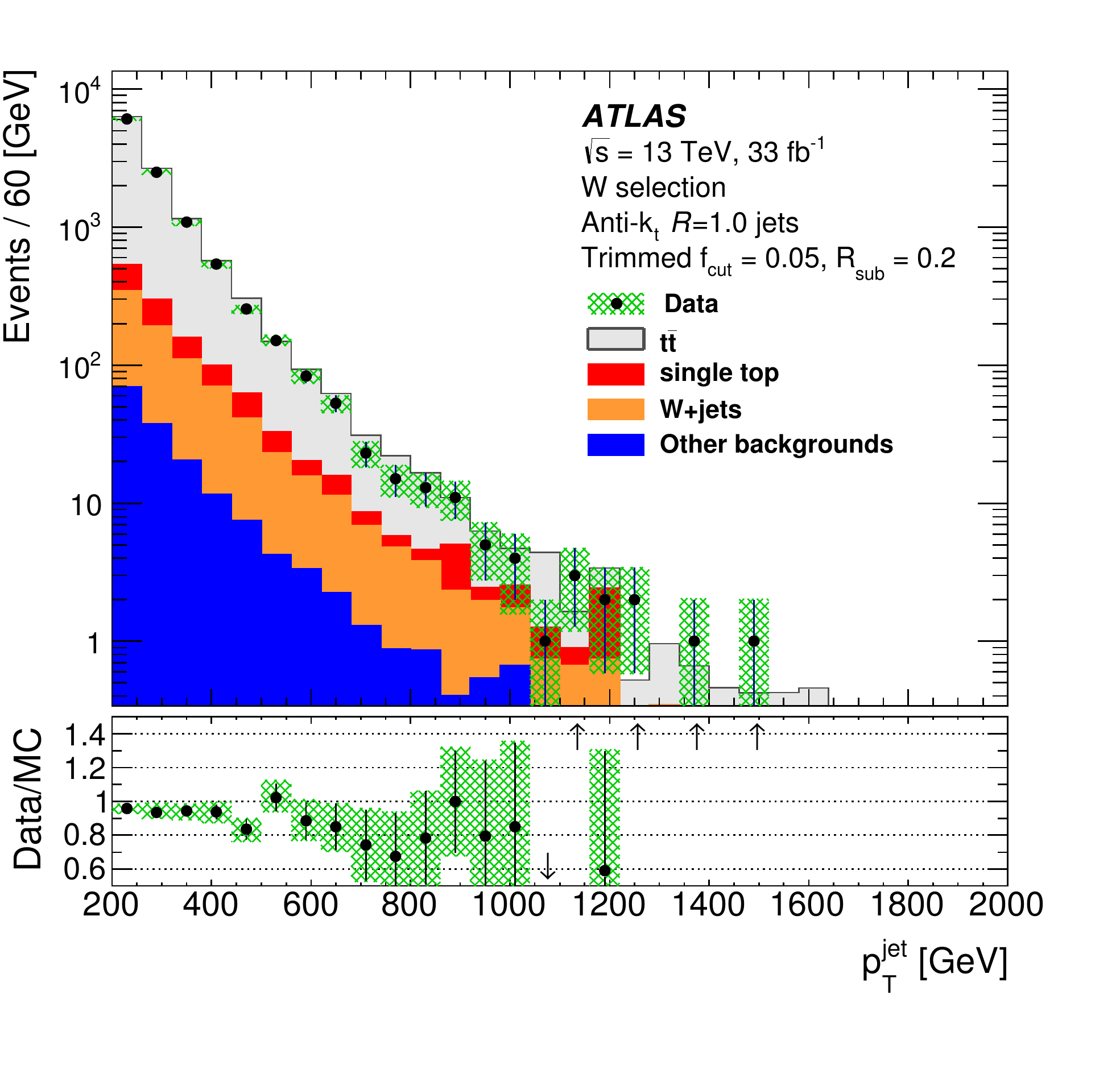}\\
\caption{Comparison of detector-level distributions in data and MC simulation for trimmed large-radius jets for dijet (top row), top (middle row),
and $W$ (bottom row) selections. For the top and $W$ selections, jet mass requirements have not been applied.
The mass is shown in the left column, while the transverse momentum
is in the right column.
The shaded bands represent the combined statistical and systematic uncertainty.
Contributions from dominant backgrounds are shown for the top and $W$ selections,
while the smaller contributions
from other processes are grouped under other backgrounds.}
\label{fig:detplots}
\end{figure}

The distributions of the trimmed large-radius jet mass and \pT at detector level are shown in Figure~\ref{fig:detplots} for dijet,
top and $W$ selections.
The peaks in the distributions due to the top and $W$ masses are clearly visible. In general,
good agreement is observed between data and simulation for the distribution of
transverse momenta, while a shift is observed for the distributions of mass.
This is a known effect~\cite{Aaboud:2018psm}, due to the lack of in situ calibrations of jet mass, and to jet mass scale
uncertainties in the detector-level plots.
 
\FloatBarrier

\section{Unfolding}
\label{sec:unf}

The measured distributions are unfolded to correct for detector effects.
The Iterative Bayesian (IB) unfolding method~\cite{DAgostini:1994zf}
with three iterations (as implemented in RooUnfold~\cite{Adye:2011gm})
is used to correct detector-level data to particle level, as defined in Section~\ref{sec:sel}.
Response matrices ($a_{ji}$) for each distribution are derived from MC simulation and used in order to estimate the
probability for a given event at particle level ($T$), contributing to bin $i$, to be reconstructed in a
given detector-level ($D$) bin $j$, also defined as $P(D_j | T_i )$.
Rather than using a simple matrix inversion, IB unfolding uses a probabilistic approach. In order to do this,
the unfolding matrix ($\theta_{ij}$) is defined such that the number of events in a particle-level bin, $T_i$, is given by
\begin{equation}
\label{eqn:unfold}
T_i = \sum_j \theta_{ij} d_j
\end{equation}
where $d_j$ is the number of data events measured in bin $j$.
Using Bayes' theorem, one can define the unfolding matrix as:
\begin{equation*}
\theta_{ij} = P(T_i | D_j) = \frac{P (D_j | T_i ) \cdot P( T_i ) } { \sum_i P (D_j | T_i ) \cdot P( T_i ) } = \frac{ a_{ji} \cdot P(T_i) } { \sum_i a_{ji} \cdot P(T_i) }.  \end{equation*}
where $P(T_i )$ is the input prior.
The unfolding matrix can therefore be constructed using the response matrix
obtained from simulation.
After corrections are
applied for detector acceptance and reconstruction efficiency,
Eq.~(\ref{eqn:unfold}) can be used to perform the unfolding.
To ensure that the final distributions are
not biased by the shape predicted by simulation the process is iterated, each subsequent
iteration using the previous estimate for the final corrected distribution as $P(T_i )$.
The number of iterations is chosen such that differences between
multiple subsequent iterations are smaller than data-driven
cross-closure uncertainties, described below.

The consistency of the unfolding procedure was tested using several closure and cross-closure tests.
 
\begin{itemize}
\item MC closure: a test where the distributions from the nominal MC generator are unfolded using the nominal method.
Uncertainties are found to be negligible.
\item Cross-closure: accounts for modelling differences between two different MC generators. The distributions from an
alternative generator are unfolded using the nominal method and the differences account for
differences in the predicted shape. These result in the largest uncertainties and
are typically around $5\%$ in the dijet selection
and around $14\%$ in the top and $W$ selections, depending on the observable and the bin.
\item Data-driven cross-closure: accounts for the sensitivity of the unfolding
method to differences between the shape of the observable seen in data and in simulation.
The particle-level substructure distributions are reweighted such
that the corresponding detector-level distributions match the data.
These reweighted distributions are unfolded using the nominal method and
uncertainties are estimated as the differences between the reweighted particle-level and unfolded distributions.
\end{itemize}
The binning of variables in the dijet selection was chosen to reduce uncertainties from the above effects by increasing
the bin purity. For the top and W selections binning was determined based on the statistical uncertainty of the dominant systematic uncertainties.
 
\section{Particle-level results}
\label{sec:res}

The results are presented in two sets of distributions:
substructure observables in data are compared with MC predictions,
and distributions measured in data corresponding to
different selections are compared with each other.
For the latter, it must be noted that
the comparisons are performed in different large-radius jet \pT ranges;
however, in each instance the most inclusive selection is used.
They are indicative of different substructures of the
large-radius jets according to their origin even with
somewhat different kinematic ranges.
All plots with soft-drop grooming are shown; the trimmed versions have very similar characteristics~\cite{hepdata}.
The dominant systematic uncertainties in the measurement are the large-radius jet uncertainties resulting
from the bottom-up approach using clusters, and modelling uncertainties affecting the unfolding closure and cross-closure.

\begin{figure}[h]
\centering
\includegraphics[width=0.4\textwidth]{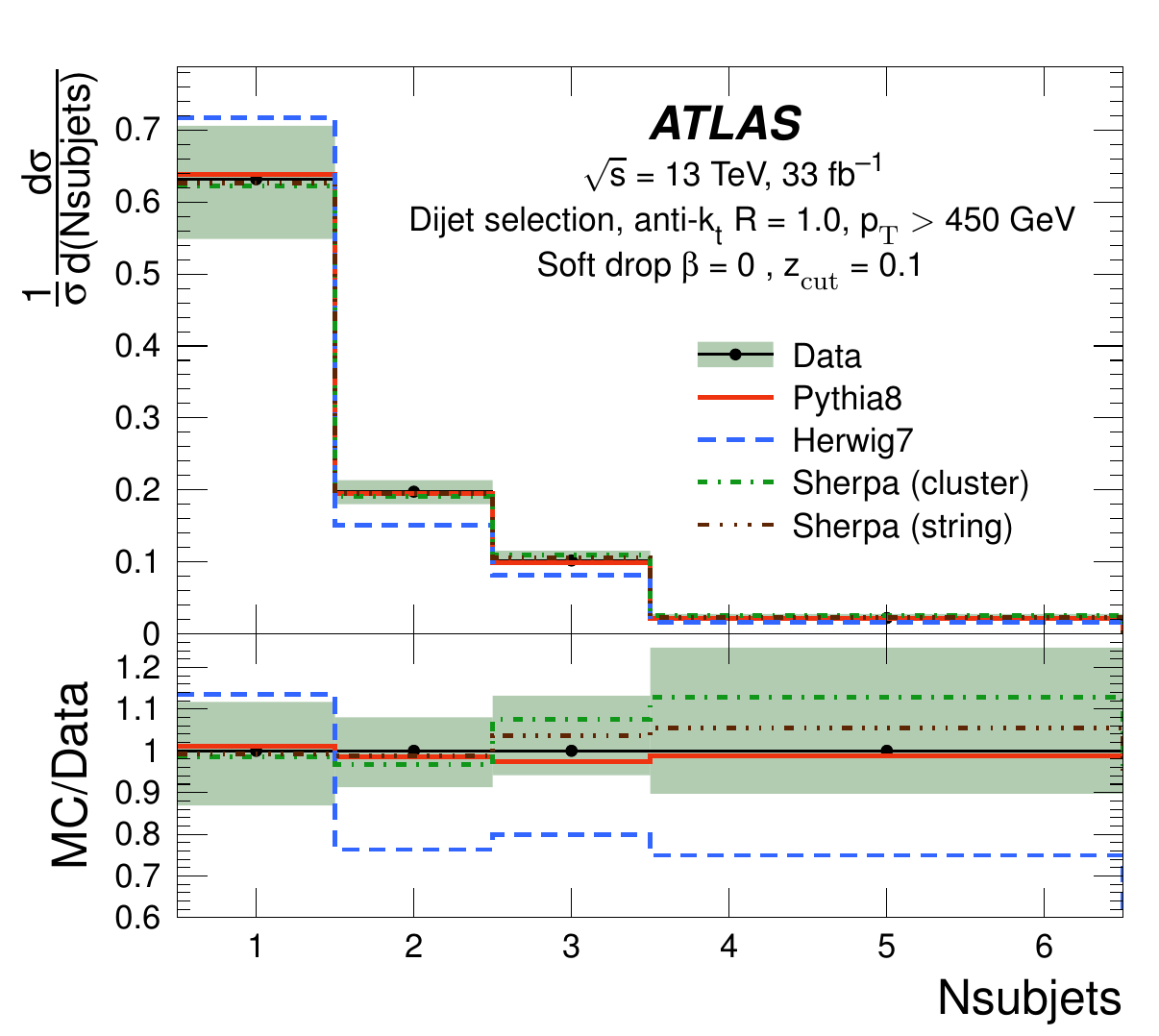}\qquad
\includegraphics[width=0.4\textwidth]{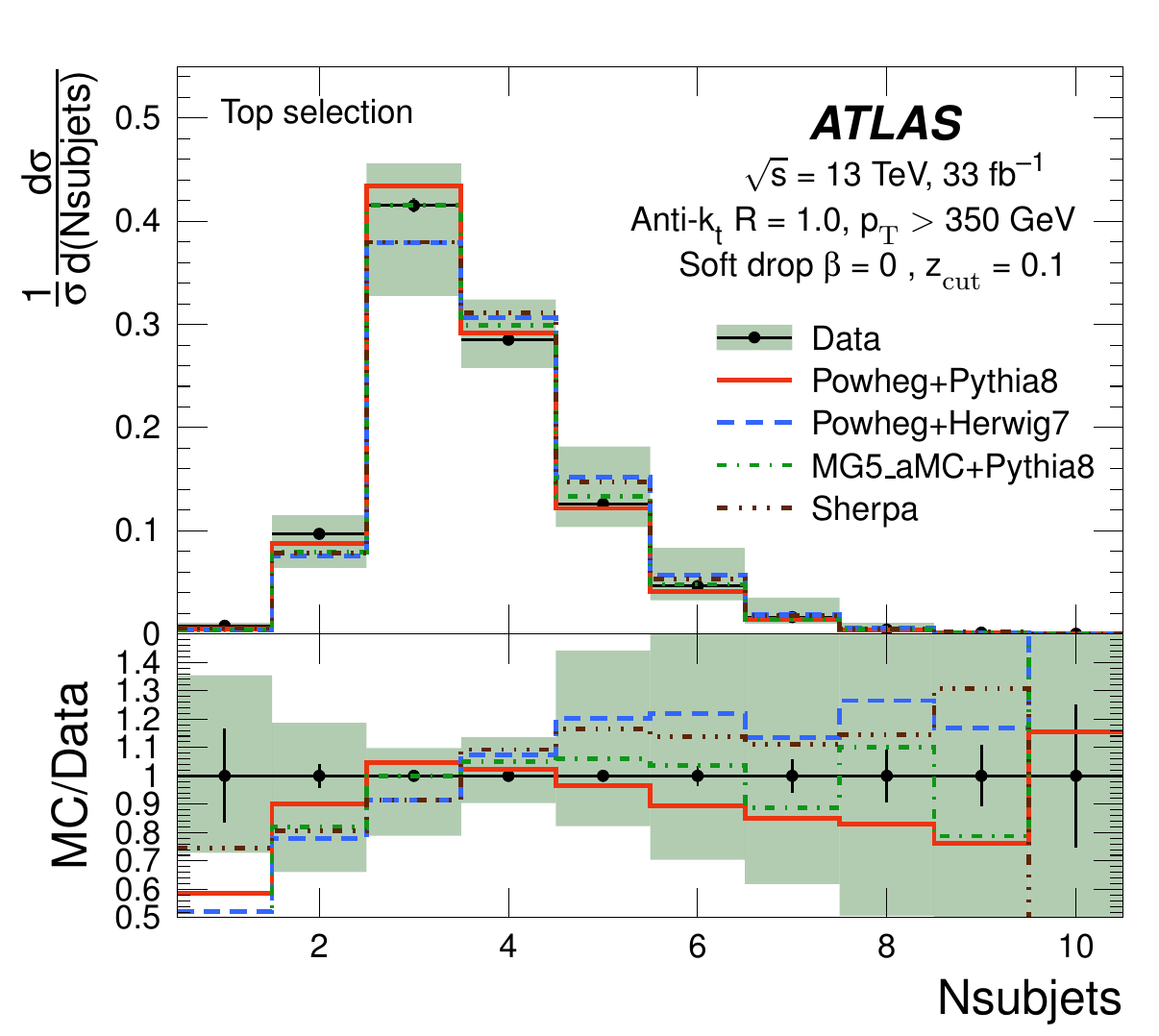}\\
\includegraphics[width=0.4\textwidth]{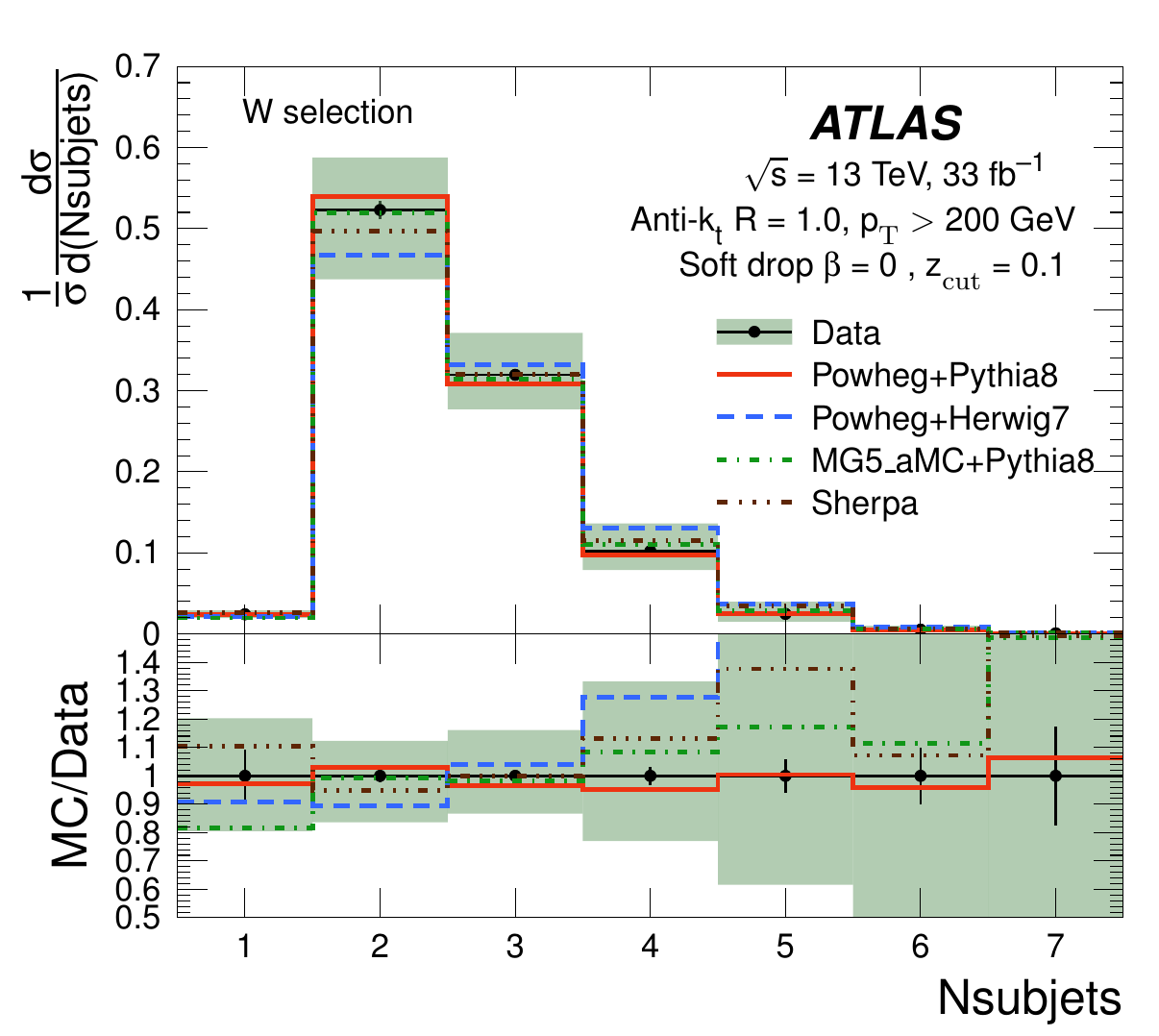}\qquad
\includegraphics[width=0.4\textwidth]{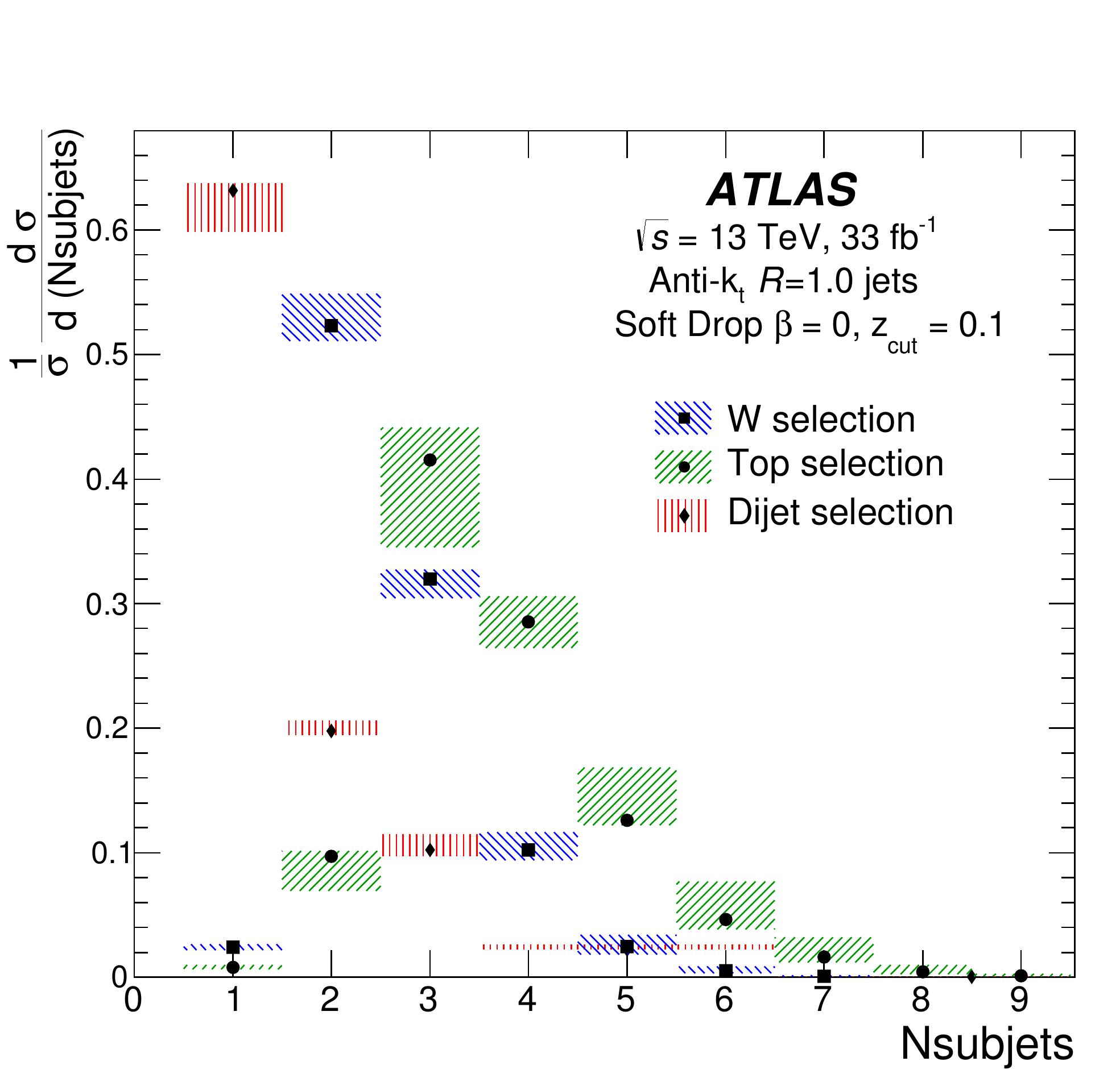}\\
\caption{Subjet multiplicity distributions compared with different MC predictions for soft-dropped
large-radius jets from dijet (top left), top (top right), and $W$ (bottom left) selections.
For the dijet selection, \sherpa is tested with two different hadronisation models.
Data are compared between the soft-dropped
large-radius jets for the three selections mentioned above (bottom right).
The shaded bands represent the total uncertainty, while
the error bars show the statistical uncertainty, except in the bottom right plot, where
the shaded areas represent the total uncertainty.
}
\label{fig:res-nsub}
\end{figure}

In Figure~\ref{fig:res-nsub}, the subjet multiplicity inside the large-radius jets from the three
different selections is compared with different MC predictions,
and the data are compared between the three selections.
While for the dijet selection most events have one subjet,
for the top selection and $W$ selection the distributions peak at three
and two subjets respectively, as expected. In both cases
a non-negligible fraction of events have more subjets, indicating the presence of semi-hard gluon radiation.
In the $W$ selection, the instances with one subjet are few,
while for the top selection, some fraction
of events have two subjets, indicating either non-containment of the top quark decay products,
or overlapping subjets that get reconstructed as a single subjet.
For the
dijet selection, \pythiaeight and \sherpa describe the data the best, while
for the top selection and $W$ selection, there is more spread among MC predictions.
Predictions from \herwigseven are very different from data for the dijet selection, a trend which
is consistent across all observables.
The difference between the different hadronisation models used in \sherpa is negligible.
Although these observables depend
on hadronisation modelling, it can be inferred that both models can be tuned to give a
good description of data.

\begin{figure}[h]
\centering
\includegraphics[width=0.4\textwidth]{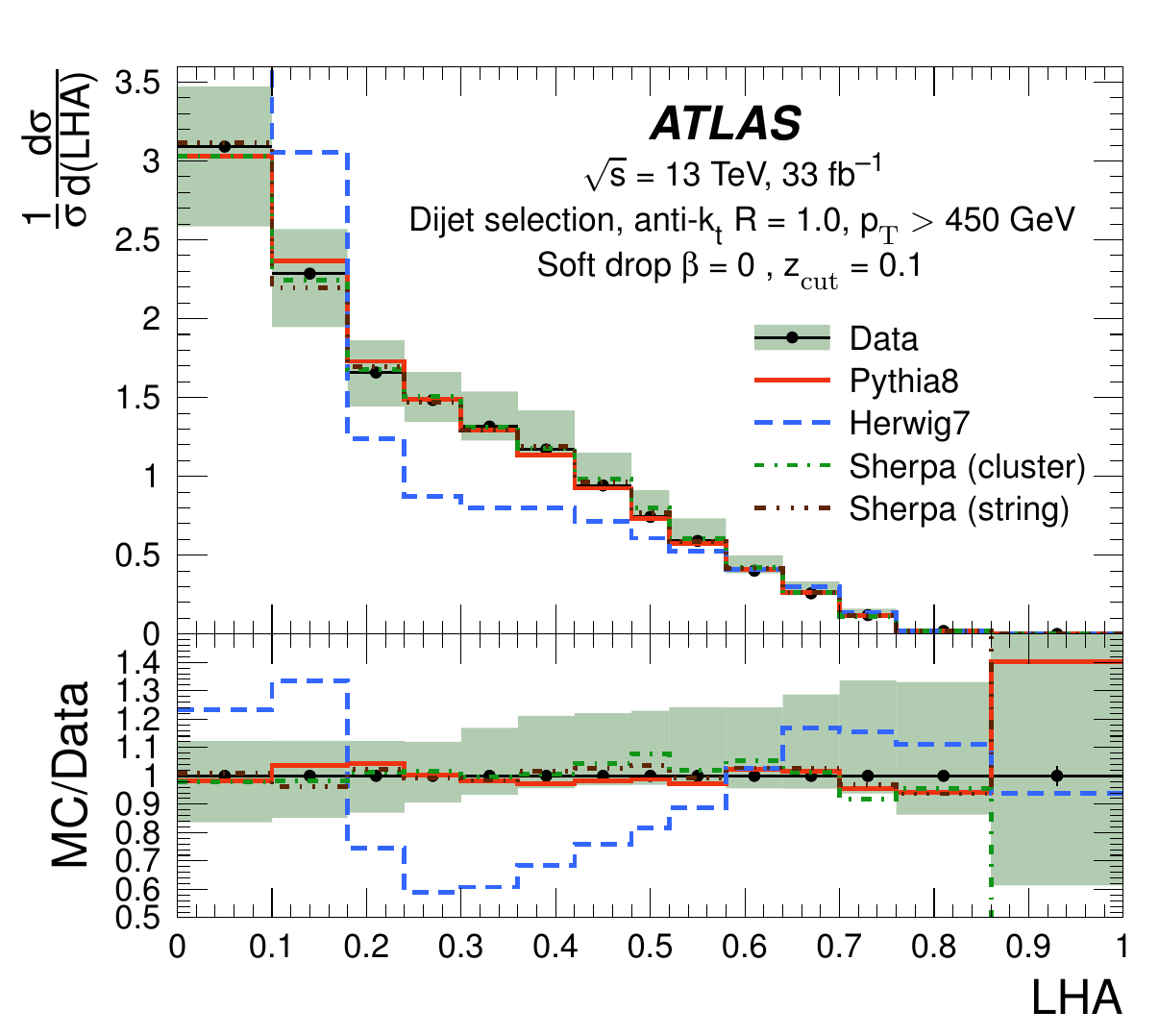}\qquad
\includegraphics[width=0.4\textwidth]{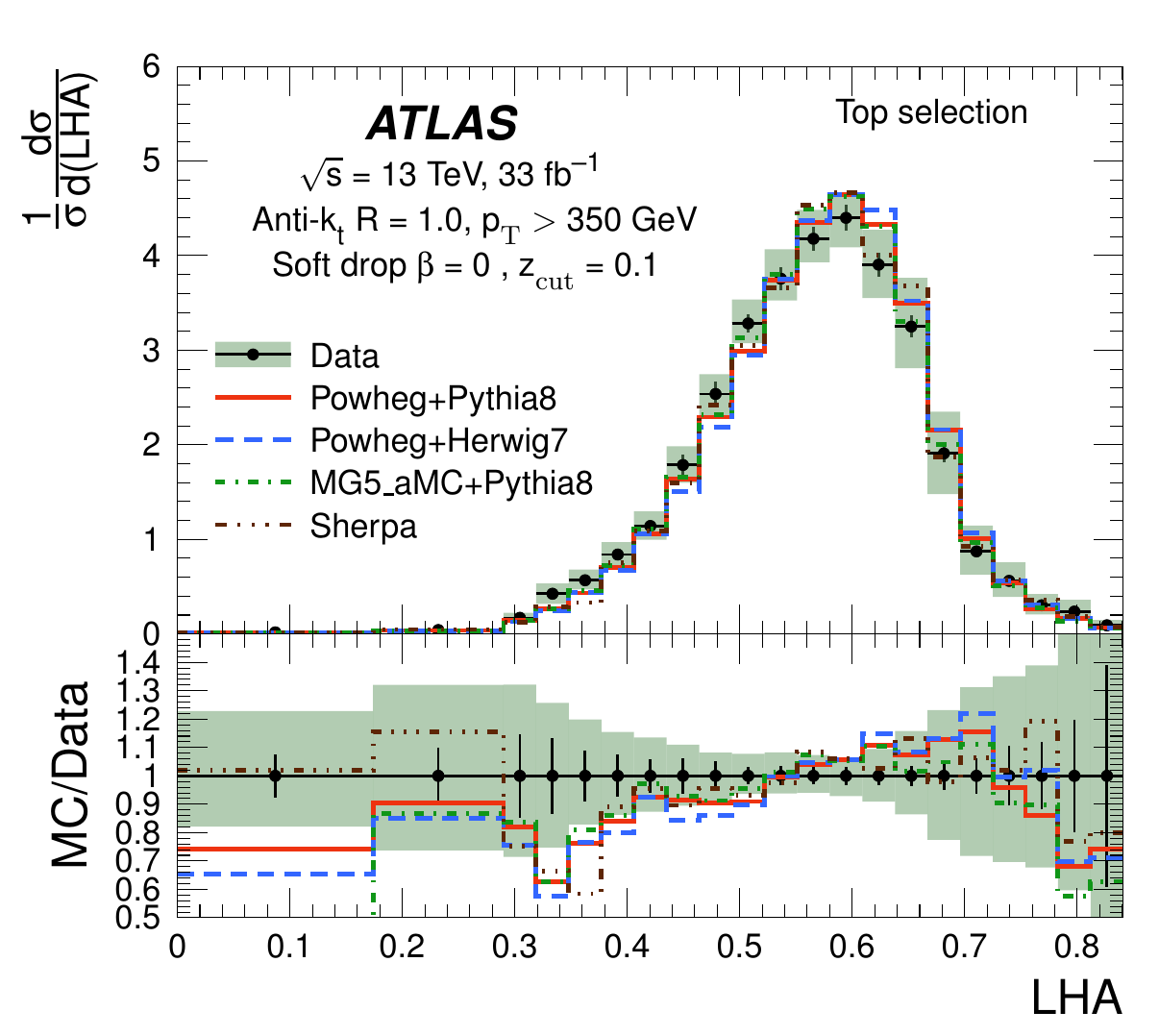}\\
\includegraphics[width=0.4\textwidth]{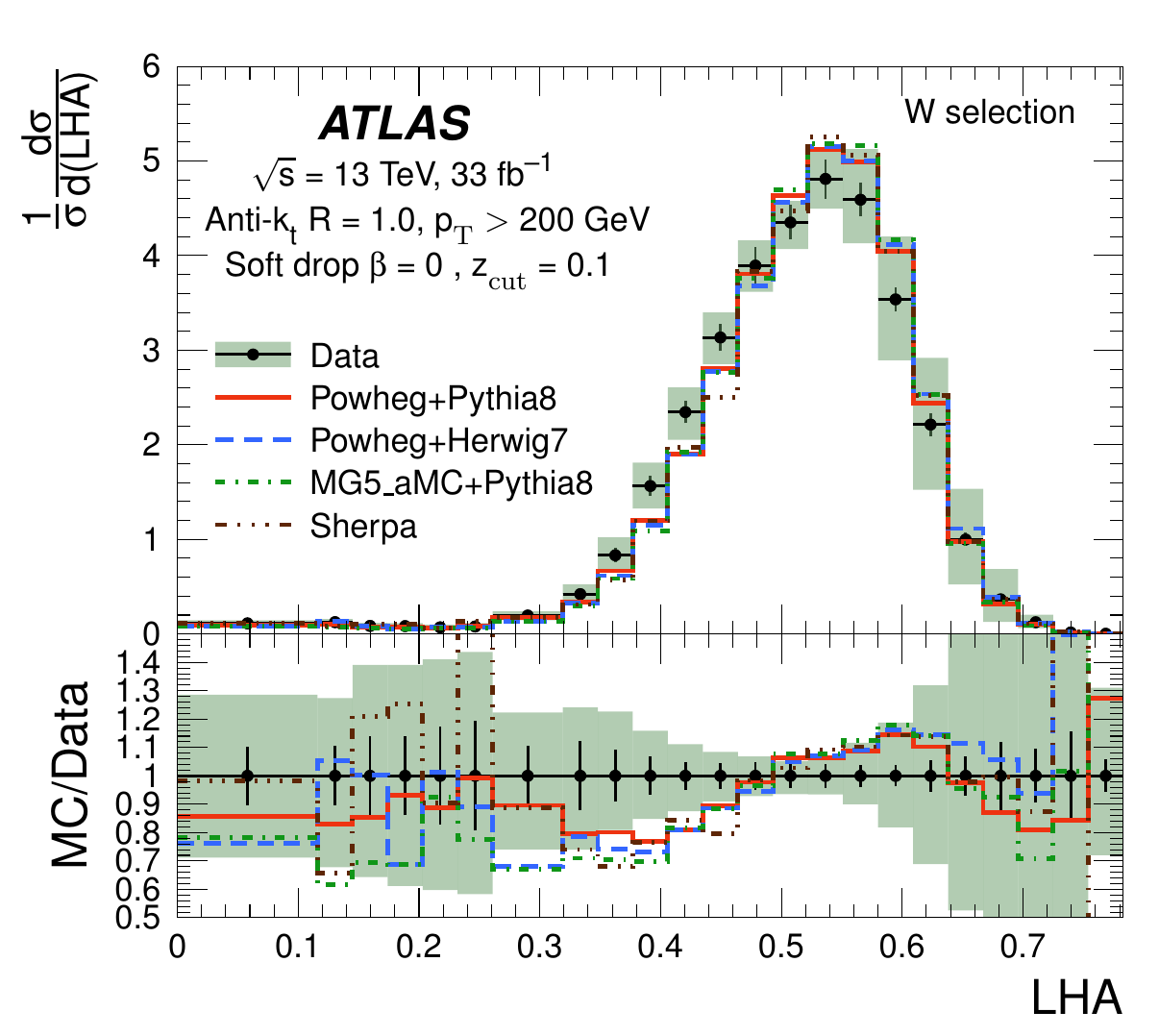}\qquad
\includegraphics[width=0.4\textwidth]{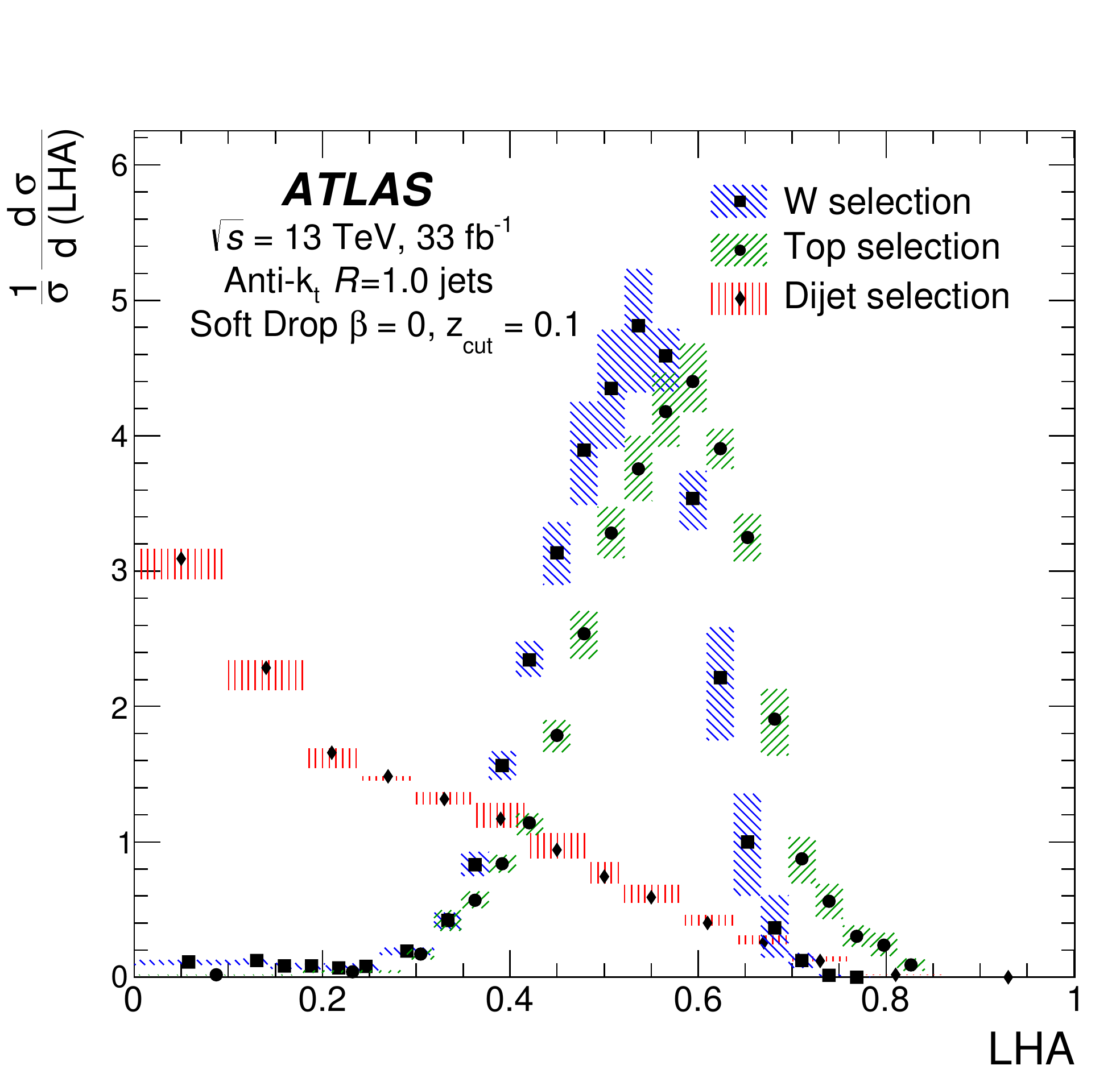}\\
\caption{Les Houches angularity is compared with different MC predictions for soft-dropped
large-radius jets from dijet (top left), top (top right), and $W$ (bottom left) selections.
For the dijet selection, \sherpa is tested with two different hadronisation models.
Data are compared between the soft-dropped
large-radius jets for the three selections mentioned above (bottom right).
The shaded bands represent the total uncertainty, while
the error bars show the statistical uncertainty,
except in the bottom right plot, where
the shaded areas represent the total uncertainty.
}
\label{fig:res-lha}
\end{figure}

In Figure~\ref{fig:res-lha}, the Les Houches angularity (LHA) is
compared between large-radius jets for the three selections
and with MC model predictions.
For the dijet selection, all models except \herwigseven describe the data, while
for the top and $W$ selections,
the level of agreement between all models and data is worse,
and the peaks of the distributions in the models are shifted relative to those in data.
While in the case of the top and $W$ selections the shapes are similar,
the distribution for the dijet selection peaks at the lowest value.
This indicates  that the additional radiation in quark/gluon jets is soft, with
little activity away from the large-radius jet axis, while for the large-radius jets from
top quarks and $W$ bosons,
there are hard emissions separated by appreciable angles.

\begin{figure}[h]
\centering
\includegraphics[width=0.4\textwidth]{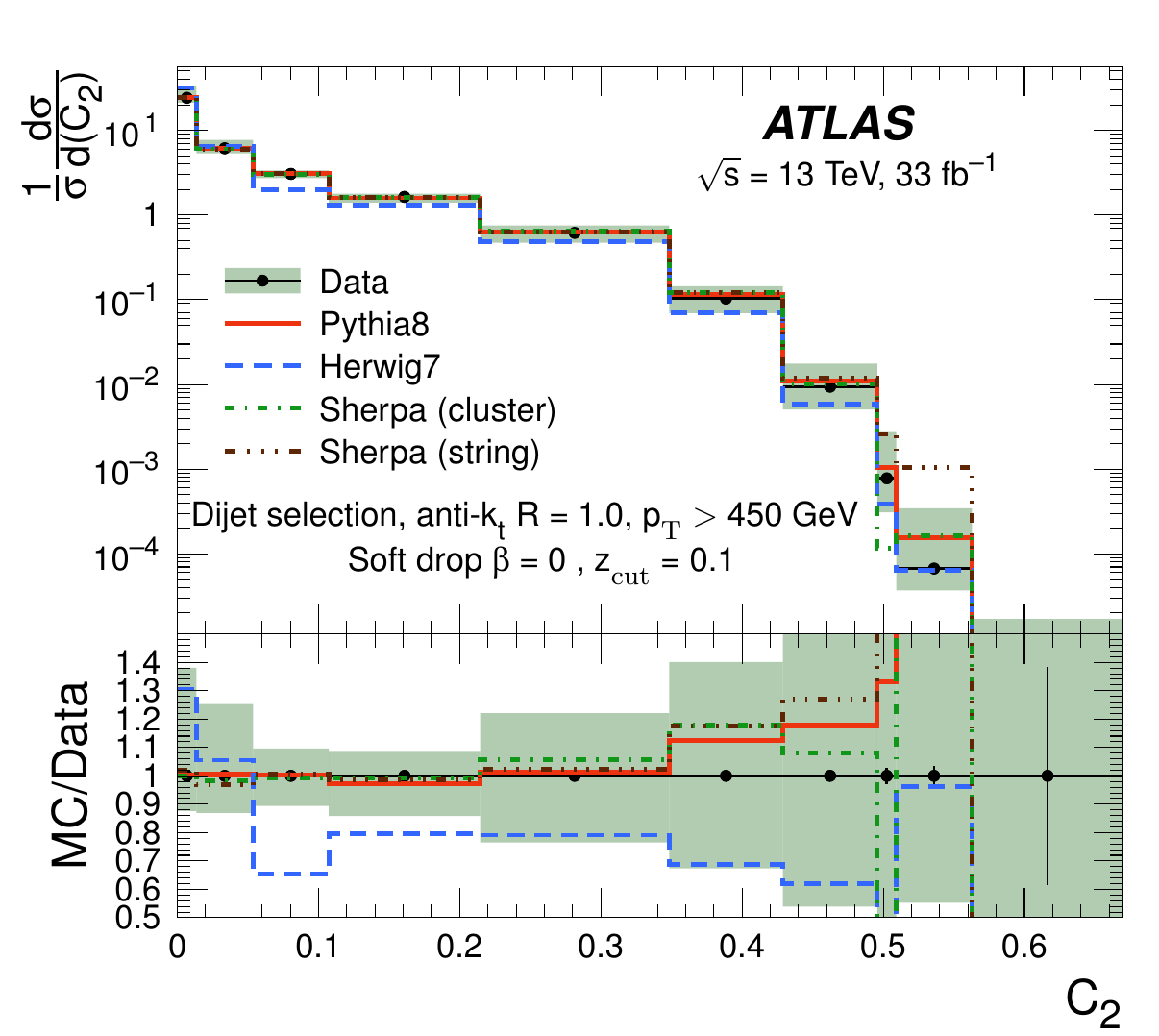}\qquad
\includegraphics[width=0.4\textwidth]{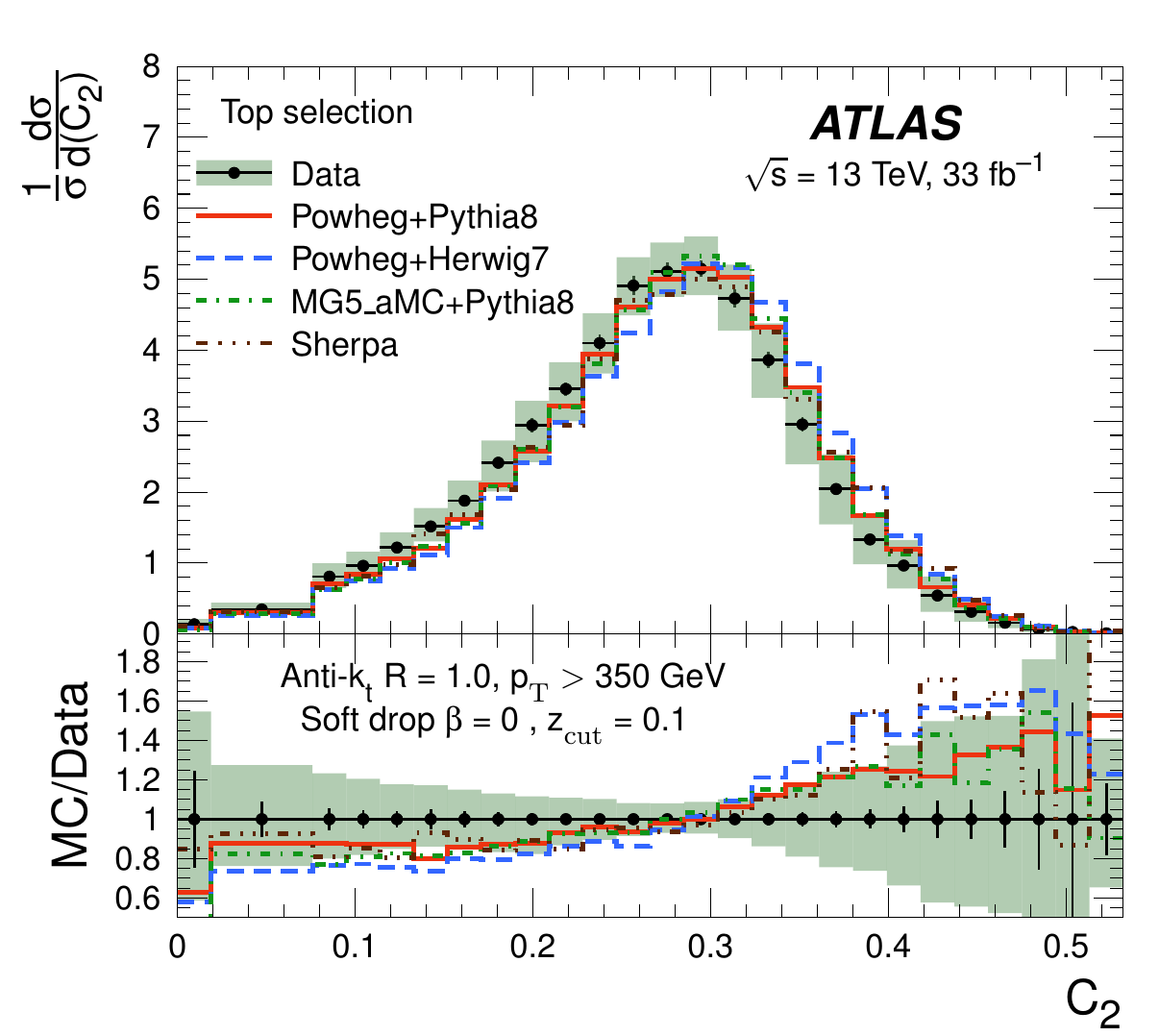}\\
\includegraphics[width=0.4\textwidth]{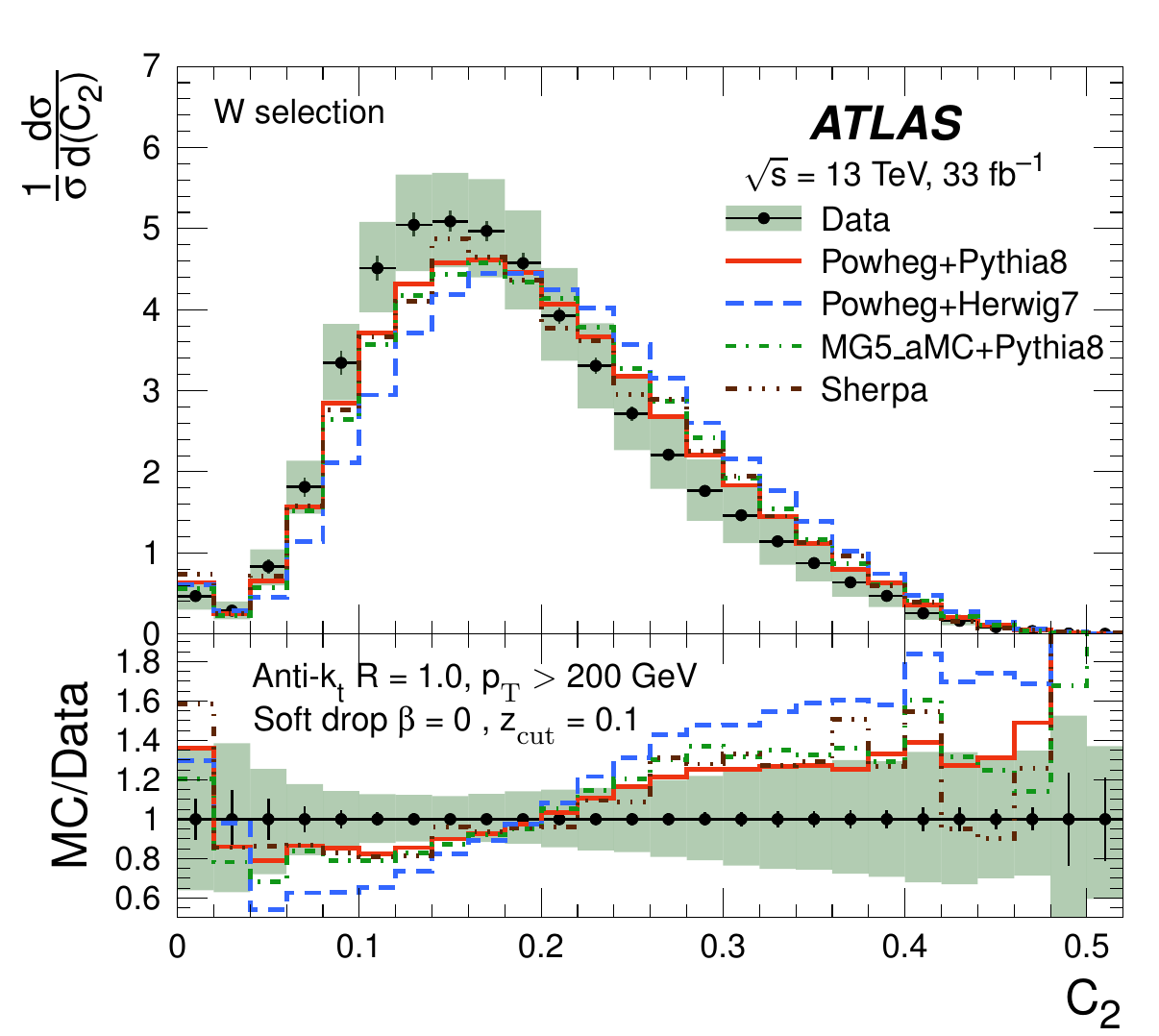}\qquad
\includegraphics[width=0.4\textwidth]{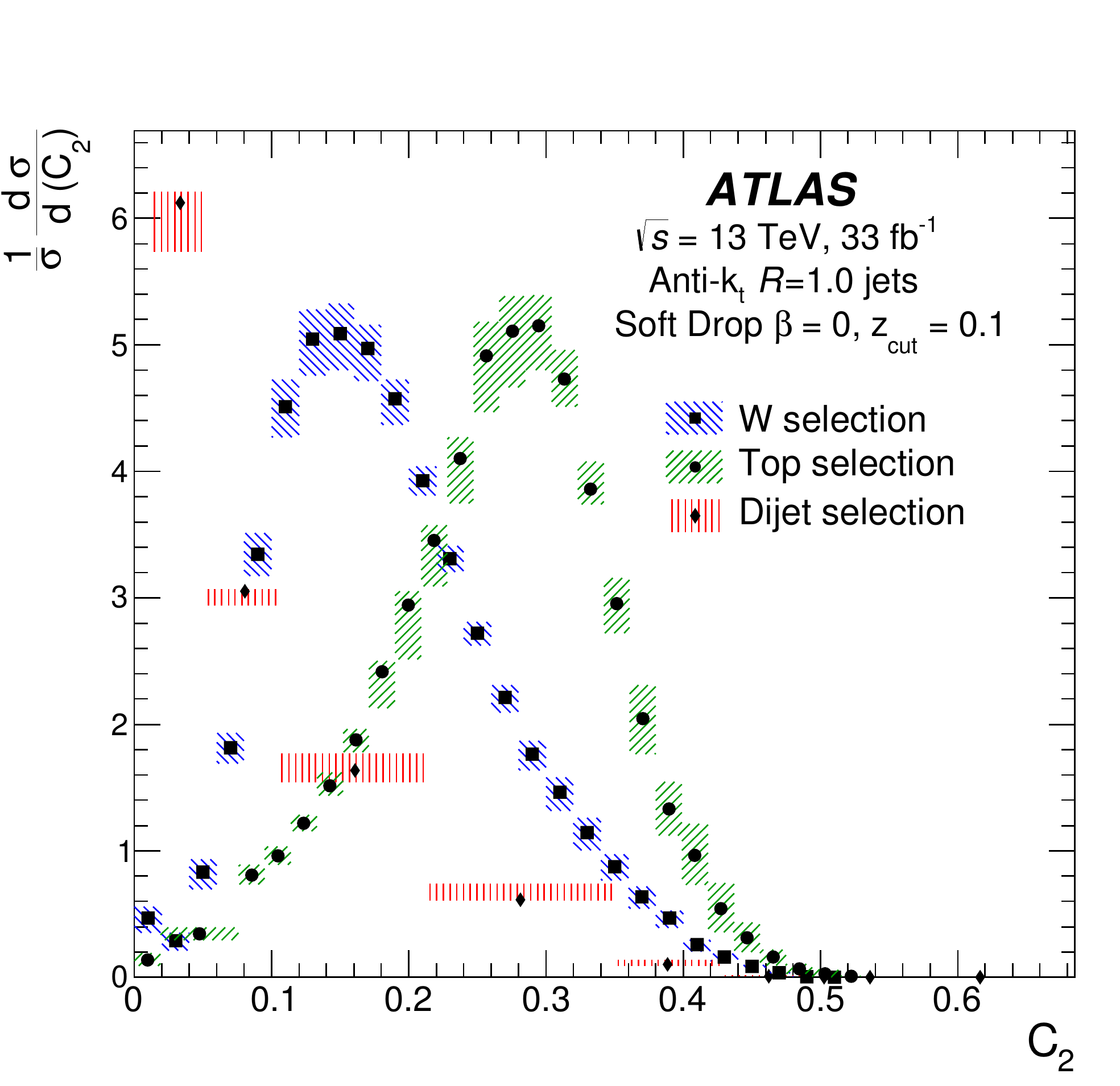}\\
\caption{
The distributions of $C_2$ compared with different MC predictions for soft-dropped
large-radius jets from dijet (top left), top (top right), and $W$ (bottom left) selections.
For the dijet selection, \sherpa is tested with two different hadronisation models.
Data are compared between the soft-dropped
large-radius jets for the three selections mentioned above (bottom right).
The shaded bands represent the total uncertainty, while
the error bars show the statistical uncertainty,
except in the bottom right plot, where
the shaded areas represent the total uncertainty.
}
\label{fig:res-c2}
\end{figure}

In Figure~\ref{fig:res-c2}, a comparison of $C_2$ among
the three different selections with MC is presented,
as well as a comparisons of data and MC predictions for each selection.
For the dijet selection, all models except \herwigseven describe the data well, while
for the top and $W$ selections, the models predict shapes that differ from
data, with {\powheg}+\herwigseven performing
somewhat worse than the rest.
The three distributions have distinct peaks, corresponding to their substructure.
The value of $C_2$ increases
as the number of subjets inside the large-radius jets increases.

\begin{figure}[h]
\centering
\includegraphics[width=0.4\textwidth]{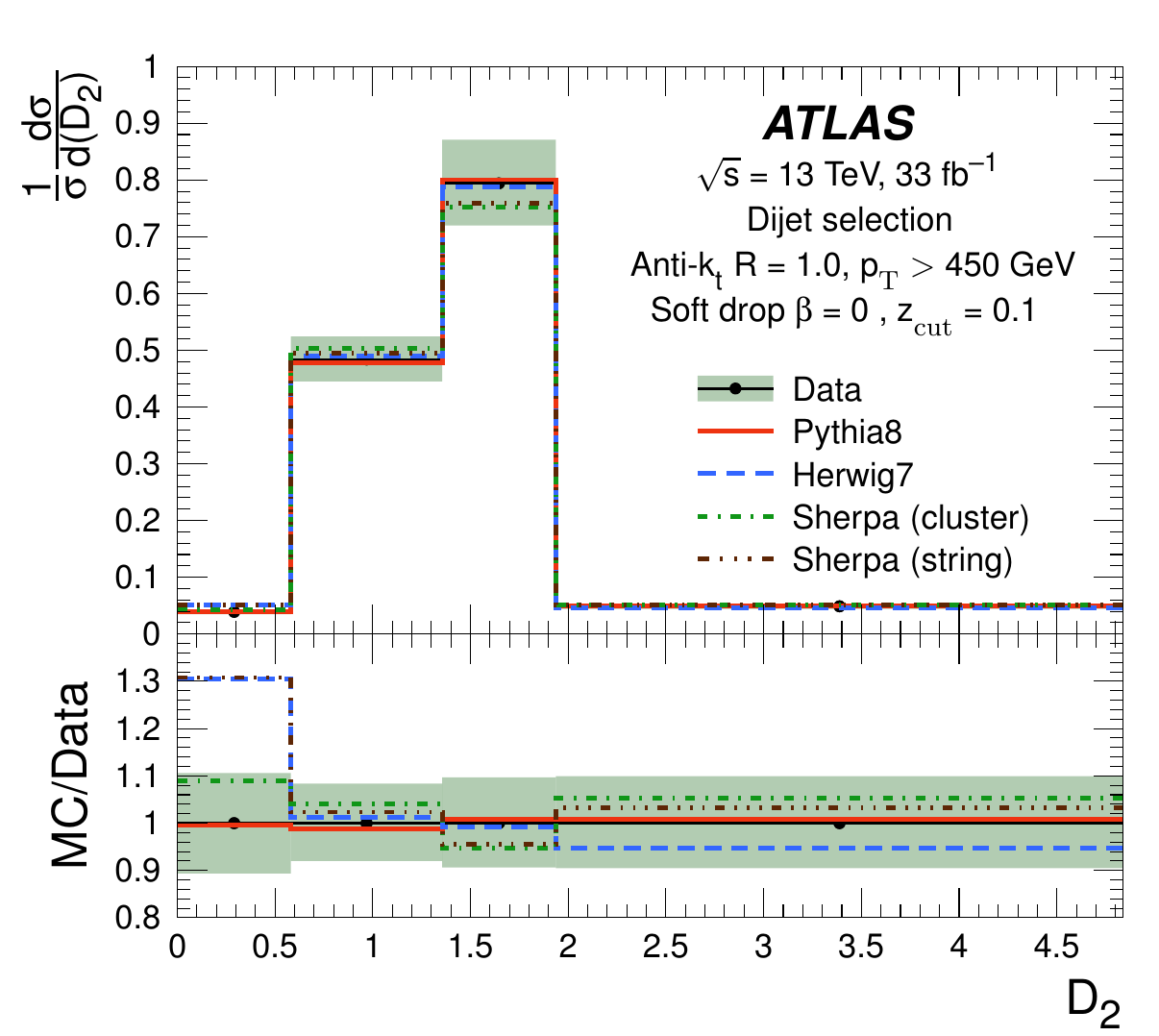}\qquad
\includegraphics[width=0.4\textwidth]{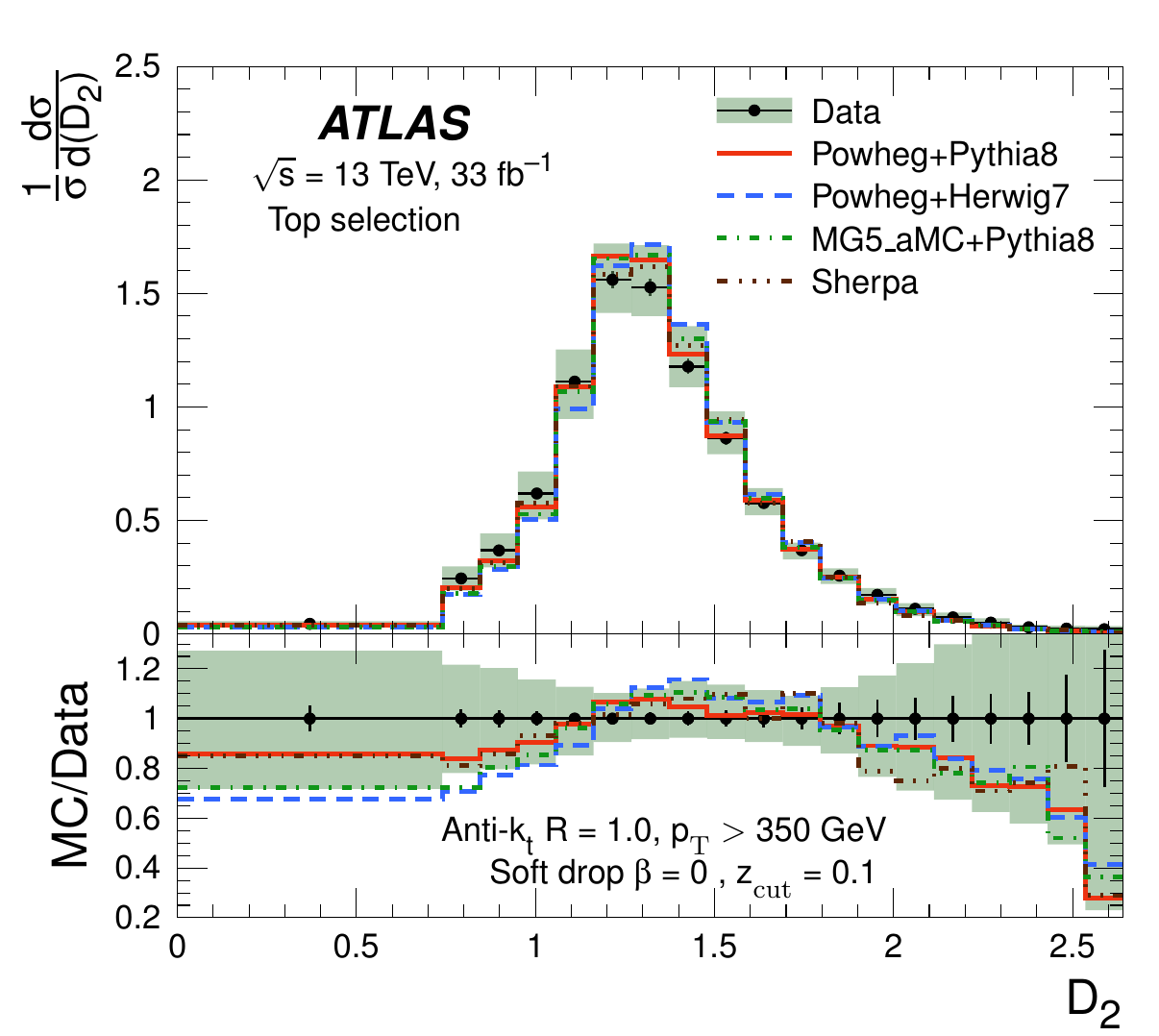}\\
\includegraphics[width=0.4\textwidth]{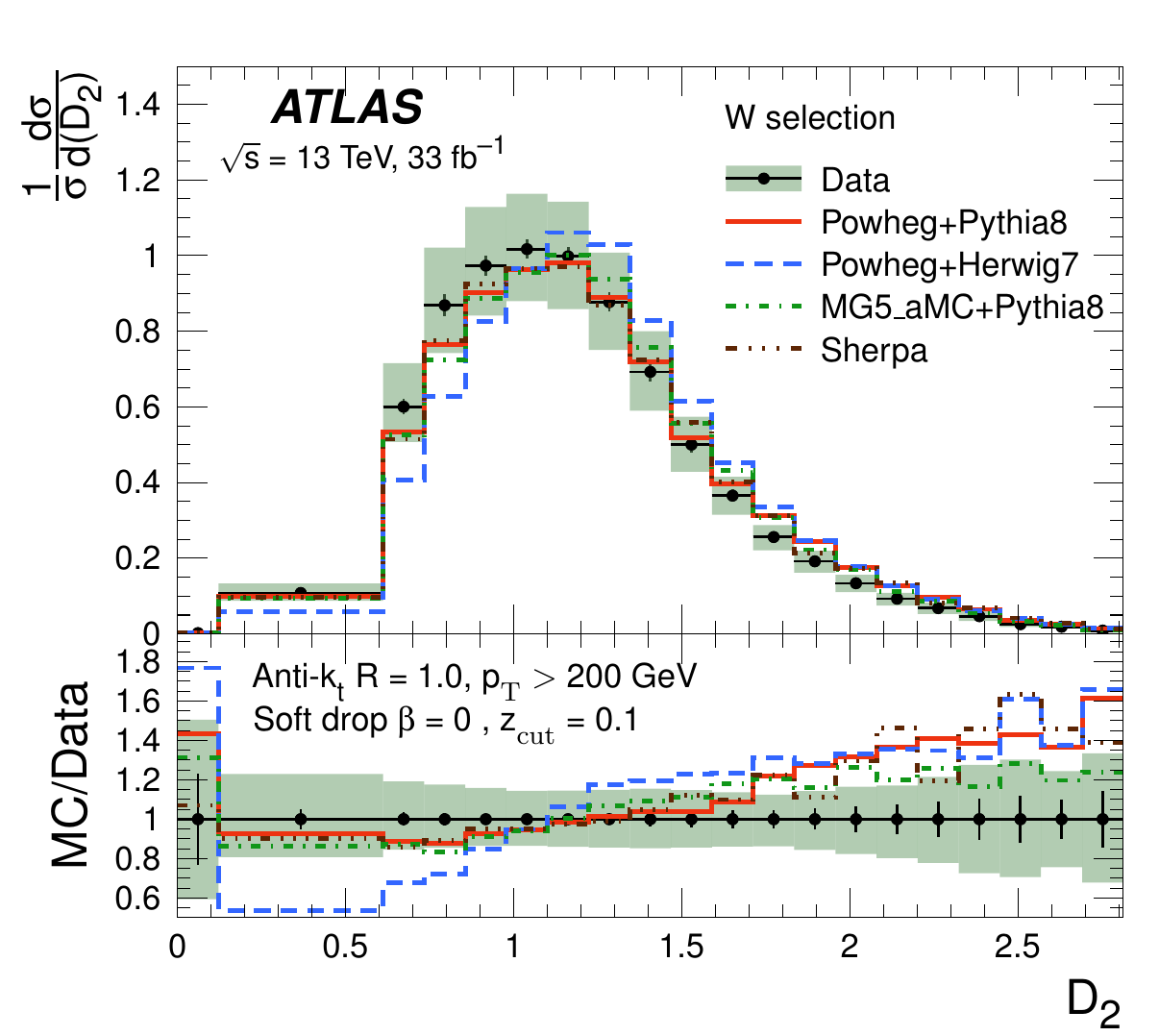}\qquad
\includegraphics[width=0.4\textwidth]{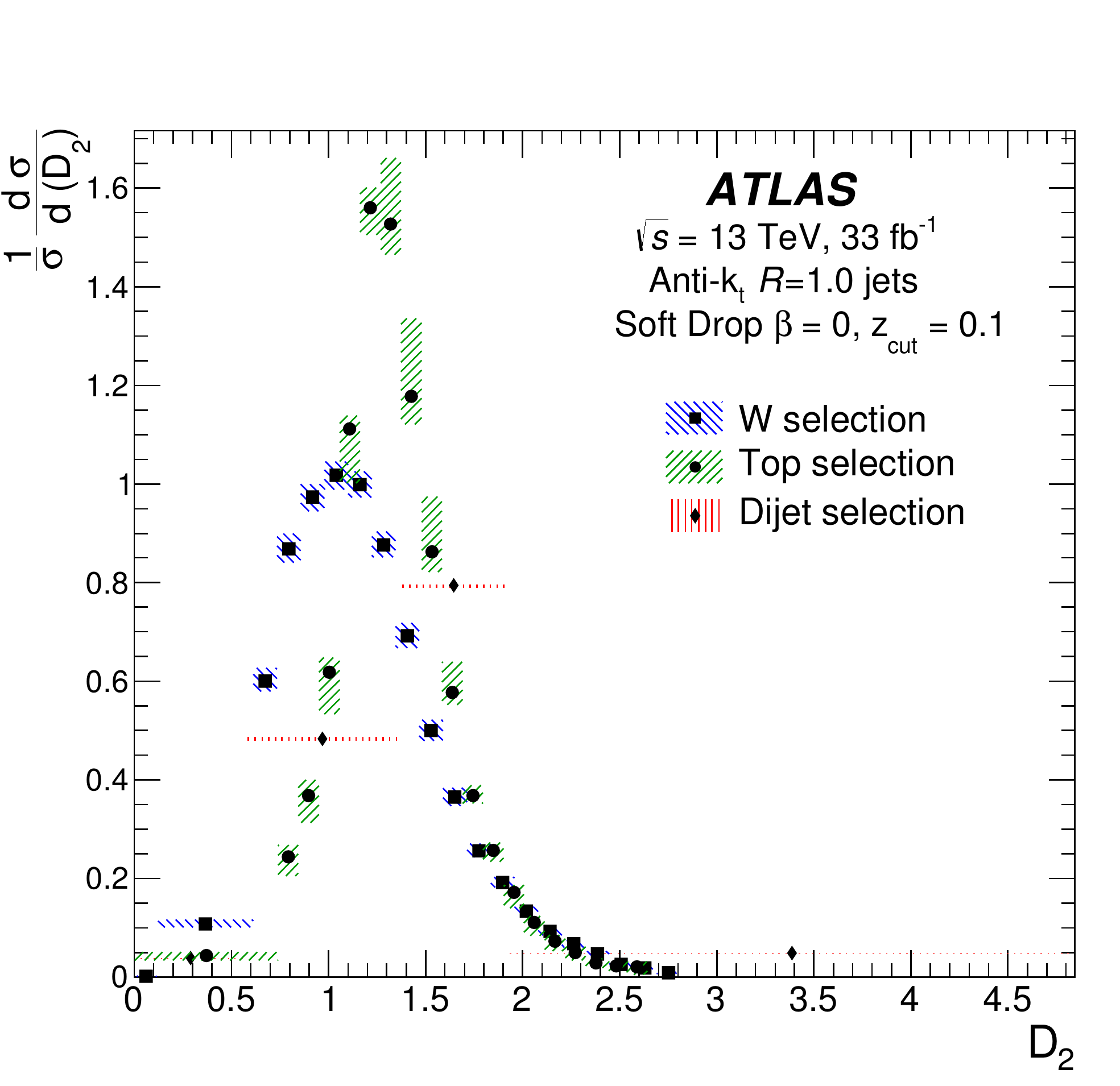}\\
\caption{
The distributions of $D_2$ compared with different MC predictions for soft-dropped
large-radius jets from dijet (top left), top (top right), and $W$ (bottom left) selections.
For the dijet selection, \sherpa is tested with two different hadronisation models.
Data are compared between the soft-dropped
large-radius jets for the three selections mentioned above (bottom right).
The shaded bands represent the total uncertainty, while
the error bars show the statistical uncertainty,
except in the bottom right plot, where
the shaded areas represent the total uncertainty.
}
\label{fig:res-d2}
\end{figure}

In Figure~\ref{fig:res-d2},
comparisons of the data with MC predictions for $D_2$ reveal some interesting features.
For the dijet selection, most of the models
describe the data well, and for the top selection the some differences can be seen.
For the $W$ selection,
all MC predictions have a peak shifted relative to data, suggesting that the models are
overestimating gluon radiation.
The distributions in data for the three selections are also compared in Figure~\ref{fig:res-d2} (bottom right),
where peaks at different values are observed.

\begin{figure}[h]
\centering
\includegraphics[width=0.4\textwidth]{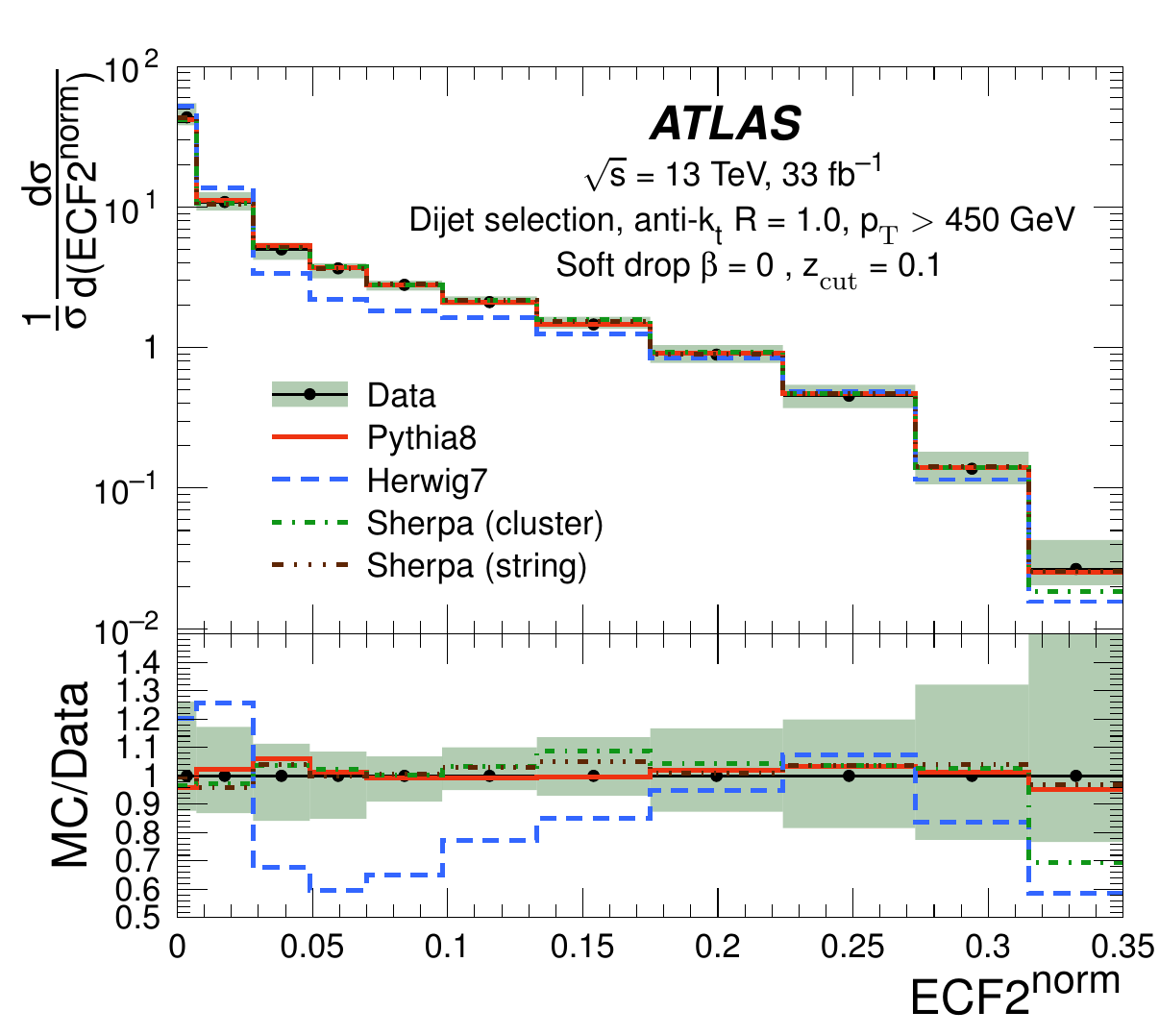}\qquad
\includegraphics[width=0.4\textwidth]{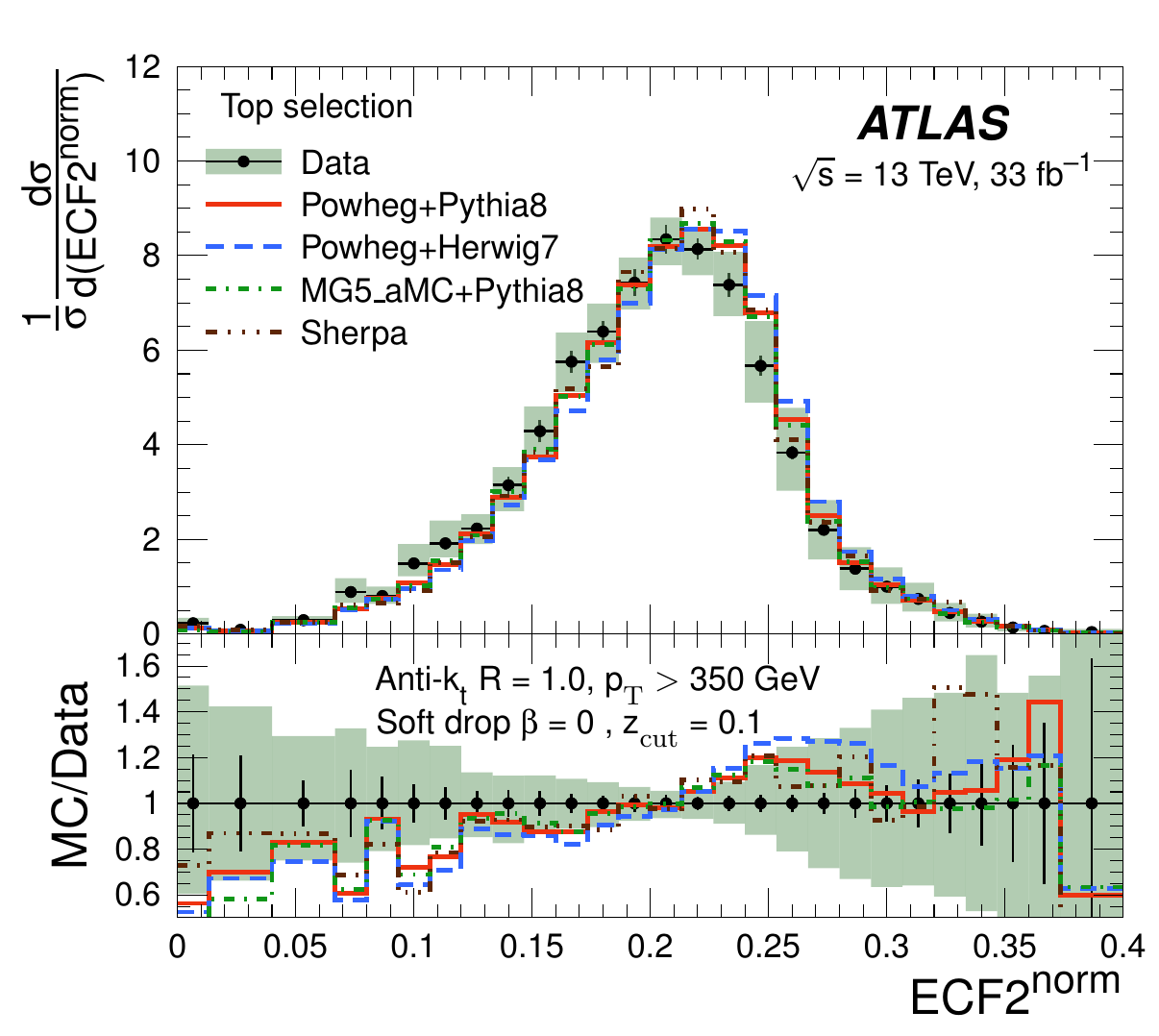}\\
\includegraphics[width=0.4\textwidth]{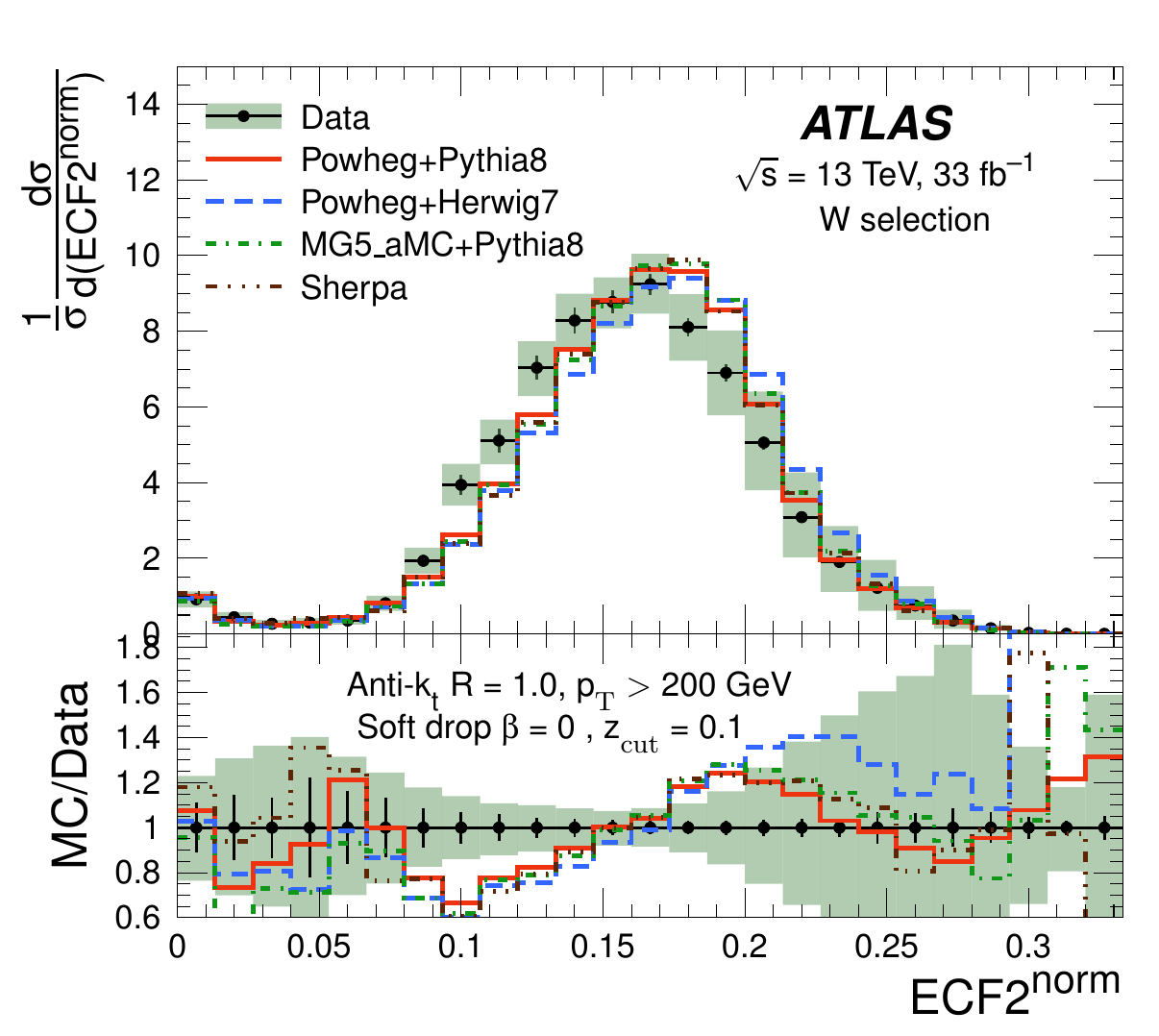}\qquad
\includegraphics[width=0.4\textwidth]{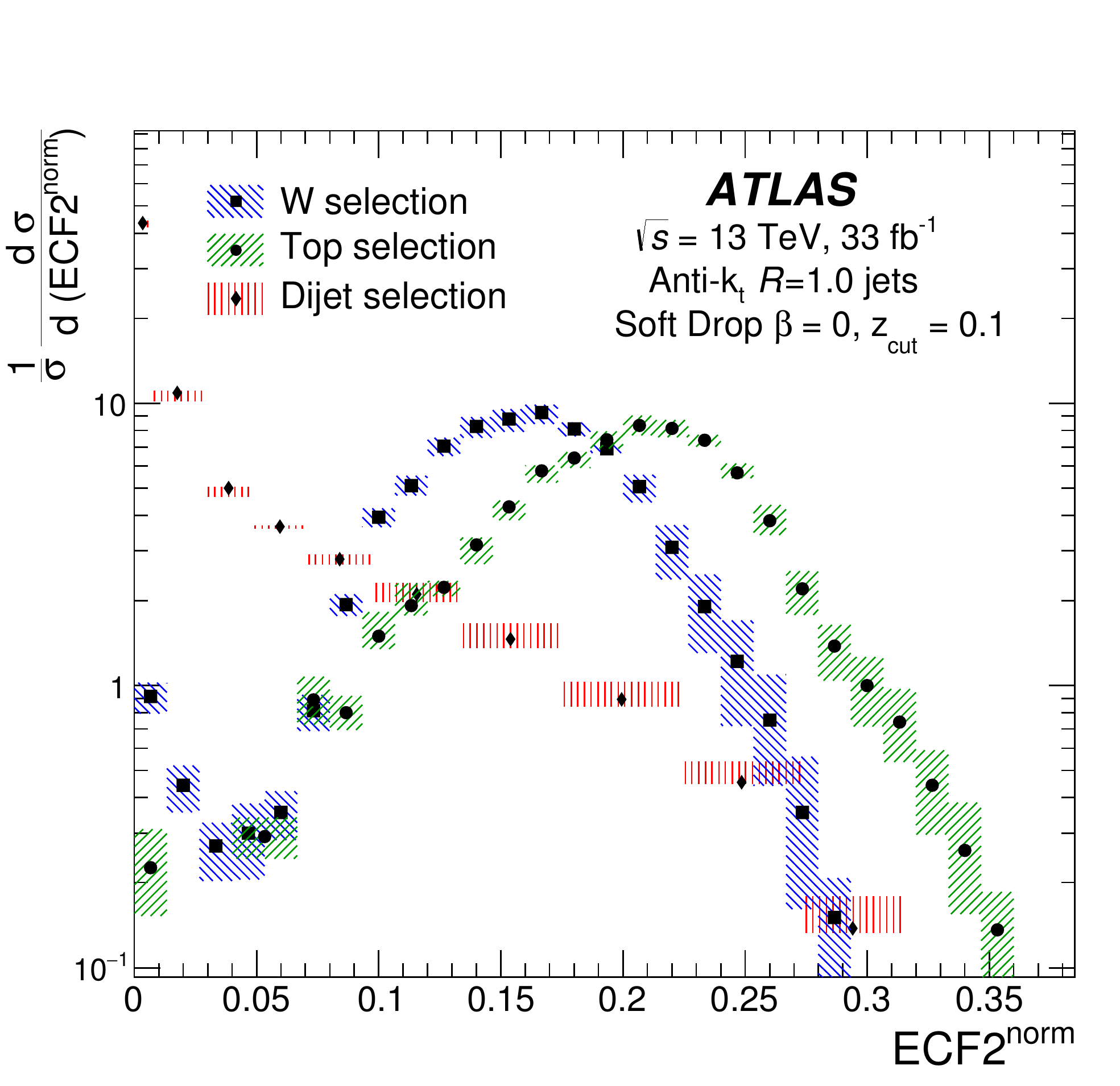}\\
\caption{
The distributions of $ECF2^{\textrm{norm}}$ compared with different MC predictions for soft-dropped
large-radius jets from dijet (top left), top (top right), and $W$ (bottom left) selections.
For the dijet selection, \sherpa is tested with two different hadronisation models.
Data are compared between the soft-dropped
large-radius jets for the three selections mentioned above (bottom right).
The subscript norm indicates that normalised versions of $ECF2^{\textrm{norm}}$ are used.
The shaded bands represent the total uncertainty, while
the error bars show the statistical uncertainty,
except in the bottom right plot, where
the shaded areas represent the total uncertainty.
}
\label{fig:res-ecf2}
\end{figure}

The distributions of $ECF2^{\textrm{norm}}$, as shown in Figure~\ref{fig:res-ecf2}
for the different selections, can discriminate between
events with two and three prong decays as opposed to one prong decay.
Similarly to $C_2$, for the
dijet selection, all models except \herwigseven describe the data well, while
for the top and $W$ selections, the models predict shapes that differ somewhat from data,
with agreement being worse for the $W$ selection case.

\begin{figure}[h]
\centering
\includegraphics[width=0.4\textwidth]{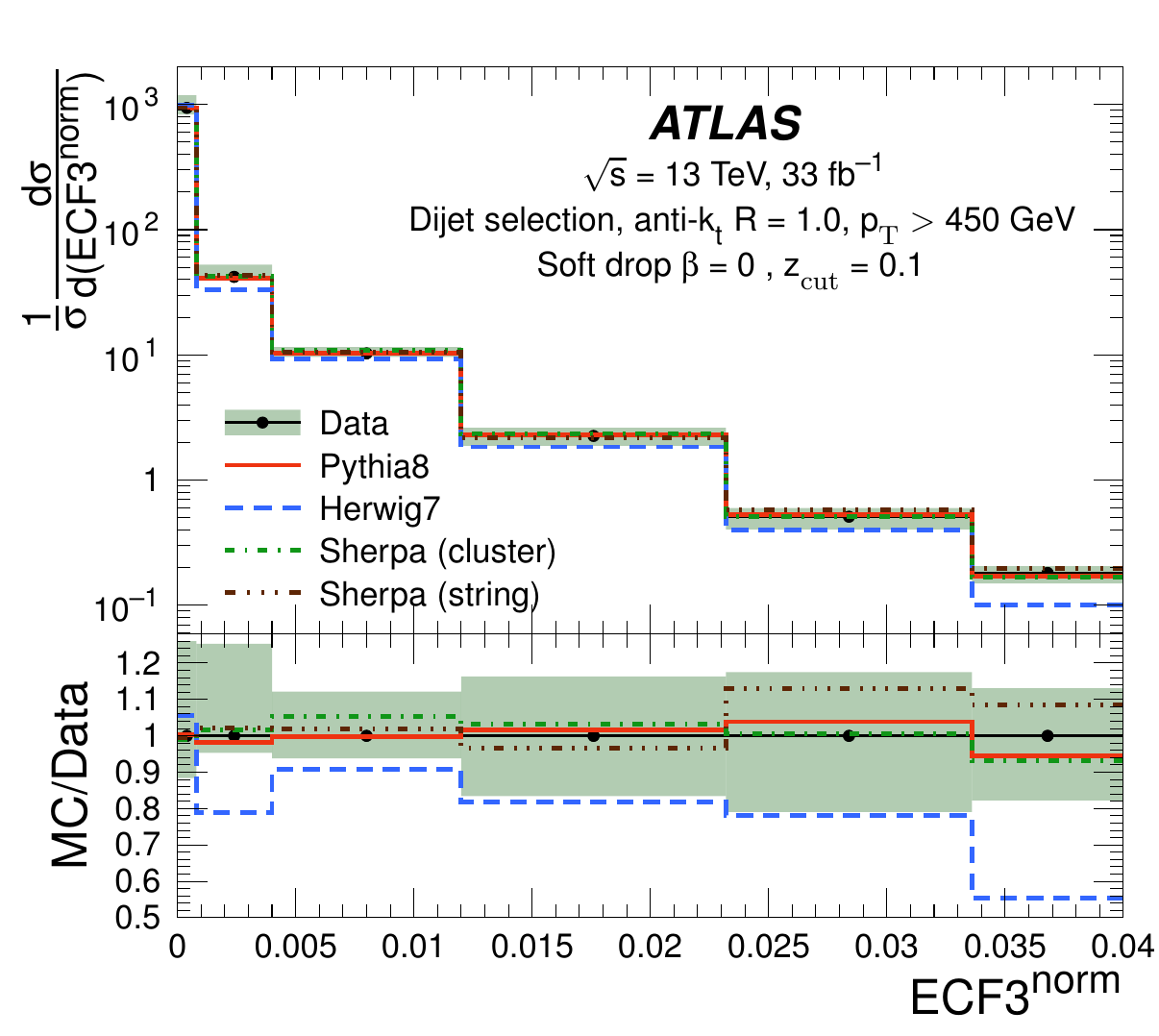}\qquad
\includegraphics[width=0.4\textwidth]{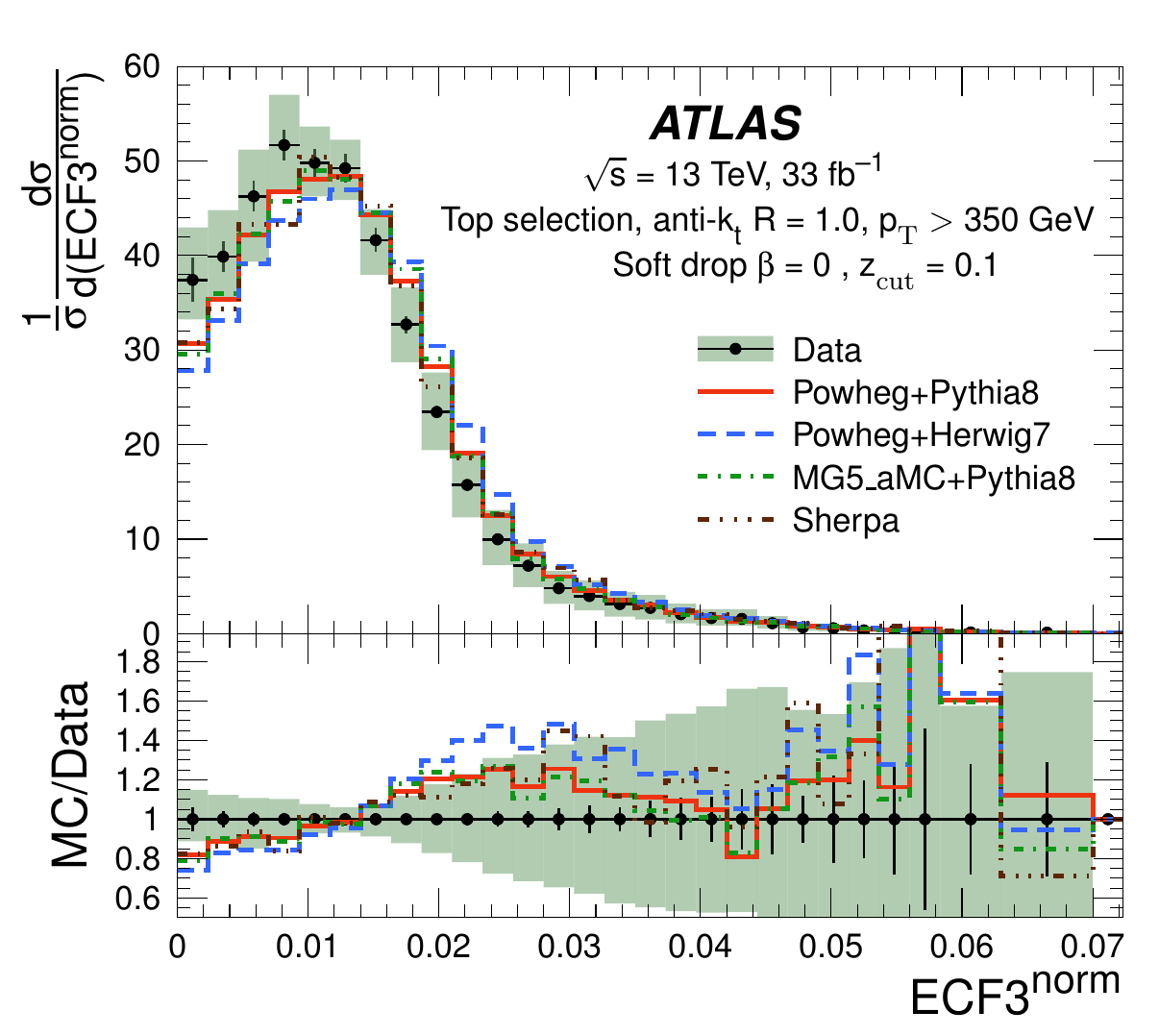}\\
\includegraphics[width=0.4\textwidth]{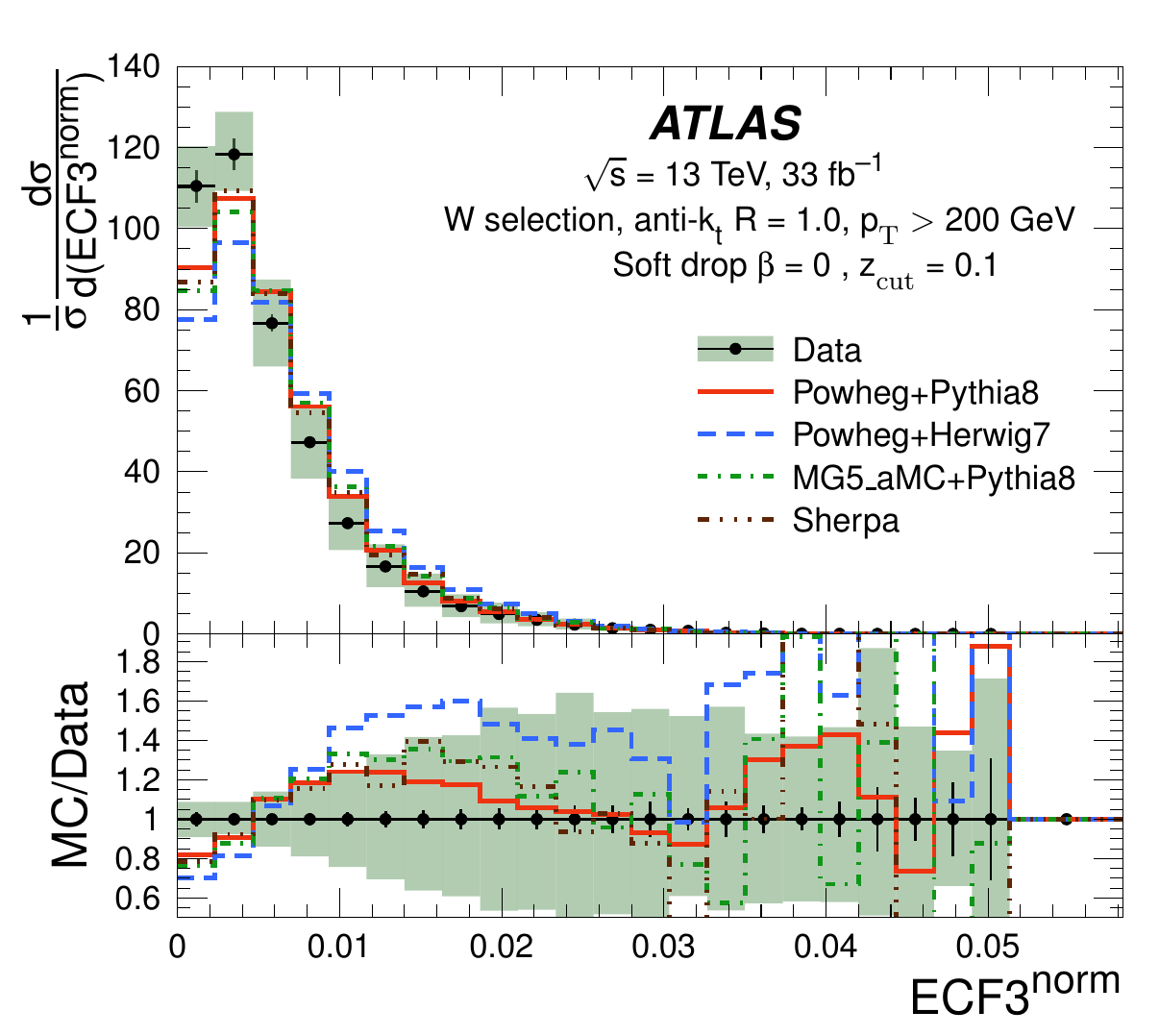}\qquad
\includegraphics[width=0.4\textwidth]{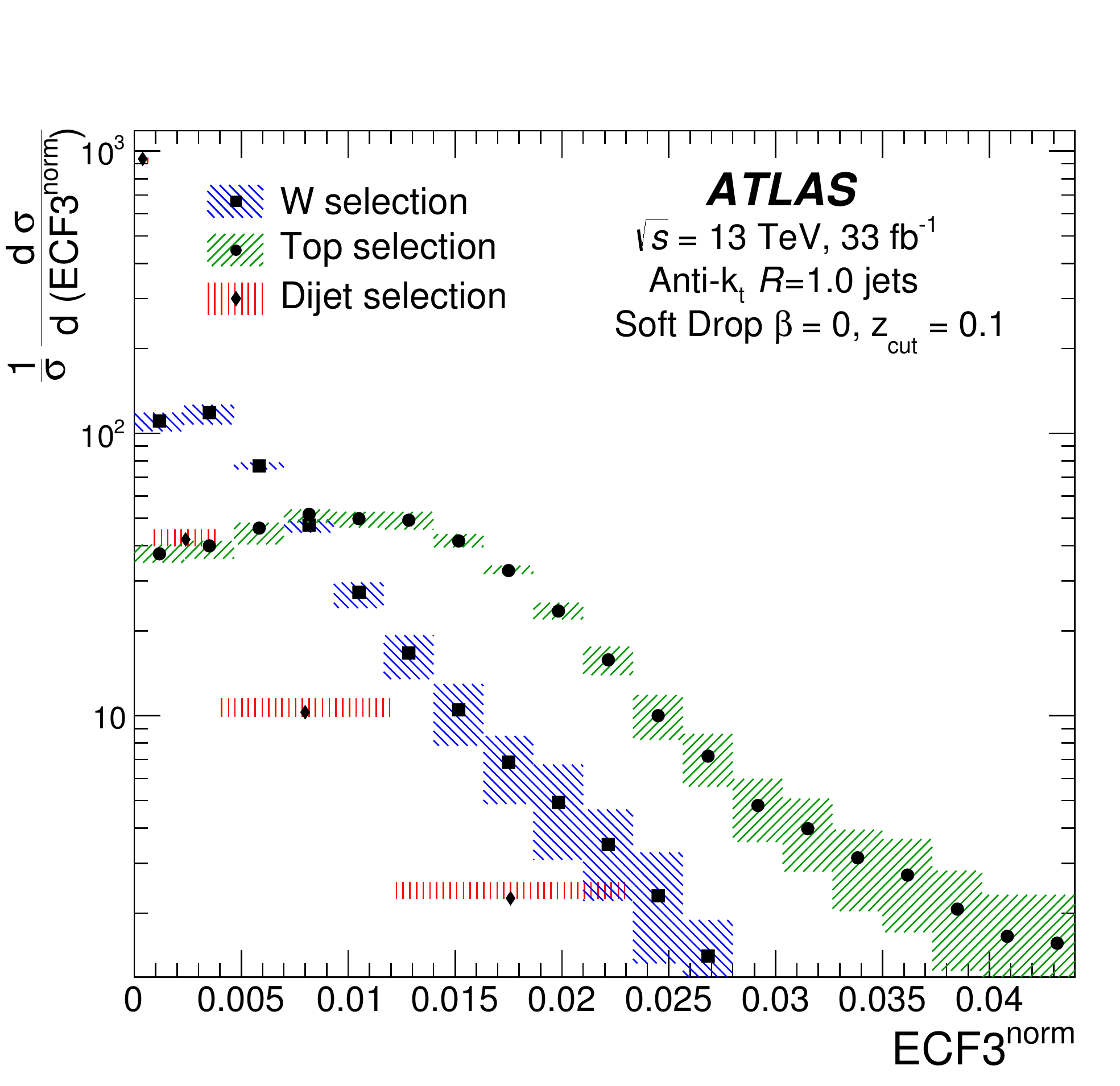}\\
\caption{
The distributions of $ECF3^{\textrm{Norm}}$ are compared with different MC predictions for soft-dropped
large-radius jets from dijet (top left), top (top right), and $W$ (bottom left) selections.
For the dijet selection, \sherpa is tested with two different hadronisation models.
Data are compared between the soft-dropped
large-radius jets for the three selections mentioned above (bottom right
The superscript ``norm'' indicates that normalised versions of $ECF3^{\textrm{Norm}}$ are used.
The shaded bands represent the total uncertainty, while
the error bars show the statistical uncertainty,
except in the bottom right plot, where
the shaded areas represent the total uncertainty.
}
\label{fig:res-ecf3}
\end{figure}

The modelling of $ECF3^{\textrm{norm}}$
in the dijet selection is better for \pythiaeight than for the other generators,
as shown in Figure~\ref{fig:res-ecf3}.
For the top and $W$ selections,
none of the models describe the shape of the data distribution well,
with noticeable differences at low values.
The three different selections again show distinct shapes.
 
\begin{figure}[h]
\centering
\includegraphics[width=0.4\textwidth]{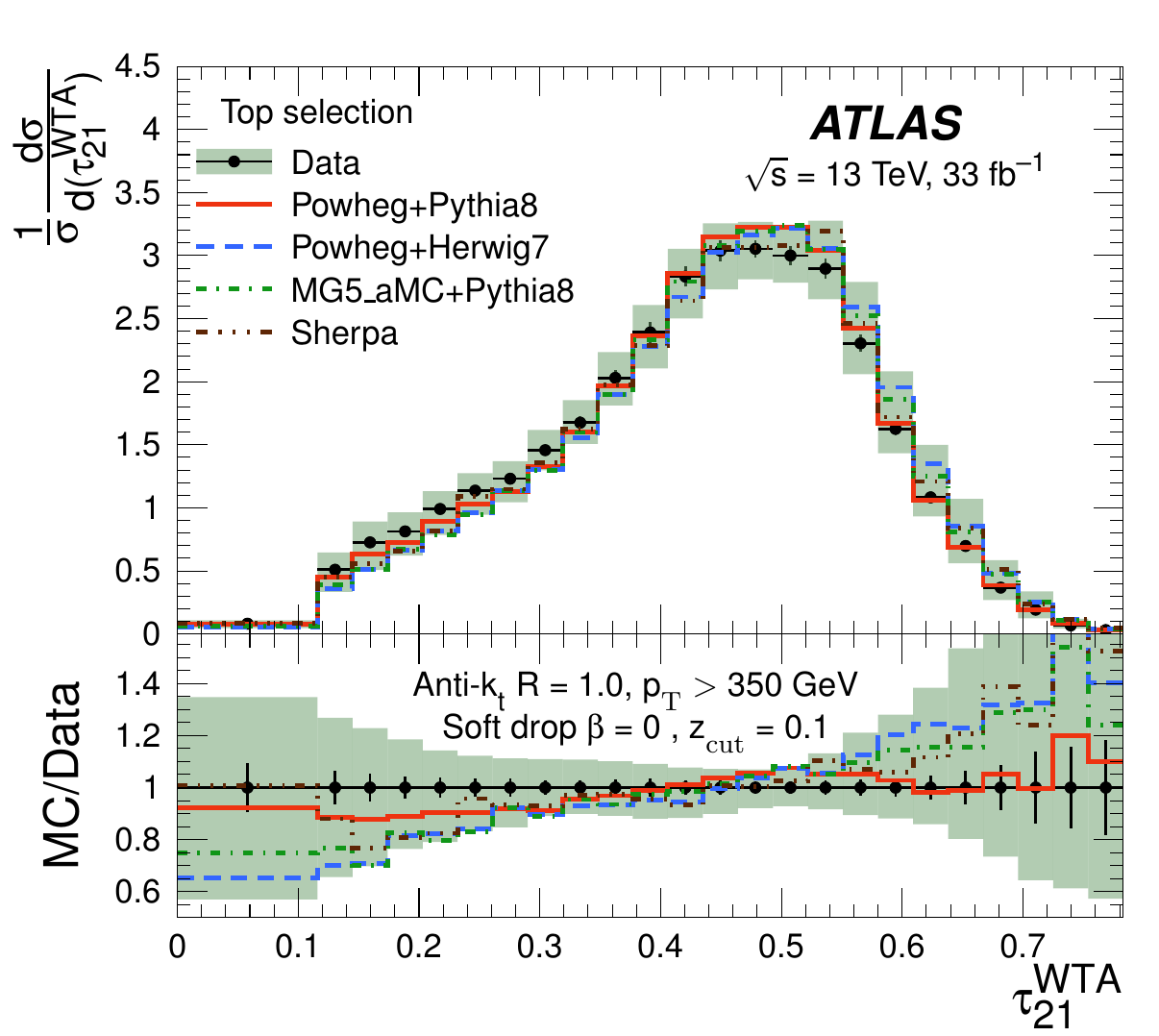}\qquad
\includegraphics[width=0.4\textwidth]{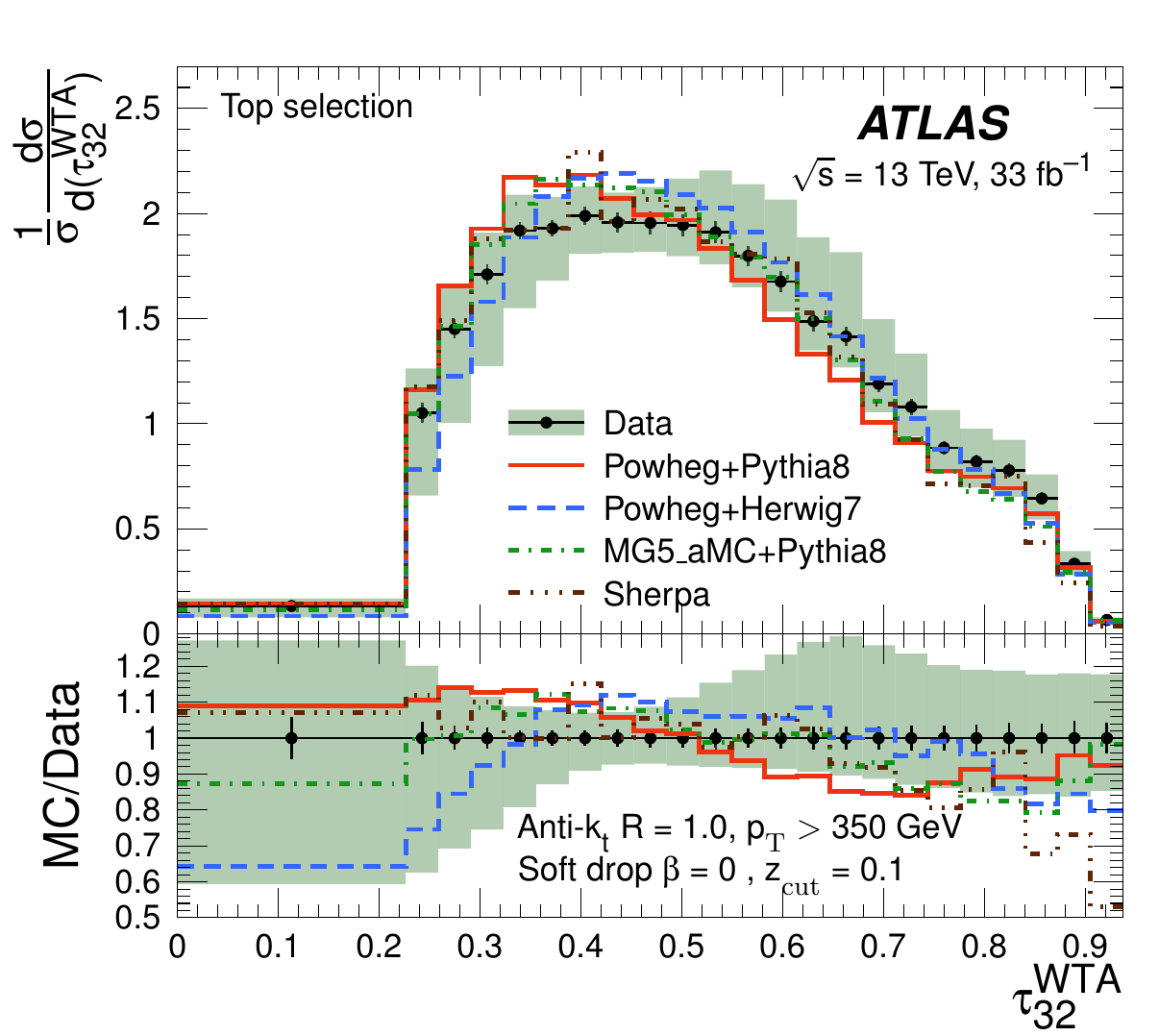}\\
\includegraphics[width=0.4\textwidth]{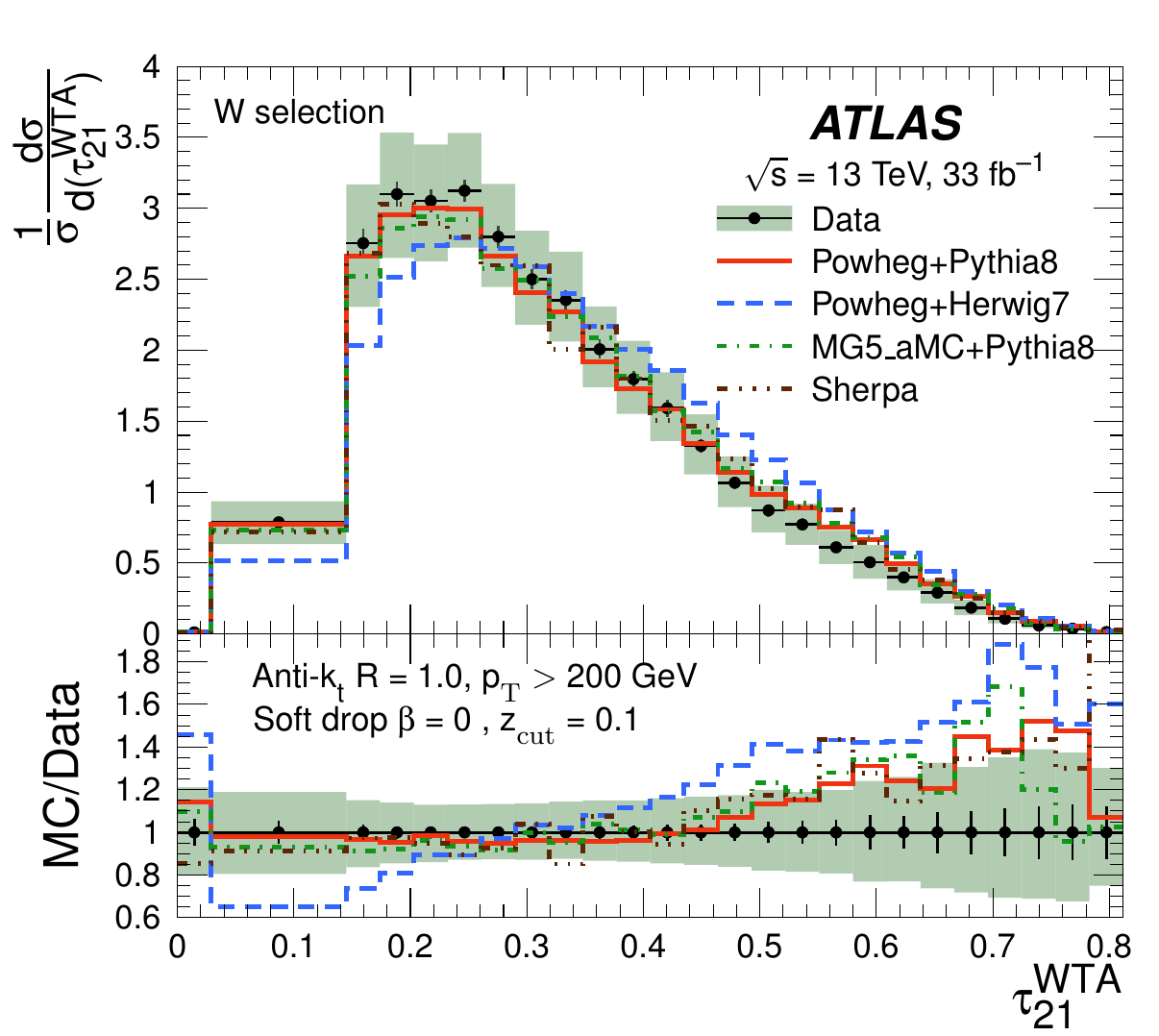}\qquad
\includegraphics[width=0.4\textwidth]{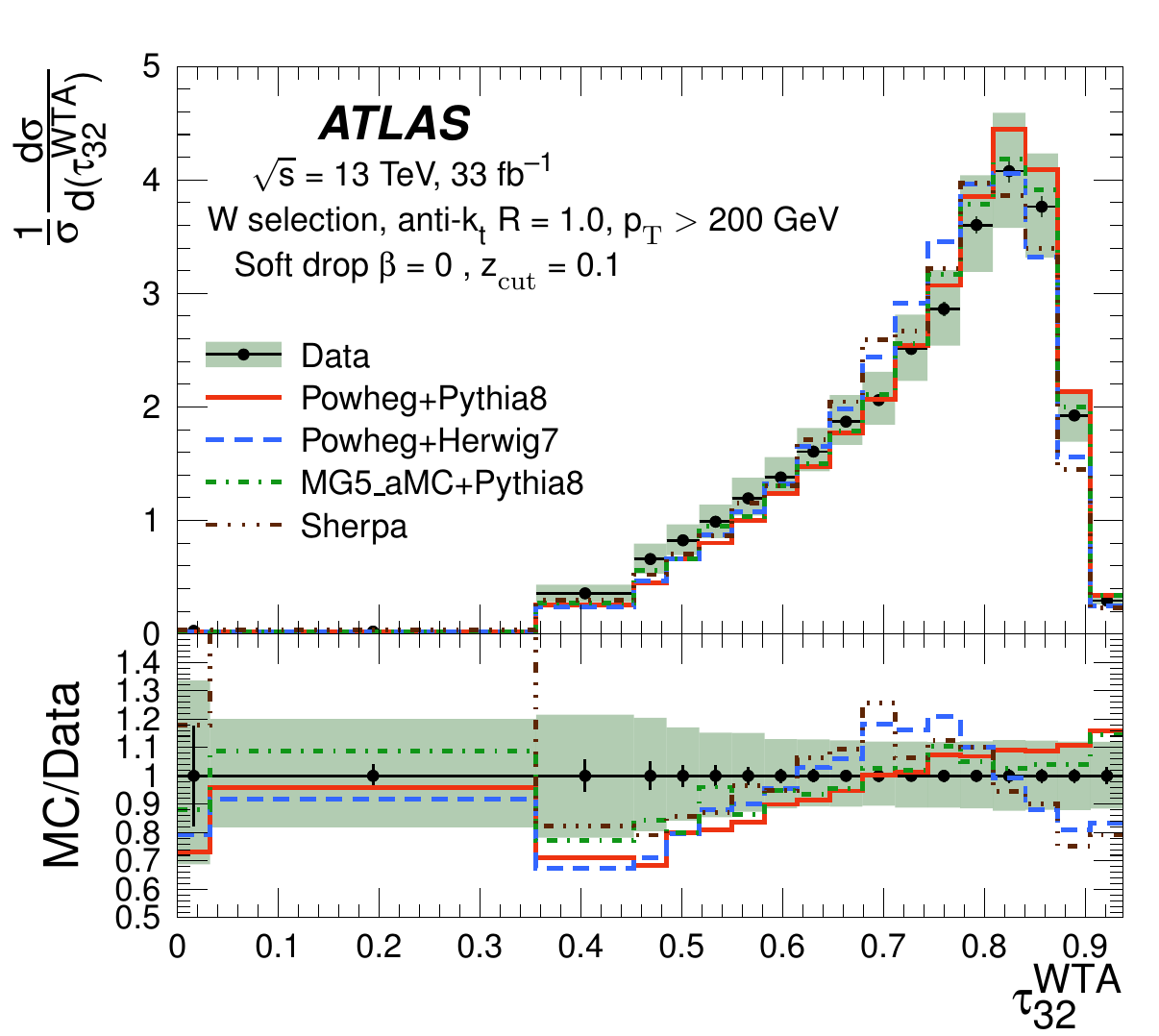}\\
\includegraphics[width=0.4\textwidth]{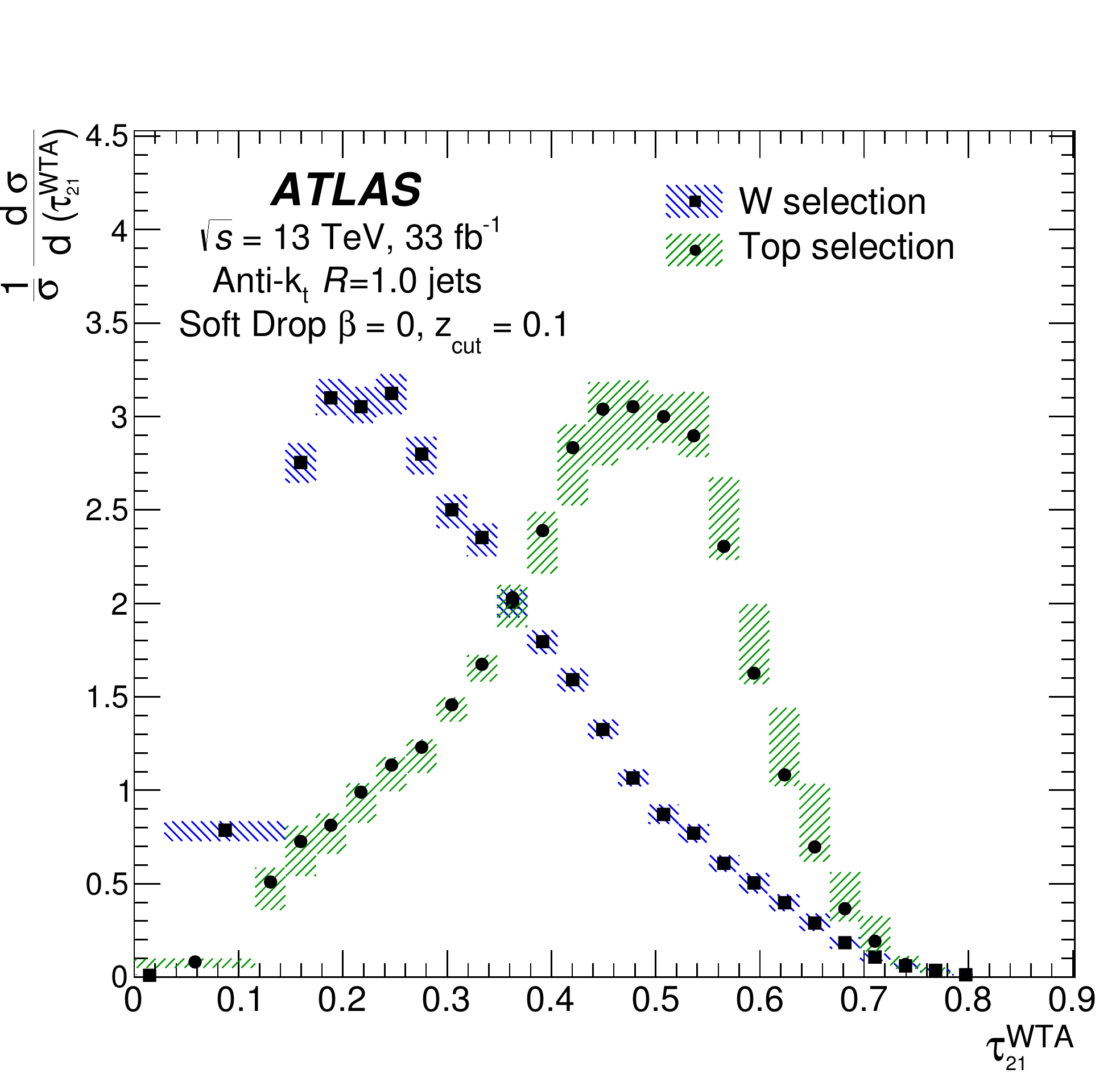}\qquad
\includegraphics[width=0.4\textwidth]{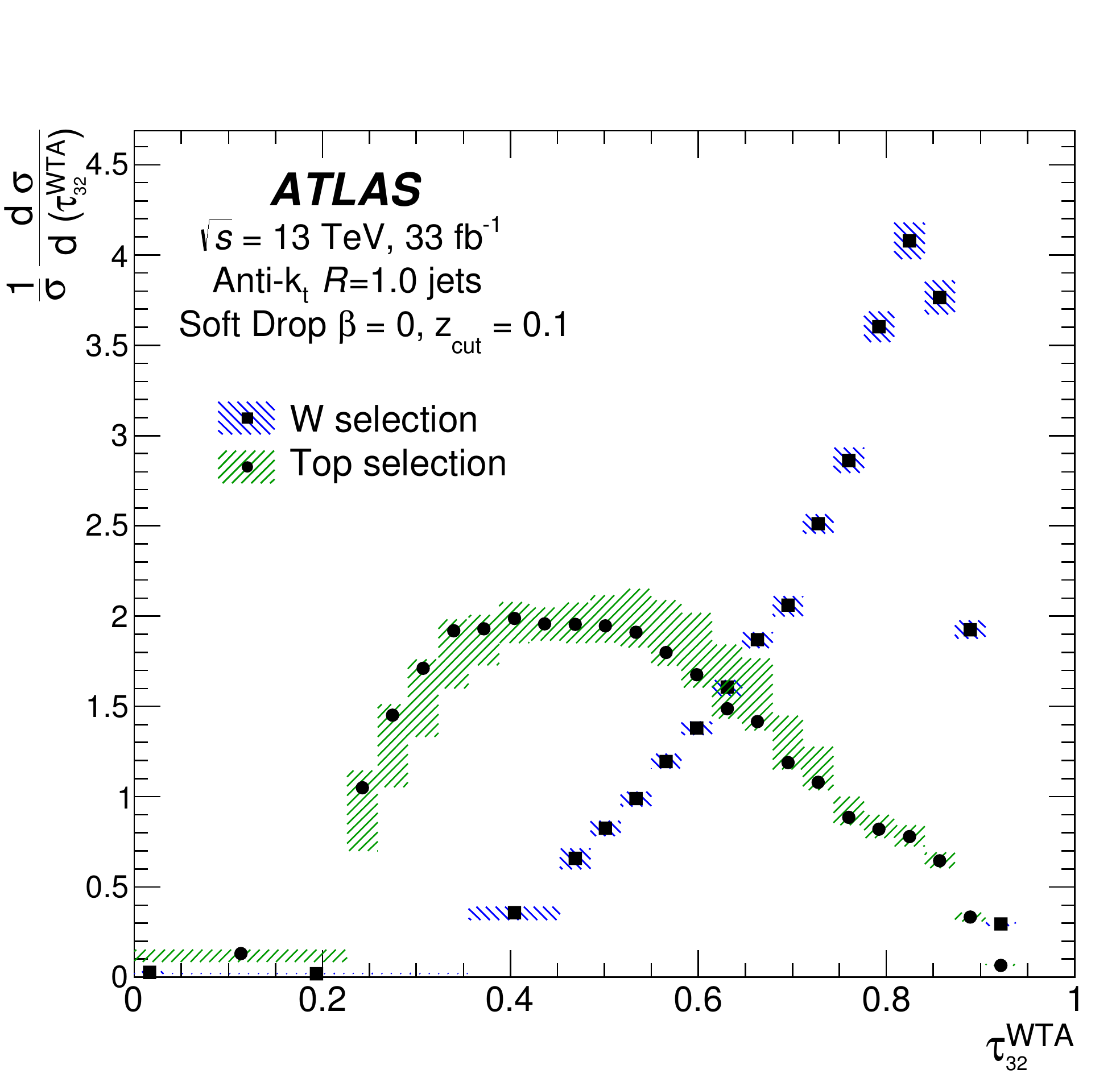}\\
\caption{
The distributions of \TauTwoOneWTA\ (left) and \TauThreeTwoWTA\ (right) are compared with different MC predictions for
large-radius jets from top (top row) and $W$ (bottom row) selections.
The distributions of \TauTwoOneWTA\ (bottom left) and \TauThreeTwoWTA\ (bottom right)
in data are
compared between the soft-dropped
large-radius jets for the two selections mentioned above.
The subscript WTA indicates that WTA axis was used in calculating these observables.
The shaded bands represent the total uncertainty, while
the error bars show the statistical uncertainty,
except in the bottom plots, where
the shaded areas represent the total uncertainty.
}
\label{fig:res-tau}
\end{figure}

Finally, in Figure~\ref{fig:res-tau}, a comparison of \TauTwoOneWTA\ and \TauThreeTwoWTA\
among top quark and $W$ selections
is presented.
The distribution of
\TauTwoOneWTA\
peaks at lower values for the $W$ selection than for the top selection, indicating
the two-prong decay of the former.
In general, \TauTwoOneWTA\ distributions are modelled well by the MC models, except \powheg+ \herwigseven.
Although most of the models also describe the \TauThreeTwoWTA\ distributions well, differences
can be observed between them, especially in the $W$ selection.

\FloatBarrier

\section{Conclusions}
\label{sec:conc}

A measurement of jet substructure observables using groomed
large-radius jets from light quarks or gluons, hadronically decaying
top quarks and $W$ bosons is
presented using $33$~fb$^{-1}$ of $\sqrt{s}=13$~\TeV\ proton--proton collision data taken with the ATLAS detector at the LHC.
The data discriminate between the various MC models probed.
In general, \pythiaeight for light-quark/gluon large-radius jet observables,
and {\powheg}+\pythiaeight, \sherpa as well as {\amcnlo}+\pythiaeight
for top quark and $W$ boson large-radius jet observables,
describe the data better than other models. The different hadronisation models in
\sherpa in the djiet selection result in similar predictions.
For most observables, \herwigseven in the dijet selection, and  {\powheg}+\herwigseven in the top and $W$ selections
do not describe the data well.
These measurements will be useful in improving the modelling of these
substructure variables in MC generators.
Since searches that utilise boosted topologies
use these observables, or combinations of them, in tagging large-radius jets,
a better modelling of them
will help to increase the sensitivity of such searches.

\section*{Acknowledgements}
 
We thank CERN for the very successful operation of the LHC, as well as the
support staff from our institutions without whom ATLAS could not be
operated efficiently.
 
We acknowledge the support of ANPCyT, Argentina; YerPhI, Armenia; ARC, Australia; BMWFW and FWF, Austria; ANAS, Azerbaijan; SSTC, Belarus; CNPq and FAPESP, Brazil; NSERC, NRC and CFI, Canada; CERN; CONICYT, Chile; CAS, MOST and NSFC, China; COLCIENCIAS, Colombia; MSMT CR, MPO CR and VSC CR, Czech Republic; DNRF and DNSRC, Denmark; IN2P3-CNRS, CEA-DRF/IRFU, France; SRNSFG, Georgia; BMBF, HGF, and MPG, Germany; GSRT, Greece; RGC, Hong Kong SAR, China; ISF and Benoziyo Center, Israel; INFN, Italy; MEXT and JSPS, Japan; CNRST, Morocco; NWO, Netherlands; RCN, Norway; MNiSW and NCN, Poland; FCT, Portugal; MNE/IFA, Romania; MES of Russia and NRC KI, Russian Federation; JINR; MESTD, Serbia; MSSR, Slovakia; ARRS and MIZ\v{S}, Slovenia; DST/NRF, South Africa; MINECO, Spain; SRC and Wallenberg Foundation, Sweden; SERI, SNSF and Cantons of Bern and Geneva, Switzerland; MOST, Taiwan; TAEK, Turkey; STFC, United Kingdom; DOE and NSF, United States of America. In addition, individual groups and members have received support from BCKDF, CANARIE, CRC and Compute Canada, Canada; COST, ERC, ERDF, Horizon 2020, and Marie Sk{\l}odowska-Curie Actions, European Union; Investissements d' Avenir Labex and Idex, ANR, France; DFG and AvH Foundation, Germany; Herakleitos, Thales and Aristeia programmes co-financed by EU-ESF and the Greek NSRF, Greece; BSF-NSF and GIF, Israel; CERCA Programme Generalitat de Catalunya, Spain; The Royal Society and Leverhulme Trust, United Kingdom.
 
The crucial computing support from all WLCG partners is acknowledged gratefully, in particular from CERN, the ATLAS Tier-1 facilities at TRIUMF (Canada), NDGF (Denmark, Norway, Sweden), CC-IN2P3 (France), KIT/GridKA (Germany), INFN-CNAF (Italy), NL-T1 (Netherlands), PIC (Spain), ASGC (Taiwan), RAL (UK) and BNL (USA), the Tier-2 facilities worldwide and large non-WLCG resource providers. Major contributors of computing resources are listed in Ref.~\cite{ATL-GEN-PUB-2016-002}.

\printbibliography

@string{JHEP="JHEP"}

@article{Aaboud:2018psm,
       author    = "{ATLAS Collaboration}",
      title          = "{Performance of top-quark and $W$-boson tagging with
                        ATLAS in Run 2 of the LHC}",
      year           = "2018",
      eprint         = "1808.07858",
      archivePrefix  = "arXiv",
      primaryClass   = "hep-ex",
      reportNumber   = "CERN-EP-2018-192",
      SLACcitation   = "%%CITATION = ARXIV:1808.07858;%%"
}

@article{Abdesselam:2010pt,
      author         = "Abdesselam, A. and others",
      title          = "{Boosted objects: a probe of beyond the standard model physics}",
      booktitle      = "{Boost 2010 Oxford, United Kingdom, June 22-25, 2010}",
      journal        = "Eur. Phys. J. C",
      volume         = "71",
      year           = "2011",
      pages          = "1661",
      doi            = "10.1140/epjc/s10052-011-1661-y",
      eprint         = "1012.5412",
      archivePrefix  = "arXiv",
      primaryClass   = "hep-ph",
      reportNumber   = "SLAC-PUB-15081, FERMILAB-PUB-10-617-CMS",
      SLACcitation   = "%%CITATION = ARXIV:1012.5412;%%"
}

@article{Krohn:2009th,
      author         = "Krohn, David and Thaler, Jesse and Wang, Lian-Tao",
      title          = "{Jet trimming}",
      journal        = "JHEP",
      volume         = "02",
      year           = "2010",
      pages          = "084",
      doi            = "10.1007/JHEP02(2010)084",
      eprint         = "0912.1342",
      archivePrefix  = "arXiv",
      primaryClass   = "hep-ph",
      SLACcitation   = "%%CITATION = ARXIV:0912.1342;%%"
}

@article{Larkoski:2014wba,
      author         = "Larkoski, Andrew J. and Marzani, Simone and Soyez,
                        Gregory and Thaler, Jesse",
      title          = "{Soft drop}",
      journal        = "JHEP",
      volume         = "05",
      year           = "2014",
      pages          = "146",
      doi            = "10.1007/JHEP05(2014)146",
      eprint         = "1402.2657",
      archivePrefix  = "arXiv",
      primaryClass   = "hep-ph",
      reportNumber   = "MIT-CTP-4531, DCPT-14-24, IPPP-14-12",
      SLACcitation   = "%%CITATION = ARXIV:1402.2657;%%"
}

@article{Dasgupta:2013ihk,
      author         = "Dasgupta, Mrinal and Fregoso, Alessandro and Marzani,
                        Simone and Salam, Gavin P.",
      title          = "{Towards an understanding of jet substructure}",
      journal        = "JHEP",
      volume         = "09",
      year           = "2013",
      pages          = "029",
      doi            = "10.1007/JHEP09(2013)029",
      eprint         = "1307.0007",
      archivePrefix  = "arXiv",
      primaryClass   = "hep-ph",
      reportNumber   = "CERN-PH-TH-2013-145, DCPT-13-86, IPPP-13-43, LPN13-036,
                        MAN-HEP-2013-12",
      SLACcitation   = "%%CITATION = ARXIV:1307.0007;%%"
}

@article{Marzani:2017mva,
      author         = "Marzani, Simone and Schunk, Lais and Soyez, Gregory",
      title          = "{A study of jet mass distributions with grooming}",
      journal        = "JHEP",
      volume         = "07",
      year           = "2017",
      pages          = "132",
      doi            = "10.1007/JHEP07(2017)132",
      eprint         = "1704.02210",
      archivePrefix  = "arXiv",
      primaryClass   = "hep-ph",
      SLACcitation   = "%%CITATION = ARXIV:1704.02210;%%"
}

@article{Marzani:2017kqd,
      author         = "Marzani, Simone and Schunk, Lais and Soyez, Gregory",
      title          = "{The jet mass distribution after Soft Drop}",
      journal        = "Eur. Phys. J. C",
      volume         = "78",
      year           = "2018",
      number         = "2",
      pages          = "96",
      doi            = "10.1140/epjc/s10052-018-5579-5",
      eprint         = "1712.05105",
      archivePrefix  = "arXiv",
      primaryClass   = "hep-ph",
      reportNumber   = "DESY-17-217",
      SLACcitation   = "%%CITATION = ARXIV:1712.05105;%%"
}

@article{Frye:2016aiz,
      author         = "Frye, Christopher and Larkoski, Andrew J. and Schwartz,
                        Matthew D. and Yan, Kai",
      title          = "{Factorization for groomed jet substructure beyond the
                        next-to-leading logarithm}",
      journal        = "JHEP",
      volume         = "07",
      year           = "2016",
      pages          = "064",
      doi            = "10.1007/JHEP07(2016)064",
      eprint         = "1603.09338",
      archivePrefix  = "arXiv",
      primaryClass   = "hep-ph",
      SLACcitation   = "%%CITATION = ARXIV:1603.09338;%%"
}

@article{Cacciari:2005hq,
      author         = "Cacciari, Matteo and Salam, Gavin P.",
      title          = "{Dispelling the $N^{3}$ myth for the $k_t$ jet-finder}",
      journal        = "Phys. Lett. B",
      volume         = "641",
      year           = "2006",
      pages          = "57-61",
      doi            = "10.1016/j.physletb.2006.08.037",
      eprint         = "hep-ph/0512210",
      archivePrefix  = "arXiv",
      primaryClass   = "hep-ph",
      reportNumber   = "LPTHE-05-32",
      SLACcitation   = "%%CITATION = HEP-PH/0512210;%%"
}

@article{Larkoski:2014uqa,
      author         = "Larkoski, Andrew J. and Neill, Duff and Thaler, Jesse",
      title          = "{Jet shapes with the broadening axis}",
      journal        = "JHEP",
      volume         = "04",
      year           = "2014",
      pages          = "017",
      doi            = "10.1007/JHEP04(2014)017",
      eprint         = "1401.2158",
      archivePrefix  = "arXiv",
      primaryClass   = "hep-ph",
      reportNumber   = "MIT--CTP-4512",
      SLACcitation   = "%%CITATION = ARXIV:1401.2158;%%"
}

@article{Larkoski:2015kga,
      author         = "Larkoski, Andrew J. and Moult, Ian and Neill, Duff",
      title          = "{Analytic boosted boson discrimination}",
      journal        = "JHEP",
      volume         = "05",
      year           = "2016",
      pages          = "117",
      doi            = "10.1007/JHEP05(2016)117",
      eprint         = "1507.03018",
      archivePrefix  = "arXiv",
      primaryClass   = "hep-ph",
      reportNumber   = "MIT-CTP-4681",
      SLACcitation   = "%%CITATION = ARXIV:1507.03018;%%"
}

@article{Larkoski:2013eya,
      author         = "Larkoski, Andrew J. and Salam, Gavin P. and Thaler,
                        Jesse",
      title          = "{Energy correlation functions for jet substructure}",
      journal        = "JHEP",
      volume         = "06",
      year           = "2013",
      pages          = "108",
      doi            = "10.1007/JHEP06(2013)108",
      eprint         = "1305.0007",
      archivePrefix  = "arXiv",
      primaryClass   = "hep-ph",
      reportNumber   = "MIT-CTP-4446, CERN-PH-TH-2013-066, LPN13-026",
      SLACcitation   = "%%CITATION = ARXIV:1305.0007;%%"
}

@article{Larkoski:2014pca,
      author         = "Larkoski, Andrew J. and Thaler, Jesse and Waalewijn,
                        Wouter J.",
      title          = "{Gaining (mutual) information about quark/gluon discrimination}",
      journal        = "JHEP",
      volume         = "11",
      year           = "2014",
      pages          = "129",
      doi            = "10.1007/JHEP11(2014)129",
      eprint         = "1408.3122",
      archivePrefix  = "arXiv",
      primaryClass   = "hep-ph",
      reportNumber   = "MIT--CTP-4572, NIKHEF-2014-026",
      SLACcitation   = "%%CITATION = ARXIV:1408.3122;%%"
}

@article{Kang:2018vgn,
      author         = "Kang, Zhong-Bo and Lee, Kyle and Liu, Xiaohui and Ringer,
                        Felix",
      title          = "{Soft drop groomed jet angularities at the LHC}",
      year           = "2018",
      eprint         = "1811.06983",
      archivePrefix  = "arXiv",
      primaryClass   = "hep-ph",
      SLACcitation   = "%%CITATION = ARXIV:1811.06983;%%"
}

@article{Thaler:2010tr,
      author         = "Thaler, Jesse and Van Tilburg, Ken",
      title          = "{Identifying boosted objects with N-subjettiness}",
      journal        = "JHEP",
      volume         = "03",
      year           = "2011",
      pages          = "015",
      doi            = "10.1007/JHEP03(2011)015",
      eprint         = "1011.2268",
      archivePrefix  = "arXiv",
      primaryClass   = "hep-ph",
      reportNumber   = "MIT-CTP-4191",
      SLACcitation   = "%%CITATION = ARXIV:1011.2268;%%"
}

@article{Butterworth:2008iy,
      author         = "Butterworth, Jonathan M. and Davison, Adam R. and Rubin,
                        Mathieu and Salam, Gavin P.",
      title          = "{Jet Substructure as a New Higgs-Search Channel at the Large Hadron Collider}",
      journal        = "Phys. Rev. Lett.",
      volume         = "100",
      year           = "2008",
      pages          = "242001",
      doi            = "10.1103/PhysRevLett.100.242001",
      eprint         = "0802.2470",
      archivePrefix  = "arXiv",
      primaryClass   = "hep-ph",
      SLACcitation   = "%%CITATION = ARXIV:0802.2470;%%"
}

@article{DAgostini:1994zf,
	author         = "D'Agostini, G.",
	title          = "{ A multidimensional unfolding method based on Bayes' theorem}",
	journal        = "Nucl. Instrum. Meth. A",
	volume         = "362",
	pages          = "487-498",
	doi            = "10.1016/0168-9002(95)00274-X",
	year           = "1995",
}

@article{Cacciari:2008gp,
      author         = "Cacciari, Matteo and Salam, Gavin P. and Soyez, Gregory",
      title          = "{The anti-$k_t$ jet clustering algorithm}",
      journal        = "JHEP",
      volume         = "04",
      year           = "2008",
      pages          = "063",
      doi            = "10.1088/1126-6708/2008/04/063",
      eprint         = "0802.1189",
      archivePrefix  = "arXiv",
      primaryClass   = "hep-ph",
      reportNumber   = "LPTHE-07-03",
      SLACcitation   = "%%CITATION = ARXIV:0802.1189;%%"
}

@inproceedings{Adye:2011gm,
      author         = "Adye, Tim",
      title          = "{Unfolding algorithms and tests using RooUnfold}",
      booktitle      = "{Proceedings, PHYSTAT 2011 Workshop on Statistical Issues
                        Related to Discovery Claims in Search Experiments and
                        Unfolding, CERN,Geneva, Switzerland 17-20 January 2011}",
      organization   = "CERN",
      publisher      = "CERN",
      address        = "Geneva",
      year           = "2011",
      pages          = "313-318",
      doi            = "10.5170/CERN-2011-006.313",
      eprint         = "1105.1160",
      archivePrefix  = "arXiv",
      primaryClass   = "physics.data-an",
      SLACcitation   = "%%CITATION = ARXIV:1105.1160;%%"
}

@article{Kidonakis:2011wy,
      author         = "Kidonakis, Nikolaos",
      title          = "{Next-to-next-to-leading-order collinear and soft gluon
                        corrections for t-channel single top quark production}",
      journal        = "Phys. Rev. D",
      volume         = "83",
      year           = "2011",
      pages          = "091503",
      doi            = "10.1103/PhysRevD.83.091503",
      eprint         = "1103.2792",
      archivePrefix  = "arXiv",
      primaryClass   = "hep-ph",
      SLACcitation   = "%%CITATION = ARXIV:1103.2792;%%"
}

@article{Kidonakis:2010ux,
      author         = "Kidonakis, Nikolaos",
      title          = "{Two-loop soft anomalous dimensions for single top quark
                        associated production with a W- or H-}",
      journal        = "Phys. Rev. D",
      volume         = "82",
      year           = "2010",
      pages          = "054018",
      doi            = "10.1103/PhysRevD.82.054018",
      eprint         = "1005.4451",
      archivePrefix  = "arXiv",
      primaryClass   = "hep-ph",
      SLACcitation   = "%%CITATION = ARXIV:1005.4451;%%"
}

@article{Kidonakis:2010tc,
      author         = "Kidonakis, Nikolaos",
      title          = "{NNLL resummation for s-channel single top quark
                        production}",
      journal        = "Phys. Rev. D",
      volume         = "81",
      year           = "2010",
      pages          = "054028",
      doi            = "10.1103/PhysRevD.81.054028",
      eprint         = "1001.5034",
      archivePrefix  = "arXiv",
      primaryClass   = "hep-ph",
      SLACcitation   = "%%CITATION = ARXIV:1001.5034;%%"
}

@article{Anastasiou:2003ds,
      author         = "Anastasiou, Charalampos and Dixon, Lance J. and Melnikov,
                        Kirill and Petriello, Frank",
      title          = "{High precision QCD at hadron colliders: Electroweak
                        gauge boson rapidity distributions at NNLO}",
      journal        = "Phys. Rev. D",
      volume         = "69",
      year           = "2004",
      pages          = "094008",
      doi            = "10.1103/PhysRevD.69.094008",
      eprint         = "hep-ph/0312266",
      archivePrefix  = "arXiv",
      primaryClass   = "hep-ph",
      reportNumber   = "SLAC-PUB-10288, UH-511-1042-03",
      SLACcitation   = "%%CITATION = HEP-PH/0312266;%%"
}

@article{Czakon:2011xx,
      author         = "Czakon, Michal and Mitov, Alexander",
      title          = "{Top++: A program for the calculation of the top-pair cross-section at hadron colliders}",
      journal        = "Comput. Phys. Commun.",
      volume         = "185",
      year           = "2014",
      pages          = "2930",
      doi            = "10.1016/j.cpc.2014.06.021",
      eprint         = "1112.5675",
      archivePrefix  = "arXiv",
      primaryClass   = "hep-ph",
      reportNumber   = "CERN-PH-TH-2011-303, TTK-11-58",
      SLACcitation   = "%%CITATION = ARXIV:1112.5675;%%"
}

@article{Larkoski:2014gra,
      author         = "Larkoski, Andrew J. and Moult, Ian and Neill, Duff",
      title          = "{Power counting to better jet observables}",
      journal        = "JHEP",
      volume         = "12",
      year           = "2014",
      pages          = "009",
      doi            = "10.1007/JHEP12(2014)009",
      eprint         = "1409.6298",
      archivePrefix  = "arXiv",
      primaryClass   = "hep-ph",
      reportNumber   = "MIT--CTP-4588",
      SLACcitation   = "%%CITATION = ARXIV:1409.6298;%%"
}

@Article{Sjostrand:2007gs,
	author    = "Sj{\"o}strand, T. and Mrenna, S. and Skands, P.",
	title     = "{A brief introduction to PYTHIA 8.1}",
	journal   = "Comput. Phys. Commun.",
	volume    = "178",
	year      = "2008",
	pages     = "852-867",
	eprint    = "0710.3820",
	archivePrefix = "arXiv",
	primaryClass  =  "hep-ph",
	doi       = "10.1016/j.cpc.2008.01.036",
	SLACcitation  = "%%CITATION = 0710.3820;%%"
}

@article{Sjostrand:2014zea,
      author         = "Sj{\"o}strand, Torbjorn and Ask, Stefan and Christiansen,
                        Jesper R. and Corke, Richard and Desai, Nishita and Ilten,
                        Philip and Mrenna, Stephen and Prestel, Stefan and
                        Rasmussen, Christine O. and Skands, Peter Z.",
      title          = "{An introduction to PYTHIA 8.2}",
      journal        = "Comput. Phys. Commun.",
      volume         = "191",
      year           = "2015",
      pages          = "159-177",
      doi            = "10.1016/j.cpc.2015.01.024",
      eprint         = "1410.3012",
      archivePrefix  = "arXiv",
      primaryClass   = "hep-ph",
      reportNumber   = "LU-TP-14-36, MCNET-14-22, CERN-PH-TH-2014-190,
                        FERMILAB-PUB-14-316-CD, DESY-14-178, SLAC-PUB-16122",
      SLACcitation   = "%%CITATION = ARXIV:1410.3012;%%"
}

@article{Bahr:2008pv,
      author         = "Bahr, M. and others",
      title          = "{Herwig++ physics and manual}",
      journal        = "Eur. Phys. J. C",
      volume         = "58",
      year           = "2008",
      pages          = "639-707",
      doi            = "10.1140/epjc/s10052-008-0798-9",
      eprint         = "0803.0883",
      archivePrefix  = "arXiv",
      primaryClass   = "hep-ph",
      reportNumber   = "CERN-PH-TH-2008-038, CAVENDISH-HEP-08-03, KA-TP-05-2008,
                        DCPT-08-22, IPPP-08-11, CP3-08-05",
      SLACcitation   = "%%CITATION = ARXIV:0803.0883;%%"
}

@article{Sjostrand:2000wi,
      author         = "Sj{\"o}strand, Torbjorn and Eden, Patrik and Friberg,
                        Christer and Lonnblad, Leif and Miu, Gabriela and others",
      title          = "{High-energy physics event generation with PYTHIA 6.1}",
      journal        = "Comput. Phys. Commun.",
      volume         = "135",
      pages          = "238-259",
      doi            = "10.1016/S0010-4655(00)00236-8",
      year           = "2001",
      eprint         = "hep-ph/0010017",
      archivePrefix  = "arXiv",
      primaryClass   = "hep-ph",
      reportNumber   = "LU-TP-00-30",
      SLACcitation   = "%%CITATION = HEP-PH/0010017;%%",
}

@article{Sjostrand:2006za,
      author         = "Sj{\"o}strand, Torbjorn and Mrenna, Stephen and Skands, Peter
                        Z.",
      title          = "{PYTHIA 6.4 Physics and Manual}",
      journal        = "JHEP",
      volume         = "05",
      year           = "2006",
      pages          = "026",
      doi            = "10.1088/1126-6708/2006/05/026",
      eprint         = "hep-ph/0603175",
      archivePrefix  = "arXiv",
      primaryClass   = "hep-ph",
      reportNumber   = "FERMILAB-PUB-06-052-CD-T, LU-TP-06-13",
      SLACcitation   = "%%CITATION = HEP-PH/0603175;%%"
}

@article{Gleisberg:2008ta,
      author         = "Gleisberg, T. and Hoeche, Stefan. and Krauss, F. and
                        Schonherr, M. and Schumann, S. and Siegert, F. and Winter, J.",
      title          = "{Event generation with SHERPA 1.1}",
      journal        = "JHEP",
      volume         = "02",
      year           = "2009",
      pages          = "007",
      doi            = "10.1088/1126-6708/2009/02/007",
      eprint         = "0811.4622",
      archivePrefix  = "arXiv",
      primaryClass   = "hep-ph",
      reportNumber   = "FERMILAB-PUB-08-477-T, SLAC-PUB-13420, ZU-TH-17-08, DCPT-08-138, IPPP-08-69, EDINBURGH-2008-30, MCNET-08-14",
      SLACcitation   = "%%CITATION = ARXIV:0811.4622;%%"
}

@article{Ball:2012cx,
      author         = "Ball, Richard D. and others",
      title          = "{Parton distributions with LHC data}",
      journal        = "Nucl. Phys. B",
      volume         = "867",
      year           = "2013",
      pages          = "244-289",
      doi            = "10.1016/j.nuclphysb.2012.10.003",
      eprint         = "1207.1303",
      archivePrefix  = "arXiv",
      primaryClass   = "hep-ph",
      reportNumber   = "EDINBURGH-2012-08, IFUM-FT-997, FR-PHENO-2012-014,
                        RWTH-TTK-12-25, CERN-PH-TH-2012-037, SFB-CPP-12-47",
      SLACcitation   = "%%CITATION = ARXIV:1207.1303;%%"
}

@article{Skands:2010ak,
      author         = "Skands, Peter Zeiler",
      title          = "{Tuning Monte Carlo generators: The Perugia tunes}",
      journal        = "Phys. Rev. D",
      volume         = "82",
      year           = "2010",
      pages          = "074018",
      doi            = "10.1103/PhysRevD.82.074018",
      eprint         = "1005.3457",
      archivePrefix  = "arXiv",
      primaryClass   = "hep-ph",
      reportNumber   = "MCNET-10-08, CERN-PH-TH-2010-113",
      SLACcitation   = "%%CITATION = ARXIV:1005.3457;%%"
}

@article{Martin:2009iq,
      author         = "Martin, A. D. and Stirling, W. J. and Thorne, R. S. and
                        Watt, G.",
      title          = "{Parton distributions for the LHC}",
      journal        = "Eur. Phys. J. C",
      volume         = "63",
      year           = "2009",
      pages          = "189-285",
      doi            = "10.1140/epjc/s10052-009-1072-5",
      eprint         = "0901.0002",
      archivePrefix  = "arXiv",
      primaryClass   = "hep-ph",
      reportNumber   = "IPPP-08-95, DCPT-08-190, CAVENDISH-HEP-08-16",
      SLACcitation   = "%%CITATION = ARXIV:0901.0002;%%"
}

@Booklet{ATL-PHYS-PUB-2014-021,
    author         = "{ATLAS Collaboration}",
    title          = "{ATLAS Pythia~8 tunes to $7\;\mbox{TeV}$ data}",
    howpublished   = "{ATL-PHYS-PUB-2014-021}",
    url            = "https://cds.cern.ch/record/1966419",
    year           = "2014",
}

@article{Alioli:2010xd,
      author         = "Alioli, Simone and Nason, Paolo and Oleari, Carlo and Re,
                        Emanuele",
      title          = "{A general framework for implementing NLO calculations in
                        shower Monte Carlo programs: the POWHEG BOX}",
      journal        = "JHEP",
      volume         = "06",
      year           = "2010",
      pages          = "043",
      doi            = "10.1007/JHEP06(2010)043",
      eprint         = "1002.2581",
      archivePrefix  = "arXiv",
      primaryClass   = "hep-ph",
      reportNumber   = "DESY-10-018, SFB-CPP-10-22, IPPP-10-11, DCPT-10-22",
      SLACcitation   = "%%CITATION = ARXIV:1002.2581;%%"
}

@article{Alwall:2014hca,
      author         = "Alwall, J. and Frederix, R. and Frixione, S. and Hirschi,
                        V. and Maltoni, F. and Mattelaer, O. and Shao, H. -S. and
                        Stelzer, T. and Torrielli, P. and Zaro, M.",
      title          = "{The automated computation of tree-level and
                        next-to-leading order differential cross sections, and
                        their matching to parton shower simulations}",
      journal        = "JHEP",
      volume         = "07",
      year           = "2014",
      pages          = "079",
      doi            = "10.1007/JHEP07(2014)079",
      eprint         = "1405.0301",
      archivePrefix  = "arXiv",
      primaryClass   = "hep-ph",
      reportNumber   = "CERN-PH-TH-2014-064, CP3-14-18, LPN14-066, MCNET-14-09,
                        ZU-TH-14-14",
      SLACcitation   = "%%CITATION = ARXIV:1405.0301;%%"
}

@article{Pumplin:2002vw,
      author         = "Pumplin, J. and Stump, D. R. and Huston, J. and Lai, H.
                        L. and Nadolsky, Pavel M. and Tung, W. K.",
      title          = "{New Generation of Parton Distributions with Uncertainties from Global QCD Analysis}",
      journal        = "JHEP",
      volume         = "07",
      year           = "2002",
      pages          = "012",
      doi            = "10.1088/1126-6708/2002/07/012",
      eprint         = "hep-ph/0201195",
      archivePrefix  = "arXiv",
      primaryClass   = "hep-ph",
      reportNumber   = "MSU-HEP-011101",
      SLACcitation   = "%%CITATION = HEP-PH/0201195;%%"
}

@article{Seymour:2013qka,
      author         = "Seymour, Michael H. and Siodmok, Andrzej",
      title          = "{Constraining MPI models using $\sigma_{eff}$ and recent
                        Tevatron and LHC Underlying Event data}",
      journal        = "JHEP",
      volume         = "10",
      year           = "2013",
      pages          = "113",
      doi            = "10.1007/JHEP10(2013)113",
      eprint         = "1307.5015",
      archivePrefix  = "arXiv",
      primaryClass   = "hep-ph",
      reportNumber   = "MAN-HEP-2013-14, MCNET-13-08",
      SLACcitation   = "%%CITATION = ARXIV:1307.5015;%%"
}

@article{Agostinelli:2002hh,
      author         = "Agostinelli, S. and others",
      title          = "{Geant4—a simulation toolkit}",
      collaboration  = "GEANT4",
      journal        = "Nucl. Instrum. Meth. A",
      volume         = "506",
      year           = "2003",
      pages          = "250-303",
      doi            = "10.1016/S0168-9002(03)01368-8",
      reportNumber   = "SLAC-PUB-9350, FERMILAB-PUB-03-339",
      SLACcitation   = "%%CITATION = NUIMA,A506,250;%%"
}

@article{Bellm:2015jjp,
      author         = "Bellm, Johannes and others",
      title          = "{Herwig 7.0/Herwig++ 3.0 release note}",
      journal        = "Eur. Phys. J. C",
      volume         = "76",
      year           = "2016",
      number         = "4",
      pages          = "196",
      doi            = "10.1140/epjc/s10052-016-4018-8",
      eprint         = "1512.01178",
      archivePrefix  = "arXiv",
      primaryClass   = "hep-ph",
      reportNumber   = "CERN-PH-TH-2015-289, MAN-HEP-2015-15, IFJPAN-IV-2015-13,
                        KA-TP-18-2015, DCPT-15-142, MCNET-15-28, IPPP-15-71,
                        HERWIG-2015-01",
      SLACcitation   = "%%CITATION = ARXIV:1512.01178;%%"
}

@article{Andersson:1983ia,
      author         = "Andersson, Bo and Gustafson, G. and Ingelman, G. and  Sj{\"o}strand, T.",
      title          = "{Parton Fragmentation and String Dynamics}",
      journal        = "Phys. Rept.",
      volume         = "97",
      year           = "1983",
      pages          = "31-145",
      doi            = "10.1016/0370-1573(83)90080-7",
      reportNumber   = "LU-TP-83-10",
      SLACcitation   = "%%CITATION = PRPLC,97,31;%%"
}

@article{Marchesini:1991ch,
      author         = "Marchesini, G. and Webber, B. R. and Abbiendi, G. and
                        Knowles, I. G. and Seymour, M. H. and Stanco, L.",
      title          = "{HERWIG: A Monte Carlo event generator for simulating
                        hadron emission reactions with interfering gluons. Version
                        5.1 - April 1991}",
      journal        = "Comput. Phys. Commun.",
      volume         = "67",
      year           = "1992",
      pages          = "465-508",
      doi            = "10.1016/0010-4655(92)90055-4",
      reportNumber   = "CAVENDISH-HEP-91-26, DESY-91-048",
      SLACcitation   = "%%CITATION = CPHCB,67,465;%%"
}

@article{Frixione:2007nw,
      author         = "Frixione, Stefano and Nason, Paolo and Ridolfi, Giovanni",
      title          = "{A Positive-weight next-to-leading-order Monte Carlo for
                        heavy flavour hadroproduction}",
      journal        = "JHEP",
      volume         = "09",
      year           = "2007",
      pages          = "126",
      doi            = "10.1088/1126-6708/2007/09/126",
      eprint         = "0707.3088",
      archivePrefix  = "arXiv",
      primaryClass   = "hep-ph",
      reportNumber   = "BICOCCA-FT-07-12, GEF-TH-19-2007",
      SLACcitation   = "%%CITATION = ARXIV:0707.3088;%%"
}

@article{Re:2010bp,
      author         = "Re, Emanuele",
      title          = "{Single-top Wt-channel production matched with parton
                        showers using the POWHEG method}",
      journal        = "Eur. Phys. J. C",
      volume         = "71",
      year           = "2011",
      pages          = "1547",
      doi            = "10.1140/epjc/s10052-011-1547-z",
      eprint         = "1009.2450",
      archivePrefix  = "arXiv",
      primaryClass   = "hep-ph",
      reportNumber   = "IPPP-10-74, DCPT-10-148",
      SLACcitation   = "%%CITATION = ARXIV:1009.2450;%%"
}

@article{Alioli:2009je,
      author         = "Alioli, Simone and Nason, Paolo and Oleari, Carlo and Re,
                        Emanuele",
      title          = "{NLO single-top production matched with shower in POWHEG:
                        s- and t-channel contributions}",
      journal        = "JHEP",
      volume         = "09",
      year           = "2009",
      pages          = "111",
      doi            = "10.1007/JHEP02(2010)011",
      note           = "[Erratum: JHEP 02 (2010)] 011",
      eprint         = "0907.4076",
      archivePrefix  = "arXiv",
      primaryClass   = "hep-ph",
      SLACcitation   = "%%CITATION = ARXIV:0907.4076;%%"
}

@article{Frederix:2012dh,
      author         = "Frederix, Rikkert and Re, Emanuele and Torrielli, Paolo",
      title          = "{Single-top t-channel hadroproduction in the four-flavour
                        scheme with POWHEG and aMC@NLO}",
      journal        = "JHEP",
      volume         = "09",
      year           = "2012",
      pages          = "130",
      doi            = "10.1007/JHEP09(2012)130",
      eprint         = "1207.5391",
      archivePrefix  = "arXiv",
      primaryClass   = "hep-ph",
      reportNumber   = "ZU-TH-14-12, IPPP-12-54, DCPT-12-108,
                        CERN-PH-TH-2012-206, MCNET-12-10, LPN12-084",
      SLACcitation   = "%%CITATION = ARXIV:1207.5391;%%"
}

@article{Cacciari:2008gn,
      author         = "Cacciari, Matteo and Salam, Gavin P. and Soyez, Gregory",
      title          = "{The catchment area of jets}",
      journal        = "JHEP",
      volume         = "04",
      year           = "2008",
      pages          = "005",
      doi            = "10.1088/1126-6708/2008/04/005",
      eprint         = "0802.1188",
      archivePrefix  = "arXiv",
      primaryClass   = "hep-ph",
      reportNumber   = "LPTHE-07-02",
      SLACcitation   = "%%CITATION = ARXIV:0802.1188;%%"
}

@techreport{Capeans:1291633,
      author        = "Capeans, M and Darbo, G and Einsweiller, K and Elsing, M
                       and Flick, T and Garcia-Sciveres, M and Gemme, C and
                       Pernegger, H and Rohne, O and Vuillermet, R",
      title         = "{ATLAS Insertable B-Layer Technical Design Report}",
      collaboration = "ATLAS Collaboration",
      number        = "CERN-LHCC-2010-013. ATLAS-TDR-19",
      month         = "Sep",
      year          = "2010",
      reportNumber  = "CERN-LHCC-2010-013",
      url           = "https://cds.cern.ch/record/1291633",
}

@article{ATLAS:2012am,
     author         = "{ATLAS Collaboration}",
      title          = "{Jet mass and substructure of inclusive jets in $\sqrt{s}=7$ TeV $pp$ collisions with the ATLAS experiment}",
      journal        = "JHEP",
      volume         = "05",
      pages          = "128",
      doi            = "10.1007/JHEP05(2012)128",
      year           = "2012",
      eprint         = "1203.4606",
      archivePrefix  = "arXiv",
      primaryClass   = "hep-ex",
      reportNumber   = "CERN-PH-EP-2012-031",
      SLACcitation   = "%%CITATION = ARXIV:1203.4606;%%",
}

@article{Aad:2012meb,
    author         = "{ATLAS Collaboration}",
    title          = "{ATLAS measurements of the properties of jets for boosted particle searches}",
    journal        = "Phys. Rev. D",
    volume         = "86",
    year           = "2012",
    pages          = "072006",
    doi            = "10.1103/PhysRevD.86.072006",
    reportNumber   = "CERN-PH-EP-2012-149",
    eprint         = "1206.5369",
    archivePrefix  = "arXiv",
    primaryClass   = "hep-ex",
}

@article{Aad:2016oit,
      author         = "{ATLAS Collaboration}",
      title          = "{Measurement of the charged-particle multiplicity inside
                        jets from $\sqrt{s}=8$ TeV $pp$ collisions with the ATLAS
                        detector}",
      journal        = "Eur. Phys. J. C",
      volume         = "76",
      year           = "2016",
      number         = "6",
      pages          = "322",
      doi            = "10.1140/epjc/s10052-016-4126-5",
      eprint         = "1602.00988",
      archivePrefix  = "arXiv",
      primaryClass   = "hep-ex",
      reportNumber   = "CERN-EP-2016-001",
      SLACcitation   = "%%CITATION = ARXIV:1602.00988;%%"
}

@article{Aaboud:2017qwh,
      author         = "{ATLAS Collaboration}",
      title          = "{Measurement of the Soft-Drop Jet Mass in pp Collisions
                        at $\sqrt{s} = 13$ TeV with the ATLAS detector}",
      journal        = "Phys. Rev. Lett.",
      volume         = "121",
      year           = "2018",
      number         = "9",
      pages          = "092001",
      doi            = "10.1103/PhysRevLett.121.092001",
      eprint         = "1711.08341",
      archivePrefix  = "arXiv",
      primaryClass   = "hep-ex",
      reportNumber   = "CERN-EP-2017-231",
      SLACcitation   = "%%CITATION = ARXIV:1711.08341;%%"
}

@article{Aaboud:2018yqu,
     author         = "{ATLAS Collaboration}",
      title          = "{Measurement of the photon identification efficiencies
                        with the ATLAS detector using LHC Run 2 data collected in
                        2015 and 2016}",
      collaboration  = "ATLAS",
      journal        = "Submitted to: Eur. Phys. J. C",
      year           = "2018",
      eprint         = "1810.05087",
      archivePrefix  = "arXiv",
      primaryClass   = "hep-ex",
      reportNumber   = "CERN-EP-2018-216",
      SLACcitation   = "%%CITATION = ARXIV:1810.05087;%%"
}

@article{Aaboud:2017jcu,
       author         = "{ATLAS Collaboration}",
      title          = "{Jet energy scale measurements and their systematic
                        uncertainties in proton-proton collisions at $\sqrt{s} =
                        13$ TeV with the ATLAS detector}",
      journal        = "Phys. Rev. D",
      volume         = "96",
      year           = "2017",
      number         = "7",
      pages          = "072002",
      doi            = "10.1103/PhysRevD.96.072002",
      eprint         = "1703.09665",
      archivePrefix  = "arXiv",
      primaryClass   = "hep-ex",
      reportNumber   = "CERN-EP-2017-038",
      SLACcitation   = "%%CITATION = ARXIV:1703.09665;%%"
}

@Article{TOPQ-2014-09,
    author         = "{ATLAS Collaboration}",
    title          = "{Measurement of colour flow with the jet pull angle in $t\bar{t}$ events using the ATLAS detector at $\sqrt{s} = 8\;\mbox{TeV}$}",
    journal        = "Phys. Lett. B",
    volume         = "750",
    year           = "2015",
    pages          = "475",
    doi            = "10.1016/j.physletb.2015.09.051",
    reportNumber   = "CERN-PH-EP-2015-128",
    eprint         = "1506.05629",
    archivePrefix  = "arXiv",
    primaryClass   = "hep-ex",
}

@article{Aaboud:2018ibj,
      author         = "{ATLAS Collaboration}",
      title          = "{Measurement of colour flow using jet-pull observables in
                        $t\bar{t}$ events with the ATLAS experiment at $\sqrt{s} =
                        13\,\hbox {TeV}$}",
      journal        = "Eur. Phys. J. C",
      volume         = "78",
      year           = "2018",
      number         = "10",
      pages          = "847",
      doi            = "10.1140/epjc/s10052-018-6290-2",
      eprint         = "1805.02935",
      archivePrefix  = "arXiv",
      primaryClass   = "hep-ex",
      reportNumber   = "CERN-EP-2018-041",
      SLACcitation   = "%%CITATION = ARXIV:1805.02935;%%"
}

@Article{STDM-2014-17,
    author         = "{ATLAS Collaboration}",
    title          = "{Measurement of jet charge in dijet events from $\sqrt{s} = 8\;\mbox{TeV}$ $pp$ collisions with the ATLAS detector}",
    journal        = "Phys. Rev. D",
    volume         = "93",
    year           = "2016",
    pages          = "052003",
    doi            = "10.1103/PhysRevD.93.052003",
    reportNumber   = "CERN-PH-EP-2015-207",
    eprint         = "1509.05190",
    archivePrefix  = "arXiv",
    primaryClass   = "hep-ex",
}

@article{Aaboud:2018mjh,
        author         = "{ATLAS Collaboration}",
      title          = "{Search for heavy particles decaying into top-quark pairs
                        using lepton-plus-jets events in proton–proton
                        collisions at $\sqrt{s} = 13$   $\text {TeV}$ with the
                        ATLAS detector}",
      journal        = "Eur. Phys. J. C",
      volume         = "78",
      year           = "2018",
      number         = "7",
      pages          = "565",
      doi            = "10.1140/epjc/s10052-018-5995-6",
      eprint         = "1804.10823",
      archivePrefix  = "arXiv",
      primaryClass   = "hep-ex",
      reportNumber   = "CERN-EP-2018-48, CERN-EP-2018-048",
      SLACcitation   = "%%CITATION = ARXIV:1804.10823;%%"
}

@Booklet{ATLAS-CONF-2015-037,
    author         = "{ATLAS Collaboration}",
    title          = "{Monte Carlo Calibration and Combination of In-situ Measurements of Jet Energy Scale, Jet Energy Resolution and Jet Mass in ATLAS}",
    howpublished   = "{ATLAS-CONF-2015-037}",
    url            = "https://cds.cern.ch/record/2044941",
    year           = "2015",
}

@Booklet{ATL-PHYS-PUB-2016-020,
    author         = "{ATLAS Collaboration}",
    title          = "{Studies on top-quark Monte Carlo modelling for Top2016}",
    howpublished   = "{ATL-PHYS-PUB-2016-020}",
    url            = "https://cds.cern.ch/record/2216168",
    year           = "2016",
}

@Article{:2008zzm,
	author         = "{ATLAS Collaboration}",
	title     = "{The ATLAS Experiment at the CERN Large Hadron Collider}",
	journal   = "JINST",
	volume    = "3",
	year      = "2008",
	pages     = "S08003",
	doi       = "10.1088/1748-0221/3/08/S08003",
	SLACcitation  = "%%CITATION = JINST,3,S08003;%%"
}

@article{Aad:2014zka,
      author         = "{ATLAS Collaboration}",
      title          = "{Measurements of normalized differential cross sections
                        for $t\bar{t}$ production in pp collisions at
                        $\sqrt{s}=7$~TeV using the ATLAS detector}",
      journal        = "Phys. Rev. D",
      volume         = "90",
      year           = "2014",
      number         = "7",
      pages          = "072004",
      doi            = "10.1103/PhysRevD.90.072004",
      eprint         = "1407.0371",
      archivePrefix  = "arXiv",
      primaryClass   = "hep-ex",
      reportNumber   = "CERN-PH-EP-2014-099",
      SLACcitation   = "%%CITATION = ARXIV:1407.0371;%%"
}

@article{Aad:2015ydr,
      author         = "{ATLAS Collaboration}",
      title          = "{Performance of b-jet identification in the ATLAS experiment}",
      journal        = "JINST",
      volume         = "11",
      year           = "2016",
      number         = "04",
      pages          = "P04008",
      doi            = "10.1088/1748-0221/11/04/P04008",
      eprint         = "1512.01094",
      archivePrefix  = "arXiv",
      primaryClass   = "hep-ex",
      reportNumber   = "CERN-PH-EP-2015-216",
      SLACcitation   = "%%CITATION = ARXIV:1512.01094;%%"
}

@article{Aaboud:2018xwy,
      author         = "{ATLAS Collaboration}",
      title          = "{Measurements of b-jet tagging efficiency with the ATLAS
                        detector using $ t\overline{t} $ events at $ \sqrt{s}=13 $
                        TeV}",
      journal        = "JHEP",
      volume         = "08",
      year           = "2018",
      pages          = "089",
      doi            = "10.1007/JHEP08(2018)089",
      eprint         = "1805.01845",
      archivePrefix  = "arXiv",
      primaryClass   = "hep-ex",
      reportNumber   = "CERN-EP-2018-047",
      SLACcitation   = "%%CITATION = ARXIV:1805.01845;%%"
}

@Booklet{ATL-PHYS-PUB-2016-012,
    author         = "{ATLAS Collaboration}",
    title          = "{Optimisation of the ATLAS $b$-tagging performance for the 2016 LHC Run}",
    howpublished   = "{ATL-PHYS-PUB-2016-012}",
    url            = "https://cds.cern.ch/record/2160731",
    year           = "2016",
}

@article{Aaboud:2019ynx,
      author         = "{ATLAS Collaboration}",
      title          = "{Electron reconstruction and identification in the ATLAS
                        experiment using the 2015 and 2016 LHC proton-proton
                        collision data at $\sqrt{s}$ = 13 TeV}",
      journal        = "Submitted to: Eur. Phys. J. C",
      year           = "2019",
      eprint         = "1902.04655",
      archivePrefix  = "arXiv",
      primaryClass   = "physics.ins-det",
      reportNumber   = "CERN-EP-2018-273",
      SLACcitation   = "%%CITATION = ARXIV:1902.04655;%%"
}

@article{Aad:2016jkr,
      author         = "{ATLAS Collaboration}",
      title          = "{Muon reconstruction performance of the ATLAS detector in
                        proton–proton collision data at $\sqrt{s}$ =13 TeV}",
      journal        = "Eur. Phys. J. C",
      volume         = "76",
      year           = "2016",
      number         = "5",
      pages          = "292",
      doi            = "10.1140/epjc/s10052-016-4120-y",
      eprint         = "1603.05598",
      archivePrefix  = "arXiv",
      primaryClass   = "hep-ex",
      reportNumber   = "CERN-EP-2016-033",
      SLACcitation   = "%%CITATION = ARXIV:1603.05598;%%"
}

@Article{:2010wqa,
     author         = "{ATLAS Collaboration}",
     title     = "{The {ATLAS} Simulation Infrastructure}",
     journal   = "Eur. Phys. J. C",
     volume    = "70",
     year      = "2010",
     pages     = "823-874",
     eprint    = "1005.4568",
     archivePrefix = "arXiv",
     primaryClass  =  "physics.ins-det",
     doi       = "10.1140/epjc/s10052-010-1429-9",
     SLACcitation  = "%%CITATION = 1005.4568;%%"
}

@article{Aaboud:2016leb,
      author         = "{ATLAS Collaboration}",
      title          = "{Performance of the ATLAS trigger system in 2015}",
      journal        = "Eur. Phys. J. C",
      volume         = "77",
      year           = "2017",
      number         = "5",
      pages          = "317",
      doi            = "10.1140/epjc/s10052-017-4852-3",
      eprint         = "1611.09661",
      archivePrefix  = "arXiv",
      primaryClass   = "hep-ex",
      reportNumber   = "CERN-EP-2016-241",
      SLACcitation   = "%%CITATION = ARXIV:1611.09661;%%"
}

@article{Aad:2015ina,
    author         = "{ATLAS Collaboration}",
      title          = "{Performance of pile-up mitigation techniques for jets in
                        $pp$ collisions at $\sqrt{s}=8$  TeV using the ATLAS
                        detector}",
      journal        = "Eur. Phys. J. C",
      volume         = "76",
      year           = "2016",
      number         = "11",
      pages          = "581",
      doi            = "10.1140/epjc/s10052-016-4395-z",
      eprint         = "1510.03823",
      archivePrefix  = "arXiv",
      primaryClass   = "hep-ex",
      reportNumber   = "CERN-PH-EP-2015-206",
      SLACcitation   = "%%CITATION = ARXIV:1510.03823;%%"
}

@article{Aaboud:2018kfi,
      author         = "{ATLAS Collaboration}",
      title          = "{In situ calibration of large-$R$ jet energy and mass in
                        13 TeV proton-proton collisions with the ATLAS detector}",
      journal        = "Eur. Phys. J. C",
      volume         = "79",
      year           = "2019",
      number         = "2",
      pages          = "135",
      doi            = "10.1140/epjc/s10052-019-6632-8",            
      eprint         = "1807.09477",
      archivePrefix  = "arXiv",
      primaryClass   = "hep-ex",
      reportNumber   = "CERN-EP-2018-191",
      SLACcitation   = "%%CITATION = ARXIV:1807.09477;%%"
}

@article{Aaboud:2018tkc,
       author         = "{ATLAS Collaboration}",
      title          = "{Performance of missing transverse momentum
                        reconstruction with the ATLAS detector using proton-proton
                        collisions at $\sqrt{s}$ = 13 TeV}",
      journal        = "Eur. Phys. J. C",
      volume         = "78",
      year           = "2018",
      number         = "11",
      pages          = "903",
      doi            = "10.1140/epjc/s10052-018-6288-9",
      eprint         = "1802.08168",
      archivePrefix  = "arXiv",
      primaryClass   = "hep-ex",
      reportNumber   = "CERN-EP-2017-274",
      SLACcitation   = "%%CITATION = ARXIV:1802.08168;%%"
}

@article{Aad:2015fna,
      author         = "{ATLAS Collaboration}",
      title          = "{A search for $ t\overline{t} $ resonances using
                        lepton-plus-jets events in proton-proton collisions at $
                        \sqrt{s}=8 $ TeV with the ATLAS detector}",
      journal        = "JHEP",
      volume         = "08",
      year           = "2015",
      pages          = "148",
      doi            = "10.1007/JHEP08(2015)148",
      eprint         = "1505.07018",
      archivePrefix  = "arXiv",
      primaryClass   = "hep-ex",
      reportNumber   = "CERN-PH-EP-2015-090",
      SLACcitation   = "%%CITATION = ARXIV:1505.07018;%%"
}

@article{Chatrchyan:2012mec,
      author         = "{CMS Collaboration}",
      title          = "{Shape, Transverse Size, and Charged Hadron Multiplicity
                        of Jets in pp Collisions at 7 TeV}",
      journal        = "JHEP",
      volume         = "06",
      year           = "2012",
      pages          = "160",
      doi            = "10.1007/JHEP06(2012)160",
      eprint         = "1204.3170",
      archivePrefix  = "arXiv",
      primaryClass   = "hep-ex",
      reportNumber   = "CMS-QCD-10-029, CERN-PH-EP-2012-079",
      SLACcitation   = "%%CITATION = ARXIV:1204.3170;%%"
}

@article{Chatrchyan:2013vbb,
      author         = "{CMS Collaboration}",
      title          = "{Studies of jet mass in dijet and W/Z + jet events}",
      journal        = "JHEP",
      volume         = "05",
      year           = "2013",
      pages          = "090",
      doi            = "10.1007/JHEP05(2013)090",
      eprint         = "1303.4811",
      archivePrefix  = "arXiv",
      primaryClass   = "hep-ex",
      reportNumber   = "CMS-SMP-12-019, CERN-PH-EP-2013-016",
      SLACcitation   = "%%CITATION = ARXIV:1303.4811;%%"
}

@article{Sirunyan:2018asm,
      author         = "{CMS Collaboration}",
      title          = "{Measurement of jet substructure observables in
                        $\mathrm{t\overline{t}}$ events from proton-proton
                        collisions at $\sqrt{s}=$ 13 TeV}",
      journal        = "Phys. Rev. D",
      volume         = "98",
      year           = "2018",
      number         = "9",
      pages          = "092014",
      doi            = "10.1103/PhysRevD.98.092014",
      eprint         = "1808.07340",
      archivePrefix  = "arXiv",
      primaryClass   = "hep-ex",
      reportNumber   = "CMS-TOP-17-013, CERN-EP-2018-214",
      SLACcitation   = "%%CITATION = ARXIV:1808.07340;%%"
}

@article{Sirunyan:2018xdh,
      author         = "{CMS Collaboration}",
      title          = "{Measurements of the differential jet cross section as a
                        function of the jet mass in dijet events from
                        proton-proton collisions at $\sqrt{s} =$ 13 TeV}",                      
      journal        = "JHEP",
      volume         = "11",
      year           = "2018",
      pages          = "113",
      doi            = "10.1007/JHEP11(2018)113",                  
      eprint         = "1807.05974",
      archivePrefix  = "arXiv",
      primaryClass   = "hep-ex",
      reportNumber   = "CMS-SMP-16-010, CERN-EP-2018-180",
      SLACcitation   = "%%CITATION = ARXIV:1807.05974;%%"
}

@article{Sirunyan:2017tyr,
      author         = "{CMS Collaboration}",
      title          = "{Measurements of jet charge with dijet events in pp
                        collisions at $\sqrt{s}=8$ TeV}",
      journal        = "JHEP",
      volume         = "10",
      year           = "2017",
      pages          = "131",
      doi            = "10.1007/JHEP10(2017)131",
      eprint         = "1706.05868",
      archivePrefix  = "arXiv",
      primaryClass   = "hep-ex",
      reportNumber   = "CMS-SMP-15-003, CERN-EP-2017-085",
      SLACcitation   = "%%CITATION = ARXIV:1706.05868;%%"
}

@article{Aaboud:2017fha,
        author         = "{ATLAS Collaboration}",
      title          = "{Measurements of top-quark pair differential
                        cross-sections in the lepton+jets channel in $pp$
                        collisions at   $\sqrt{s}=13$ TeV using the ATLAS
                        detector}",
      journal        = "JHEP",
      volume         = "11",
      year           = "2017",
      pages          = "191",
      doi            = "10.1007/JHEP11(2017)191",
      eprint         = "1708.00727",
      archivePrefix  = "arXiv",
      primaryClass   = "hep-ex",
      reportNumber   = "CERN-EP-2017-058",
      SLACcitation   = "%%CITATION = ARXIV:1708.00727;%%"
}

@article{Aaboud:2018eqg,
       author         = "{ATLAS Collaboration}",
      title          = "{Measurements of $t\bar{t}$ differential cross-sections
                        of highly boosted top quarks decaying to all-hadronic
                        final states in $pp$ collisions at $\sqrt{s}=13\,$ TeV
                        using the ATLAS detector}",
      journal        = "Phys. Rev. D",
      volume         = "98",
      year           = "2018",
      number         = "1",
      pages          = "012003",
      doi            = "10.1103/PhysRevD.98.012003",
      eprint         = "1801.02052",
      archivePrefix  = "arXiv",
      primaryClass   = "hep-ex",
      reportNumber   = "CERN-EP-2017-226",
      SLACcitation   = "%%CITATION = ARXIV:1801.02052;%%"
}

@article{Aaboud:2017mae,
      author         = "{ATLAS Collaboration}",
      title          = "{Top-quark mass measurement in the all-hadronic $
                        t\overline{t} $ decay channel at $ \sqrt{s}=8 $ TeV with
                        the ATLAS detector}",
      journal        = "JHEP",
      volume         = "09",
      year           = "2017",
      pages          = "118",
      doi            = "10.1007/JHEP09(2017)118",
      eprint         = "1702.07546",
      archivePrefix  = "arXiv",
      primaryClass   = "hep-ex",
      reportNumber   = "CERN-EP-2016-264",
      SLACcitation   = "%%CITATION = ARXIV:1702.07546;%%"
}

@article{Aaboud:2019roo,
     author         = "{ATLAS Collaboration}",
      title          = "{Search for heavy particles decaying into a top-quark
                        pair in the fully hadronic final state in $pp$ collisions
                        at $\sqrt{s} =13$ TeV with the ATLAS detector}",
      journal        = "Phys. Rev. D",
      volume         = "99",
      year           = "2019",
      number         = "9",
      pages          = "092004",
      doi            = "10.1103/PhysRevD.99.092004",
      eprint         = "1902.10077",
      archivePrefix  = "arXiv",
      primaryClass   = "hep-ex",
      reportNumber   = "CERN-EP-2018-350",
      SLACcitation   = "%%CITATION = ARXIV:1902.10077;%%"
}

@article{Aaboud:2018juj,
      author         = "{ATLAS Collaboration}",
      title          = "{Search for $W' \rightarrow tb$ decays in the hadronic
                        final state using pp collisions at $\sqrt{s}=13$ TeV with
                        the ATLAS detector}",
      journal        = "Phys. Lett. V",
      volume         = "781",
      year           = "2018",
      pages          = "327-348",
      doi            = "10.1016/j.physletb.2018.03.036",
      eprint         = "1801.07893",
      archivePrefix  = "arXiv",
      primaryClass   = "hep-ex",
      reportNumber   = "CERN-EP-2017-340",
      SLACcitation   = "%%CITATION = ARXIV:1801.07893;%%"
}

@article{Aaboud:2017rss,
          author         = "{ATLAS Collaboration}",
      title          = "{Search for the standard model Higgs boson produced in
                        association with top quarks and decaying into a $b\bar{b}$
                        pair in $pp$ collisions at $\sqrt{s}$ = 13  TeV with the
                        ATLAS detector}",
      journal        = "Phys. Rev. D",
      volume         = "97",
      year           = "2018",
      number         = "7",
      pages          = "072016",
      doi            = "10.1103/PhysRevD.97.072016",
      eprint         = "1712.08895",
      archivePrefix  = "arXiv",
      primaryClass   = "hep-ex",
      reportNumber   = "CERN-EP-2017-291",
      SLACcitation   = "%%CITATION = ARXIV:1712.08895;%%"
}

@article{Aaboud:2018zhk,
       author         = "{ATLAS Collaboration}",
      title          = "{Observation of $H \rightarrow b\bar{b}$ decays and $VH$
                        production with the ATLAS detector}",
      journal        = "Phys. Lett. B",
      volume         = "786",
      year           = "2018",
      pages          = "59-86",
      doi            = "10.1016/j.physletb.2018.09.013",
      eprint         = "1808.08238",
      archivePrefix  = "arXiv",
      primaryClass   = "hep-ex",
      reportNumber   = "CERN-EP-2018-215",
      SLACcitation   = "%%CITATION = ARXIV:1808.08238;%%"
}

@article{Aaboud:2018knk,
      author         = "{ATLAS Collaboration}",
      title          = "{Search for pair production of Higgs bosons in the
                        $b\bar{b}b\bar{b}$ final state using proton-proton
                        collisions at $\sqrt{s} = 13$ TeV with the ATLAS
                        detector}",
      journal        = "JHEP",
      volume         = "01",
      year           = "2019",
      pages          = "030",
      doi            = "10.1007/JHEP01(2019)030",
      eprint         = "1804.06174",
      archivePrefix  = "arXiv",
      primaryClass   = "hep-ex",
      reportNumber   = "CERN-EP-2018-029",
      SLACcitation   = "%%CITATION = ARXIV:1804.06174;%%"
}

@article{Aaboud:2018zhh,
     author         = "{ATLAS Collaboration}",
      title          = "{Search for Higgs boson pair production in the
                        $b\bar{b}WW^{*}$ decay mode at $\sqrt{s}=13$ TeV with the
                        ATLAS detector}",
      journal        = "JHEP",
      volume         = "04",
      year           = "2019",
      pages          = "092",
      doi            = "10.1007/JHEP04(2019)092",
      eprint         = "1811.04671",
      archivePrefix  = "arXiv",
      primaryClass   = "hep-ex",
      reportNumber   = "CERN-EP-2018-237",
      SLACcitation   = "%%CITATION = ARXIV:1811.04671;%%"
}

@article{Gras:2017jty,
      author         = "Gras, Philippe and Höche, Stefan and Kar, Deepak and
                        Larkoski, Andrew and Lönnblad, Leif and Plätzer, Simon
                        and Siódmok, Andrzej and Skands, Peter and Soyez, Gregory
                        and Thaler, Jesse",
      title          = "{Systematics of quark/gluon tagging}",
      journal        = "JHEP",
      volume         = "07",
      year           = "2017",
      pages          = "091",
      doi            = "10.1007/JHEP07(2017)091",
      eprint         = "1704.03878",
      archivePrefix  = "arXiv",
      primaryClass   = "hep-ph",
      reportNumber   = "MIT-CTP-4885, COEPP-MN-17-2, MCNET-17-04",
      SLACcitation   = "%%CITATION = ARXIV:1704.03878;%%"
}

@misc{hepdata,
  howpublished = {\url{https://doi.org/10.17182/hepdata.89324}},
}

@Booklet{ATL-GEN-PUB-2016-002,
    author         = "{ATLAS Collaboration}",
    title          = "{ATLAS Computing Acknowledgements 2016--2017}",
    howpublished   = "{ATL-GEN-PUB-2016-002}",
    url            = "https://cds.cern.ch/record/2202407",
}

@Booklet{ATL-PHYS-PUB-2011-014,
    author         = "{ATLAS Collaboration}",
    title          = "{Further ATLAS tunes of \textsc{Pythia} 6 and Pythia 8}",
    howpublished   = "{ATL-PHYS-PUB-2011-014}",
    url            = "https://cds.cern.ch/record/1400677",
    year           = "2011",
}

\clearpage
 
\begin{flushleft}
{\Large The ATLAS Collaboration}

\bigskip

M.~Aaboud$^\textrm{\scriptsize 35d}$,    
G.~Aad$^\textrm{\scriptsize 101}$,    
B.~Abbott$^\textrm{\scriptsize 128}$,    
D.C.~Abbott$^\textrm{\scriptsize 102}$,    
O.~Abdinov$^\textrm{\scriptsize 13,*}$,    
A.~Abed~Abud$^\textrm{\scriptsize 70a,70b}$,    
D.K.~Abhayasinghe$^\textrm{\scriptsize 93}$,    
S.H.~Abidi$^\textrm{\scriptsize 167}$,    
O.S.~AbouZeid$^\textrm{\scriptsize 40}$,    
N.L.~Abraham$^\textrm{\scriptsize 156}$
% ...
\bigskip
\\

% ...
$^{13}$Institute of Physics, Azerbaijan Academy of Sciences, Baku; Azerbaijan.\\
% ...
$^{35}$$^{(a)}$Facult\'e des Sciences Ain Chock, R\'eseau Universitaire de Physique des Hautes Energies - Universit\'e Hassan II, Casablanca;$^{(b)}$Facult\'{e} des Sciences, Universit\'{e} Ibn-Tofail, K\'{e}nitra;$^{(c)}$Facult\'e des Sciences Semlalia, Universit\'e Cadi Ayyad, LPHEA-Marrakech;$^{(d)}$Facult\'e des Sciences, Universit\'e Mohamed Premier and LPTPM, Oujda;$^{(e)}$Facult\'e des sciences, Universit\'e Mohammed V, Rabat; Morocco.\\
% ...
$^{40}$Niels Bohr Institute, University of Copenhagen, Copenhagen; Denmark.\\
% ...
$^{70}$$^{(a)}$INFN Sezione di Pavia;$^{(b)}$Dipartimento di Fisica, Universit\`a di Pavia, Pavia; Italy.\\
% ...
$^{93}$Department of Physics, Royal Holloway University of London, Egham; United Kingdom.\\
% ...
$^{101}$CPPM, Aix-Marseille Universit\'e, CNRS/IN2P3, Marseille; France.\\
$^{102}$Department of Physics, University of Massachusetts, Amherst MA; United States of America.\\
% ...
$^{128}$Homer L. Dodge Department of Physics and Astronomy, University of Oklahoma, Norman OK; United States of America.\\
% ...
$^{156}$Department of Physics and Astronomy, University of Sussex, Brighton; United Kingdom.\\
% ...
$^{167}$Department of Physics, University of Toronto, Toronto ON; Canada.\\
% ...
$^{a}$ Also at Borough of Manhattan Community College, City University of New York, New York NY; United States of America.\\
$^{b}$ Also at Centre for High Performance Computing, CSIR Campus, Rosebank, Cape Town; South Africa.\\
$^{c}$ Also at CERN, Geneva; Switzerland.\\
$^{d}$ Also at CPPM, Aix-Marseille Universit\'e, CNRS/IN2P3, Marseille; France.\\
$^{e}$ Also at D\'epartement de Physique Nucl\'eaire et Corpusculaire, Universit\'e de Gen\`eve, Gen\`eve; Switzerland.\\
% ...
$^{*}$ Deceased

\end{flushleft}

 
\end{document}